\documentclass[aps, prd, twocolumn, showpacs, superscriptaddress,
               longbibliography, floatfix ]{revtex4-2}

\usepackage[utf8]{inputenc}
\usepackage[T1]{fontenc}
\usepackage{float}
\usepackage{graphicx}
\usepackage{amsmath}
\usepackage{amssymb}
\usepackage{dcolumn}
\usepackage{bm}
\usepackage{url}
\usepackage{subcaption}
\usepackage{ragged2e}
\usepackage[usenames]{color}
\usepackage{marginnote}
\usepackage{xcolor}
\usepackage{soul}
\usepackage[normalem]{ulem}

\DeclareCaptionJustification{justified}{\justifying}
\begin{document}
\setstcolor{red}
\title{Analysis of the kinematic boundaries of the quasi-elastic 
neutrino-nucleus cross section in the superscaling model with a relativistic
effective mass} 
\author{I. Ruiz Simo} 
\email{ruizsig@ugr.es}
\affiliation{Universidad de Granada and Instituto Interuniversitario
  Carlos I de F\'isica Te\'orica y Computacional, E-18071, Granada,
  Spain} 
\author{I. D. Kakorin}
\affiliation{Bogoliubov Laboratory of Theoretical Physics, \\
  Joint Institute for Nuclear Research, RU-141980 Dubna, Russia} 
\author{V. A. Naumov}
\affiliation{Bogoliubov Laboratory of Theoretical Physics, \\
	Joint Institute for Nuclear Research, RU-141980 Dubna, Russia} 
\author{K. S. Kuzmin}
\affiliation{Bogoliubov Laboratory of Theoretical Physics, \\
	Joint Institute for Nuclear Research, RU-141980 Dubna, Russia}
\affiliation{Institute for Theoretical and Experimental Physics, 
	RU-117259 Moscow, Russia}
\author{J. E. Amaro}
\affiliation{Universidad de Granada and Instituto Interuniversitario
  Carlos I de F\'isica Te\'orica y Computacional, E-18071, Granada,
  Spain}

\begin{abstract}
In this work we obtain the analytical expressions for the boundaries
of the charged current quasi-elastic double differential cross section
in terms of dimensionless energy and momentum transfers, for the
Relativistic Fermi Gas (RFG) and the Super-Scaling approach with
relativistic effective mass (SuSAM*) models, within the scaling
formalism.  In addition, we show that this double differential cross
section in the scaling formalism has very good properties to be
implemented in the Monte Carlo (MC) neutrino event generators,
particularly because its peak is almost flat with the (anti)neutrino
energy. This makes it especially well-suited for the event generation
by the acceptance-rejection method usually used in the neutrino
generators.  Finally, we analyze the total charged current
quasi-elastic (CCQE) cross section $\sigma(E_{\nu})$ for both models
and attribute the enhancement observed in the SuSAM* total cross
section to the high-momentum components which are present, in a
phenomenological way, in its scaling function, while these are absent
in the RFG model.
\end{abstract}

\maketitle

\section{Introduction}\label{sec:introduction}
The measurement of neutrino/antineutrino-nucleus cross sections is a
fundamental topic of research, not only in itself because it can
provide knowledge on the fundamental interaction and on the nuclear
properties and modeling, but also for its importance in other special
fields in particle physics such as the mixing of neutrino flavors, the
extraction of the CP-violating phase in the lepton sector and the
origin of the asymmetry between matter and antimatter in the Universe.
In particular, in the last years many reviews and works have been
dedicated to these topics
\cite{Alvarez-Ruso:2014bla,Balasi:2015dba,Mosel:2016cwa,
  Katori:2016yel,Alvarez-Ruso:2017oui,Benhar:2015wva,
  Giusti:2019cup,Amaro:2019zos,SajjadAthar:2020nvy,Coloma:2020nhf}.

The total integrated Charged Current Quasi-elastic (CCQE)
neutrino/antineutrino cross section $\sigma_\text{CCQE}(E_{\nu})$ is
an important quantity to be known for the neutrino scattering and
oscillation experiments
\cite{Kitagaki:1983px,Belikov:1983kg,Abe:2011ks,Abe:2014ugx,
  Abe:2014iza,Abe:2015oar,Abe:2015biq,Nakajima:2010fp,
  AlcarazAunion:2009ku,Aguilar-Arevalo:2018ylq,Aguilar-Arevalo:2013dva,
  Aguilar-Arevalo:2010zc,Aguilar-Arevalo:2007ab,
  Adamson:2014pgc,Carneiro:2019jds,Ruterbories:2018gub,
  Patrick:2018gvi,Wolcott:2015hda,
  Fiorentini:2013ezn,Fields:2013zhk,Abratenko:2020acr,Acciarri:2020lhp,
  Acciarri:2014isz,Adamson:2017gxd,Ankowski:2005jf}.  In particular,
the knowledge of this observable is crucial for choosing of CCQE
channel among others to generate appropriate final lepton event
kinematics in neutrino event generators, that usually use the
acceptance-rejection method to generate the events with a probability
distribution given by differential cross section.

In addition, the importance of a precise knowledge of the total CCQE
cross section and particularly its ratio between the electron and muon
neutrinos species is of great importance in order to reduce the
systematic uncertainties for the determination of the CP violating
phase in the lepton sector, as it has been shown in Refs.
\cite{Ankowski:2019yll,Abe:2019vii,*Abe:2019viiErratum,Acciarri:2015uup,
  Ankowski:2016jdd, Ankowski:2017yvm, Martini:2016eec,
  Nikolakopoulos:2019qcr}.
  
Our aim in this work is to perform a thorough study of the analytical
boundaries of the phase space of the CCQE double differential cross
section $\frac{d^2\sigma}{dT_{\mu}\,d\cos\theta_{\mu}}$ for the
relativistic Fermi gas (RFG) \cite{Alberico:1981xd,
  Smith:1972xh,*Smith:1972xhErratum, Moniz:1971mt, Moniz:1969sr,
  Kuzmin:2007kr} and Super-scaling with relativistic effective mass
(SuSAM*) models
\cite{Amaro:2015zja,*Amaro:2015zjaErratum,Amaro:2017pkd,
  Martinez-Consentino:2017ryk,RuizSimo:2018kdl,Amaro:2018xdi} within
the scaling formalism \cite{Alberico:1988bv,Barbaro:2003ie,
  Barbaro:1998gu,Day:1990mf,Caballero:2007tz}, where the boundaries
are easier to be obtained.  To this end, we will study the double
differential $\frac{d^2\sigma}{d\kappa\,d\lambda}$ CCQE cross section,
where $\kappa$ and $\lambda$ are the dimensionless momentum and energy
transfer variables in the scaling formalism. This new double
differential cross section has also the very good property, for the
generation of the final charged lepton kinematics in the MC event
generators, of an almost flat peak, i.e, very weak dependent on the
neutrino/antineutrino energy.  This important feature makes it
specially well-suited for the generation of these events by the
acceptance-rejection method. It seems that this fact was already known
by some scientists working in the implementation of theoretical models
in some MC event generators\ \cite{SanchezNieto:2021}, but it was not
familiar to us until this article was completed.

The paper is organized as follows: In sect.\ \ref{sec:formalism} we
review in brief the general formalism for the description of the CCQE
double differential cross section; in sect.\ \ref{RFG_model} we
perform a thorough discussion about the analytical boundaries of the
phase space in the RFG model, lately extended to the SuSAM* model in
sect.\ \ref{susam_model}. In sect.\ \ref{sec:results} we show our main
results for the double differential cross section and the integrated
total one, and finally, in sect.\ \ref{conclusions} we draw our
conclusions and outline our future plans or prospects related to the
conclusions of the present work.

\section{General formalism}\label{sec:formalism}
In this section, we are going to discuss in brief the elementary
ingredients to calculate the double differential CCQE
$\frac{d^2\sigma}{dT_{\mu}\,d\cos\theta_{\mu}}$ cross section and its
transformation into the easier to work, for our purposes within the
scaling formalism, $\frac{d^2\sigma}{d\kappa\,d\lambda}$ cross
section.  The expression for the first double differential cross
section is given by
\cite{Amaro:2004bs,Amaro:2005dn,Amaro:2015lga,RuizSimo:2018kdl}:
\begin{align}\label{d2sigma_dtmu_dcosmu}
\nonumber
\frac{d^2\sigma}{dT_{\mu}\,d\cos\theta_{\mu}}=&\
\frac{G^2_F\,\cos^2\theta_c}{4\pi}\,\frac{k^\prime}{E_{\nu}}\,v_0
\left(V_{CC} R_{CC} + 2 V_{CL} R_{CL} \right. \\
+&\ \left.V_{LL} R_{LL} +V_{T} R_{T} \pm
2 V_{T^\prime} R_{T^\prime} \right),
\end{align}
where $G_F=1.116\times 10^{-11}$ MeV$^{-2}$ is the Fermi coupling
constant, $\theta_c$ is the Cabibbo angle ($\cos\theta_c=0.975$),
$k^\prime$ is the value of the final charged lepton momentum,
$\vec{k}^\prime$, $E_{\nu}$ is the neutrino/antineutrino energy in the
lab frame, and $v_0=(E_{\nu} + T_{\mu} + m_{\mu})^2 - q^2$, with $q^2$
being the squared three-momentum transfer, $\vec{q}$, to the
nucleus \footnote{Notice that in Eq.\ \eqref{d2sigma_dtmu_dcosmu} we
have particularized the general expression for the scattering of muon
neutrinos/antineutrinos, but the expression is general for other kind
of neutrino species, just by changing the final lepton
mass.}. Finally, it is worth noting that the $\pm$ sign in the
$T^\prime$ contribution of Eq.\ \eqref{d2sigma_dtmu_dcosmu} applies
for neutrino and antineutrino CCQE scattering, respectively.

The other ingredients appearing in Eq.\ \eqref{d2sigma_dtmu_dcosmu}
are the lepton kinematic factors $V_{K}$ and the nuclear response
functions $R_{K}$, the last ones depending only on the energy and
momentum transfer from the leptons to the nucleus, $\omega$ and $q$,
respectively.  These factors come mainly from the contraction of the
lepton tensor with the hadron one, and each of them are suitable
combinations of the tensors in a frame where the $Z$-axis is defined
by the direction of the three-momentum transfer,
$\vec{q}=\vec{k}-\vec{k}^\prime$. Their explicit expressions can be
found, for instance, in Refs.
\cite{RuizSimo:2018kdl,Amaro:2015lga,Amaro:2005dn,Amaro:2004bs}, and
particularly in sect.\ IIIA and appendices B and C of the recent
review \cite{Amaro:2019zos}, where an exhaustive discussion and
derivation of the response functions and scaling in the RFG model are
given.

It is quite general that the nuclear response functions $R_K$ can be
written in factorized form as a product of an integrated
single-nucleon response ($U_K$ or $G_K$ in the nomenclature of Ref.
\cite{Amaro:2019zos}) times a scaling function which depends on the
nuclear model. Nonetheless, in other nuclear models different from
those discussed in this work, several different scaling functions can
appear for the different nuclear response functions. Examples of these
are the models for the description of the QE response in the inclusive
$(e,e^\prime)$ or $(\nu,\mu)$ scattering, where different scaling
functions appear for each one of the nuclear responses (see
Refs.\ \cite{Maieron:2001it, Gonzalez-Jimenez:2014eqa,
  Barbaro:1998gu}, just to cite a few of them).

However, in the two models discussed in this work, a single scaling
function appears as a common factor in all the nuclear response
functions $R_K$ and factorizes in the cross section given in
Eq.\ \eqref{d2sigma_dtmu_dcosmu}:
\begin{align}\label{d2sigma_dtmu_dcosmu_v2}
\nonumber
\frac{d^2\sigma}{dT_{\mu}\,d\cos\theta_{\mu}}=&\
\frac{G^2_F\,\cos^2\theta_c}{4\pi}\,\frac{k^\prime}{E_{\nu}}\,v_0
\left(V_{CC} U_{CC} + 2 V_{CL} U_{CL} \right. \\
+&\left.V_{LL} U_{LL} +V_{T} U_{T} \pm
2 V_{T^\prime} U_{T^\prime} \right)f_\text{scal}(\psi),
\end{align}
where $\psi$ is the scaling variable and it is, in general, a function
of $\omega$ and $q$. For instance, in the particular case of the RFG
model, its expression is given by
\begin{equation}\label{rfg_scalfunc}
f_\text{RFG}(\psi)=\frac34\left(
1-\psi^2\right) \; \theta(1-\psi^2),
\end{equation}
where $\theta(x)$ is the step function, while in the SuSAM* model its
expression is discussed in Sect. \ref{susam_model}.

However, the differential cross section of
Eq.\ \eqref{d2sigma_dtmu_dcosmu} is given with respect to the final
lepton kinematic variables, its kinetic energy $T_{\mu}$ and the
cosine of its scattering angle with respect to the incident neutrino
direction, $\theta_{\mu}$. In the scaling formalism it is not very
difficult to find the relevant boundaries where the differential cross
section of Eq.\ \eqref{d2sigma_dtmu_dcosmu} is different from zero (as
it will be shown in Sects.\ \ref{RFG_model} and \ref{susam_model}, and
appendices\ \ref{appendix-a} and \ref{appendix-b}), but using the
relevant scaling variables, namely, the dimensionless energy and
momentum transfers, $\lambda=\omega/(2m_N)$ and $\kappa=q/(2m_N)$,
with $m_N$ the nucleon mass.

Therefore, in order to inspect the behavior of the double differential
cross section along its phase space and efficiently integrate it to
obtain the total CCQE cross section, it is better to work with the
$\frac{d^2\sigma}{d\kappa\,d\lambda}$ cross section, which can be
obtained from that in Eq.\ \eqref{d2sigma_dtmu_dcosmu} using the
Jacobian transformation from $(T_{\mu},\cos\theta_{\mu})$ variables to
$(\kappa,\lambda)$ ones:
\begin{align*}
\nonumber
\frac{d^2\sigma}{d\kappa\,d\lambda}=&\ \left|
\frac{\partial(T_{\mu},\cos\theta_{\mu})}{\partial(\kappa,\lambda)}
\right| \frac{d^2\sigma}{dT_{\mu}\,d\cos\theta_{\mu}} \\
=&\ \frac{4\,
  m^2_N\, q}{E_{\nu}\, k^\prime}\;
\frac{d^2\sigma}{dT_{\mu}\,d\cos\theta_{\mu}},
\end{align*}
where the Jacobian has been calculated knowing the relationships
between both sets of independent variables,
\begin{eqnarray}
\lambda&=&\frac{E_{\nu}-T_{\mu}-m_{\mu}}{2m_N} \nonumber \\
\kappa&=&\dfrac{\sqrt{E^2_{\nu}+P^2_{\mu}-
2 E_{\nu} P_{\mu}\cos\theta_{\mu}}}{2m_N}, \nonumber
\end{eqnarray}
with $E_{\mu}=T_{\mu}+m_{\mu}$ and $P_{\mu}\equiv
k^\prime=\sqrt{E^2_{\mu}-m^2_{\mu}}$.

\section{RFG model case}\label{RFG_model}

\subsection{Analytical boundaries due to the scaling model}
\label{subsec: analytic-boundaries-RFG}

With the definitions of the dimensionless variables in the scaling
formalism, where the electroweak probe transfers energy $\omega$ and
momentum $q$ to the nucleus:
\begin{eqnarray}
 \lambda&=&\frac{\omega}{2m_N}, 
 \qquad \kappa=\frac{q}{2m_N} \label{lambda_def}\\
 \tau&=&\kappa^2-\lambda^2=
 \frac{Q^2}{4m^2_N}\ge0 \nonumber \\
 \eta_F&=&\frac{k_F}{m_N},
 \qquad \epsilon_F=
 \sqrt{1+\eta^2_F}\ge1 \label{etafermi_def}\\
 \psi&=&\sqrt{\frac{\epsilon_0-1}{\epsilon_F-1}}\;\text{sign}(\lambda-\tau)
 \label{psi_def},
 \end{eqnarray}
 where $\epsilon_0$ is defined as
 \begin{equation}\label{e0_definition}
  \epsilon_0=\max\left(\kappa
  \sqrt{1+\frac{1}{\tau}}-\lambda,\epsilon_F-2\lambda
  \right),
 \end{equation}
 and $k_F$ is the Fermi momentum of the nucleus. The definition of
 $\epsilon_0$ given in Eq.\ \eqref{e0_definition} represents the
 minimum energy of the initial nucleon, in units of the nucleon mass
 $m_N$, that can contribute to a quasi-elastic (QE) scattering event
 for given energy and momentum transfers $(\lambda,\kappa)$
 (cf. Eq.\ (C11) of Ref.\ \cite{Amaro:2019zos}).
  
  In the RFG model, see Eq.\ (\ref{rfg_scalfunc}), the scaling
  variable $\psi$ is restricted to lie between $-1$ and $+1$ to get a
  non-vanishing contribution to the cross section.  The scaling
  variable is zero when $\epsilon_0=1$, i.e, when $\lambda=\tau$ (see
  appendix\ \ref{appendix-a}) in the non Pauli blocking (NPB) region.
  This condition is equivalent to
    \begin{eqnarray}
   \tau&=&\lambda \Longleftrightarrow
   \kappa^2=\lambda^2+\lambda
   \Longleftrightarrow \kappa=
   \sqrt{\lambda(\lambda+1)},
   \label{psi0_curve}
  \end{eqnarray}
and hence Eq.\ \eqref{psi0_curve} corresponds to where the scaling
variable is always zero, and where the QE peak appears.  For this
reason, we call this curve in the $(\lambda,\kappa)$ plane as
$\kappa_\text{QE}(\lambda)= \sqrt{\lambda(\lambda+1)}$.

The boundaries of the RFG scaling variable ($-1$, $+1$) are reached
when $\epsilon_0=\epsilon_F$ as it follows from
Eq.~\eqref{psi_def}. Solving the equation $\epsilon_0=\kappa
\sqrt{1+1/\tau}-\lambda=\epsilon_F$ in the NPB region (corresponding
to $\kappa\ge\eta_F$) we get two different curves in the
$(\lambda,\kappa)$ plane. These curves are labelled as
$\kappa^\text{NPB}_{\pm}(\lambda)$, and along them the scaling
variable is always $\psi=\mp1$, respectively. For a more detailed
derivation the reader is referred to appendices\ \ref{subsec:easy} and
\ \ref{difficult-form-boundaries}.

\begin{figure}[htb]
\centering
\includegraphics[width=\linewidth]{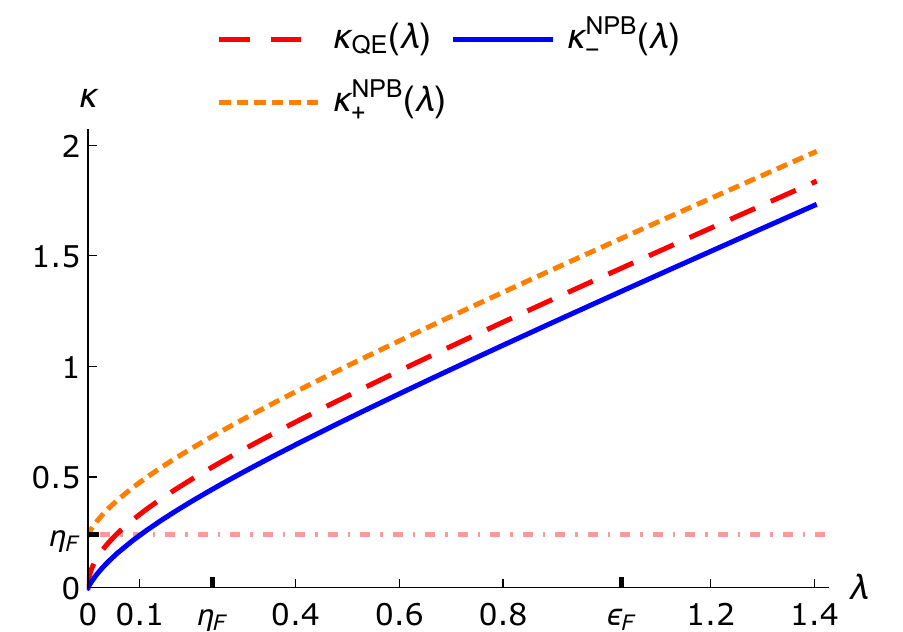}
\caption{Plot of the two limiting curves
  $\kappa^\text{NPB}_{\pm}(\lambda)$ as a function of $\lambda$ in the
  RFG model in the NPB region, i.e, for $\kappa\ge\eta_F$ (notice that
  for $\kappa<\eta_F$ we are entering in the Pauli blocking (PB)
  region).  In this figure, we have taken $\eta_F=0.239$.  The
  long-dashed curve corresponds to $\kappa_\text{QE}(\lambda)$.}
\label{figure1}
\end{figure}

The expressions of these two curves are given by
\begin{equation}\label{eq: npb_pm_kappa_curves}
\kappa^\text{NPB}_{\pm}(\lambda)=
\frac12\sqrt{(\epsilon_F+2\lambda)^2-1}\pm \frac{\eta_F}{2},
\end{equation}
and they are depicted in Fig.\ \ref{figure1}.

\begin{figure}[htb]
\centering
\includegraphics[width=\linewidth]{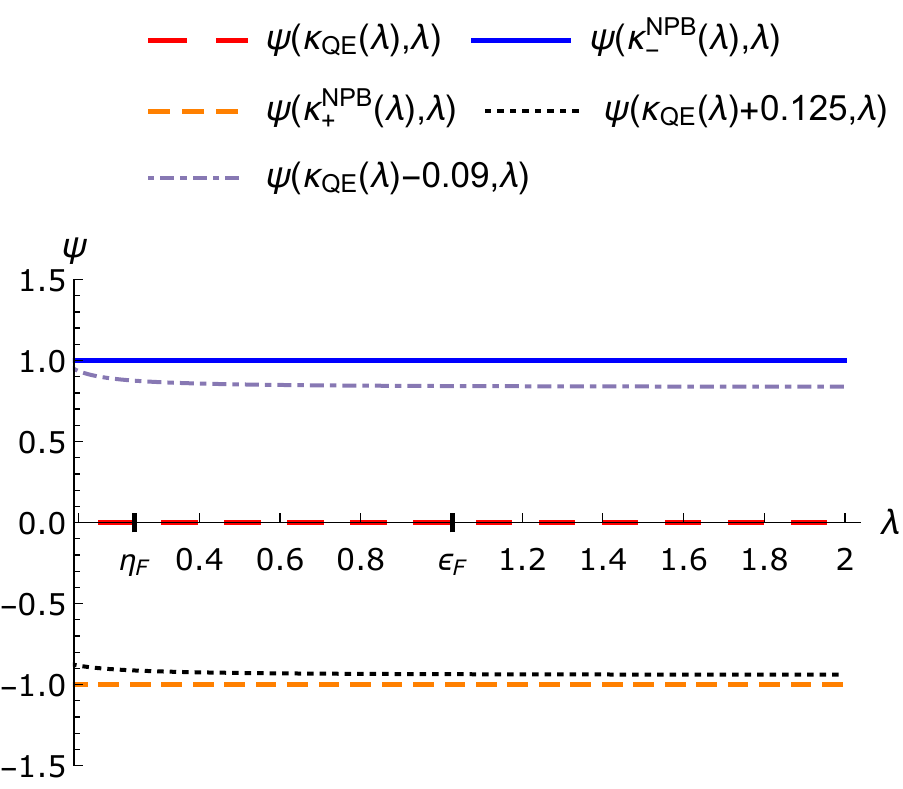}
\caption{Plot of the values taken by the scaling variable $\psi$,
  defined on Eq.\ \eqref{psi_def}, as a function of $\lambda$, along
  different curves in the $(\lambda,\kappa)$-plane.  The long-dashed
  line corresponds to the case $\lambda=\tau$, the solid and
  short-dashed lines correspond, respectively, to the lower and upper
  bounds of the RFG, given by
  $\kappa=\kappa^{\text{NPB}}_{\mp}(\lambda)$ in
  Fig.\ \ref{figure1}. Two additional curves are shown for comparison
  (see main text for discussion).}
\label{figure2}
\end{figure}

In Fig.\ \ref{figure2} we show the values taken by the scaling
variable $\psi\left( \kappa(\lambda),\lambda \right)$, as a function
of $\lambda$, along different curves in the NPB region. It can be seen
that the scaling variable is zero along the curve
$\kappa_\text{QE}(\lambda)$, i.e, the position of the QE peak. Along
the curves $\kappa^{\text{NPB}}_{\mp}(\lambda)$, the scaling variable
always takes its limiting values in the RFG model, $\psi=\pm 1$,
respectively. These values are shown by the solid and short-dashed
lines in Fig.\ \ref{figure2}, respectively.  Any curve lying in
between $\kappa^\text{NPB}_{-}(\lambda)$ and
$\kappa_\text{QE}(\lambda)$ in Fig.\ \ref{figure1}, as the dot-dashed
one indicates in Fig.\ \ref{figure2}, has a positive value for the
scaling variable; while those curves lying in between
$\kappa_{\text{QE}}(\lambda)$ and $\kappa^\text{NPB}_{+}(\lambda)$,
always have negative values for the scaling variable, as it can be
inspected from the dotted line of Fig.\ \ref{figure2}.

When $\kappa\ge\eta_F$ we are in the NPB region and $\epsilon_0$ is
\emph{always} equal to the first argument of the maximum function
given in Eq.\ (\ref{e0_definition}). However, when $\kappa<\eta_F$,
there are some regions in the $(\lambda,\kappa)$-plane where
$\epsilon_0$ is equal to the second argument of the maximum function
of Eq.\ \eqref{e0_definition} and we call this region as the Pauli
blocking (PB) region; while there are other regions where $\epsilon_0$
is still equal to the first argument of the maximum function.  For a
detailed derivation of the boundaries of these regions and other
proofs, the reader is referred to appendix\ \ref{PB-region}. Here we
only provide the final results.

According to the derivation discussed in appendix\ \ref{PB-region}, we
can conclude that the region where PB makes $\epsilon_0$ to be equal
to the second argument of Eq.\ \eqref{e0_definition},
$\epsilon_F-2\lambda$, corresponds to the region
$\kappa^\text{PB}_{-}(\lambda)\le \kappa \le
\kappa^\text{PB}_{+}(\lambda)$ in the range where
$0\le\lambda\le\lambda_{-}$, with
\begin{equation*}
 \kappa^\text{PB}_{\pm}(\lambda)=
 \sqrt{\frac{\rho \pm  
 \sqrt{\rho^2-4\left(\lambda
 \epsilon_F-\lambda^2
 \right)^2}}{2}},
\end{equation*}
where $\rho=2\lambda^2-2\lambda\epsilon_F+\eta^2_F$ and
$\lambda_{-}=\frac{\epsilon_F-1}{2}$.

\begin{figure}[htb]
\centering
\includegraphics[width=\linewidth]{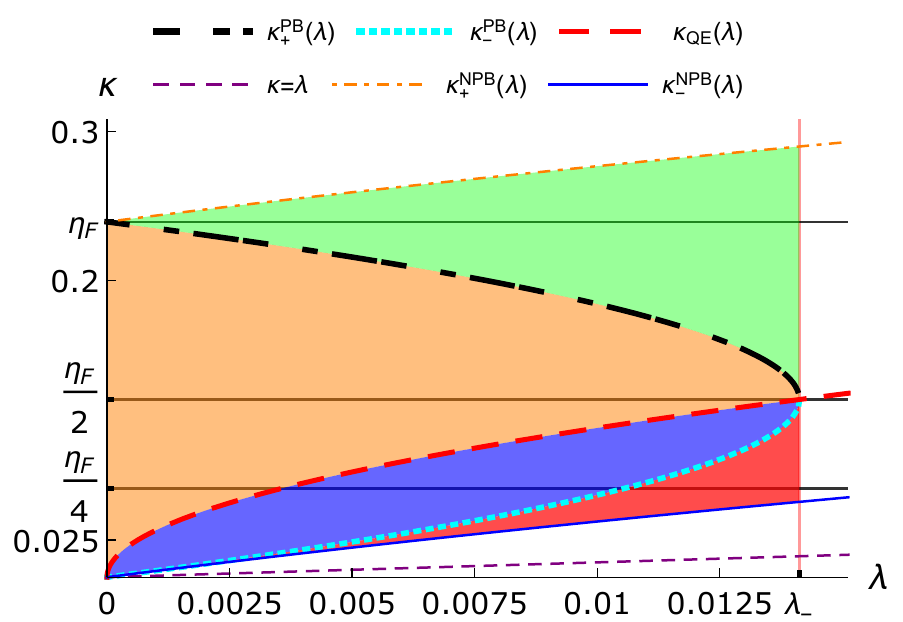}
\caption{Plot of the $(\lambda,\kappa)$-plane in the PB region, i.e,
  for $0\le\lambda\le\lambda_{-}$ and $0< \kappa < \eta_F$.  The
  three-fold dashed thick curve corresponds to the
  $\kappa^{\text{PB}}_{+}(\lambda)$ curve, while the dotted thick line
  is for the $\kappa^\text{PB}_{-}(\lambda)$ boundary. All the region
  surrounded by these two curves corresponds to the PB region.  We
  have also displayed the previously shown (in Fig.\ \ref{figure1})
  $\kappa^\text{NPB}_{\pm}(\lambda)$ curves as dot-dashed thin and
  solid lines, respectively. The curves $\kappa_{\text{QE}}(\lambda)$
  and $\kappa=\lambda$ are shown as long-dashed thick and short-dashed
  thin lines, respectively.}
  \label{figure3}
\end{figure}

In Fig.\ \ref{figure3} we show the different regions filled with
colors for $\kappa<\eta_F$, where PB effect occurs or not. The shaded
regions between $\kappa^{\text{PB}}_{+}(\lambda)$ and $\kappa=\eta_F$,
and between $\kappa^{\text{PB}}_{-}(\lambda)$ and
$\kappa^{\text{NPB}}_{-}(\lambda)$, respectively, correspond to those
zones of the allowed phase space of the RFG where there is no PB, i.e,
where $\epsilon_0=\kappa\sqrt{1+1/\tau}-\lambda$.  On the other hand,
the shaded regions between $\kappa^{\text{PB}}_{+}(\lambda)$ and
$\kappa_{\text{QE}}(\lambda)$, and between this last curve and the
dotted $\kappa^{\text{PB}}_{-}(\lambda)$ one, respectively, correspond
to zones where $\epsilon_0=\epsilon_F-2\lambda$, i.e, where there is
PB. It is worth noting that in this region, delimited by that kind of
inverted parabola formed by joining together the dotted and the
three-fold dashed curves of Fig.\ \ref{figure3}, the variable
$\epsilon_0$ and, consequently, the scaling variable $\psi$ only
depend on $\lambda$ and not at all on $\kappa$. The only important
issue to select the sign of $\psi$ is whether the points in these
regions are above the long-dashed thick line corresponding to the
curve $\kappa_\text{QE}(\lambda)$ (in whose case the scaling variable
is negative); or if on the contrary, the points are below this line,
in whose case the scaling variable is positive.

\begin{figure}[htb]
\centering
\includegraphics[width=\linewidth]{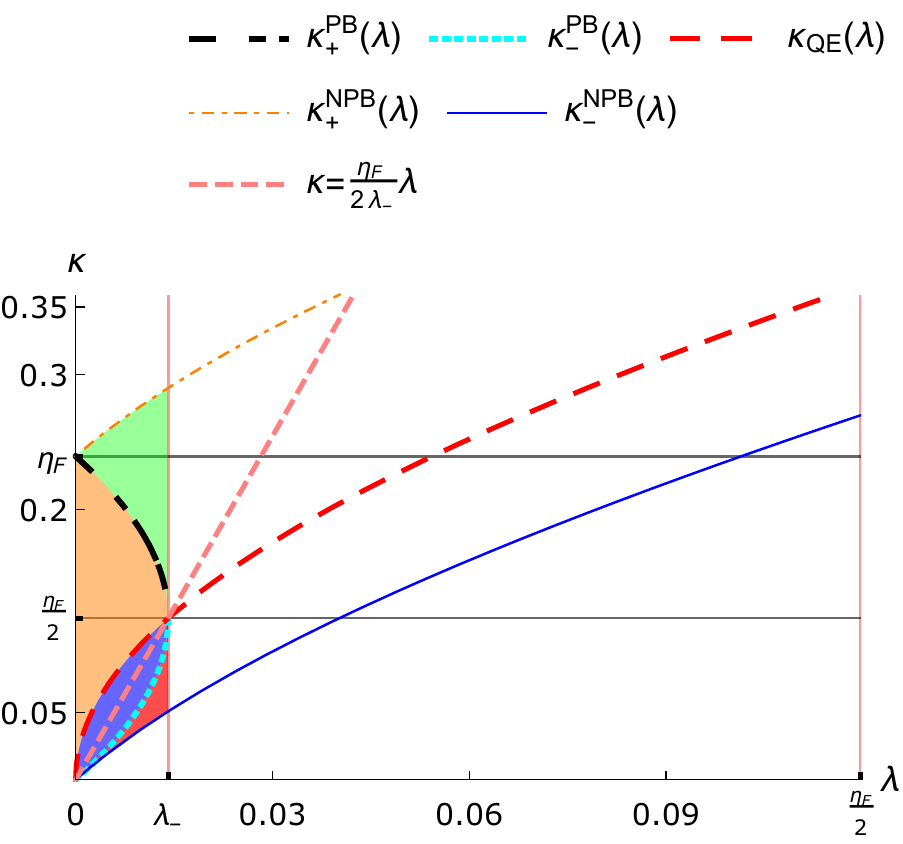}
\caption{Same plot as in Fig.\ \ref{figure3}, but highlighting the
  smallness of the region where PB plays a role. Notice that
  $\lambda_{-}\ll \eta_F$. Also shown a new straight line,
  $\kappa=\eta_F\,\lambda/(2\lambda_{-})$, in short-dashed style, that
  is entirely contained in the filled region between the long-dashed
  and dotted curves, for the range of values $0\le \lambda\le
  \lambda_{-}$. The purpose of this line and that corresponding to the
  horizontal line $\kappa=\eta_F/2$ will be clear in
  Fig.\ \ref{figure5}.}
\label{figure4}
\end{figure}

The purpose of Fig.\ \ref{figure4} is to highlight, in general, the
smallness of the region of the ($\lambda$, $\kappa$)-space where PB
plays a role. Note that $\lambda_{-}\ll \eta_F/2$ and that $\eta_F$ is
not visible in the $\lambda$-axis due to units, while in the vertical
axis it appears. Also notice that the horizontal straight line
$\kappa=\eta_F/2$ is entirely contained in the filled PB region above
the curve of the QE peak (as far as $0\le\lambda\le\lambda_{-}$), just
as it happens for the $\kappa^{\text{PB}}_{+}(\lambda)$ curve. The
same can be said for the straight line
$\kappa=\eta_F\,\lambda/(2\lambda_{-})$ and the
$\kappa^\text{PB}_{-}(\lambda)$ curve in the filled PB region just
below the QE peak position curve (long-dashed line).  The purposes of
these two straight lines will be clear in the following discussion.

\begin{figure}[htb]
\centering
\includegraphics[width=\linewidth]{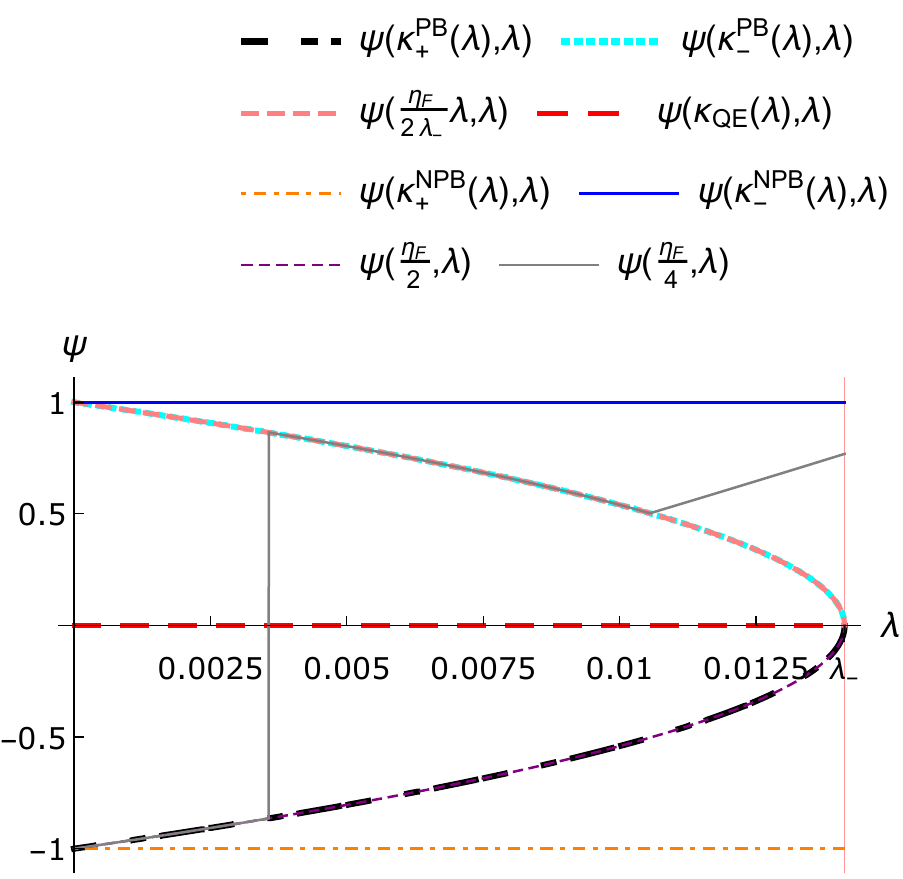}
\caption{Values taken by the scaling variable
  $\psi\left(\kappa(\lambda),\lambda\right)$ along the different
  curves shown in Figs.\ \ref{figure3} and \ref{figure4} in the PB
  region, i.e, when $0\le\lambda\le\lambda_{-}$.  For an exhaustive
  explanation see the main text.}
\label{figure5}
\end{figure}

In Fig.\ \ref{figure5} we show the values taken by the scaling
variable $\psi$ in the RFG model along different curves
$\kappa=\kappa(\lambda)$ in the $(\lambda,\kappa)$-plane in the region
where $0\le\lambda\le\lambda_{-}$. Of course, we have shown the
limiting boundaries $\psi=\pm 1$ given by the curves
$\kappa^{\text{NPB}}_{\mp}(\lambda)$, respectively.  They correspond
to the medium-thick solid and dot-dashed horizontal straight lines in
Fig.\ \ref{figure5}, respectively; and to the curves of the same
styles in Figs.\ \ref{figure3} and \ref{figure4}. Along the
long-dashed thick $\kappa_\text{QE}(\lambda)$ curve of
Figs.\ \ref{figure3} and \ref{figure4}, the scaling variable is equal
to zero in Fig.\ \ref{figure5} because this is the curve where
$\lambda=\tau$ and the sign function vanishes (see
Eq.\ \eqref{psi_def}).

The rest of curves shown, especially in Fig.\ \ref{figure4}, remains
inside the PB region for $0\le\lambda\le\lambda_{-}$. In this region,
remarked by the filled region between
$\kappa^{\text{PB}}_{+}(\lambda)$ and
$\kappa^{\text{PB}}_{-}(\lambda)$ curves in Figs.\ \ref{figure3} and
\ref{figure4}, the scaling variable
$\psi\left(\kappa(\lambda),\lambda\right)$ does not depend at all on
the $\kappa$ value taken by any point or curve inside the region,
except for the sign of $\psi$.  This can be viewed in different
forms. For instance, taking a look at the values taken by $\psi$ along
the curves $\kappa^{\text{PB}}_{+}(\lambda)$ (three-fold dashed thick
line) and along the straight line $\kappa=\eta_F/2$ (very short-dashed
thin curve in Fig.\ \ref{figure5}): both curves are totally inside the
filled PB region above the QE peak position curve of
Figs.\ \ref{figure3} and \ref{figure4}, however, their values of
$\kappa$ along the curves are totally different, and still the scaling
variable takes the same values in Fig.\ \ref{figure5}, i.e, it starts
equaling $\psi=-1$ for $\lambda=0$ because then
$\epsilon_0=\epsilon_F-2\lambda=\epsilon_F$ and both curves are above
the $\kappa=\kappa_\text{QE}(\lambda)$ curve, thus having negative
values for the scaling variable. Finally, for $\lambda=\lambda_{-}$,
the scaling variable is zero along both paths because it is the
intersection point with the $\kappa_\text{QE}(\lambda)$ curve (see
especially Fig.\ \ref{figure3}).

Something similar occurs along the paths defined by
$\kappa^{\text{PB}}_{-}(\lambda)$ (dotted thick line) and
$\kappa=\eta_F\,\lambda/(2\lambda_{-})$ (medium-dashed thick line),
but in this case for positive values of the scaling variable, because
in this case both paths are entirely in the filled PB region below the
$\kappa_\text{QE}(\lambda)$ curve of Fig.\ \ref{figure4}, thus in the
region of positive values for the scaling variable, as it can be seen
again in Fig.\ \ref{figure5}.
 
 The final example is a mixed case, a straight line $\kappa=\eta_F/4$
 (see Fig.\ \ref{figure3}) that starts in the filled PB region of
 negative values of the scaling variable $\psi$, passes across the
 $\kappa_{\text{QE}}(\lambda)$ curve, enters in the filled PB region
 below the $\kappa_{\text{QE}}(\lambda)$ line and, finally, it gets
 out of the PB region by entering entirely in the NPB region of
 positive values of $\psi$.  In this case (corresponding to the solid
 thin line in Fig.\ \ref{figure5}), the initial behavior of the
 scaling variable is the same as that corresponding to the other
 curves lying entirely in the PB region above the
 $\kappa_\text{QE}(\lambda)$ curve (negative values for $\psi$), until
 the point where $\kappa_{\text{QE}}(\lambda)=\eta_F/4$ (corresponding
 approximately to $\lambda\simeq 0.0036$), where the $\kappa=\eta_F/4$
 horizontal line enters in the PB region below the
 $\kappa_\text{QE}(\lambda)$ curve, and it suddenly changes the sign
 of $\psi$ along this crossing point, as it can be seen in
 Fig.\ \ref{figure5} as the vertical solid thin line. Now the values
 of $\psi$ roam along those of any curve entirely contained in the PB
 region below the $\kappa_\text{QE}(\lambda)$ curve (corresponding to
 positive values of $\psi$) until the new point where
 $\kappa^{\text{PB}}_{-}(\lambda)=\eta_F/4$ ($\lambda \simeq 0.011$),
 where the line $\kappa=\eta_F/4$ enters finally in the NPB region.
 In this last region, however, $\epsilon_0$ is no longer equal to
 $\epsilon_F -2\lambda$, but to $\kappa\sqrt{1+1/\tau}-\lambda$, and
 then, while still having positive values, the scaling variable now
 approaches $\psi=+1$, what will happen when
 $\kappa^\text{NPB}_{-}(\lambda)=\eta_F/4$ (corresponding to a value
 of $\lambda\simeq0.017 > \lambda_{-}$, and therefore out of the range
 of Fig.\ \ref{figure5}).

\subsection{Analytical boundaries 
coming from the lepton kinematics}\label{lepton-kinem-bound}

Up to now, nothing has been imposed from the lepton kinematics, but we
know that the final lepton scattering angle must be a physical one. We
discuss in this section that imposing constraints from lepton
kinematics (for a fixed initial neutrino/antineutrino energy) further
restricts the available phase space for the RFG model, for a detailed
derivation of some formulae relevant for this and future sections, the
reader is referred to appendix\ \ref{appendix-a4}. Here we only
provide the relevant results.

The lepton kinematics' restrictions come from the allowed maximum and
minimum final lepton energies (we will assume muon neutrinos, and so
the final lepton will be a muon \footnote{If the kind of neutrino is a
distinct one, one can use the formulae developed in
sects.\ \ref{lepton-kinem-bound} and appendix\ \ref{appendix-a4} and
change them for other final charged lepton masses.}) for a given
initial neutrino/antineutrino energy,
\begin{eqnarray}
\omega&=&E_{\nu}-E_{\mu} \Longleftrightarrow
E_{\mu}=E_{\nu}-2m_N \lambda \label{eq:energy_transfer}\\
q^2&=&(\vec{k}-\vec{k}^\prime)^2=E^2_{\nu}+k^{\prime\,2}-
2E_{\nu}k^\prime \cos\theta_{\mu},\label{eq:momentum_transfer2}
\end{eqnarray}
where $E_{\nu}$ is the initial neutrino energy, $\theta_{\mu}$ is the
muon scattering angle with respect to the direction of the incident
neutrino, and $k^\prime=\sqrt{E^2_{\mu}-m^2_{\mu}}$ is the final muon
momentum with energy $E_{\mu}$ and mass $m_{\mu}$.

The minimal muon energy is its mass and from this condition we can
obtain from Eq.\ \eqref{eq:energy_transfer} the, in principle, maximum
allowed value for $\lambda$,
\begin{equation}\label{eq:lambda_max}
\lambda_{\max}=\frac{E_{\nu}-m_{\mu}}{2m_N}=
\epsilon_{\nu}-\widetilde{m}_{\mu},
\end{equation}
where we have introduced ``reduced'' and dimensionless neutrino energy
and muon mass variables, defined as
\begin{equation*}
\epsilon_{\nu}\equiv \frac{E_{\nu}}{2m_N}, \qquad
\widetilde{m}_{\mu}\equiv \frac{m_{\mu}}{2m_N}.
\end{equation*}

From Eq.\ \eqref{eq:momentum_transfer2} we can write that the absolute
value of the cosine of the muon scattering angle must be lesser or
equal to 1:
\begin{align}\label{eq:inequality_cosine_muon}
\nonumber
&\ \left| \cos\theta_{\mu} \right| \leqslant 1
\Longleftrightarrow 
\left| \frac{E^2_{\nu}+k^{\prime\,2}-q^2}{2E_{\nu}k^\prime} 
\right| \leqslant 1 \\
&\ \Longleftrightarrow 
-2E_{\nu}k^\prime \leqslant E^2_{\nu}+k^{\prime\,2}-q^2
 \leqslant 2E_{\nu}k^\prime.
\end{align}
Notice that Eq.\ \eqref{eq:inequality_cosine_muon} gives two
additional inequalities for $\kappa$ in terms of $\lambda$ (the
variable $\lambda$ is hidden in $k^\prime$ via its dependence on
$E_{\mu}$ and the dependence of the latter on $\lambda$ through
Eq.\ \eqref{eq:energy_transfer}).  From the first inequality, and
using the ``reduced'' and dimensionless variables, we obtain:
\begin{align}
\nonumber
&\ q^2\leqslant\ (E_{\nu}+k^\prime)^2 \\
&\ \Longleftrightarrow
\frac{q}{2m_N}\leqslant \frac{E_{\nu}}{2m_N}+
\frac{\sqrt{(E_{\nu}-2m_N\lambda)^2-m^2_{\mu}}}{2m_N}\nonumber\\
&\ \Longleftrightarrow
\kappa \leqslant \epsilon_{\nu}+
\sqrt{(\epsilon_{\nu}-\lambda)^2-\widetilde{m}^2_{\mu}}
\equiv \kappa^\text{lepton}_{\max}(\lambda) \label{kappa_max_lepton}
\end{align}

Analogously with the other inequality of expression
\eqref{eq:inequality_cosine_muon}, we obtain the lower bound for
$\kappa$ constrained from the lepton kinematics alone:
\begin{equation}\label{kappa_min_lepton}
\kappa \geqslant \epsilon_{\nu}-
\sqrt{(\epsilon_{\nu}-\lambda)^2-\widetilde{m}^2_{\mu}}
\equiv \kappa^\text{lepton}_{\min}(\lambda).
\end{equation}
As the PB region (filled domains in Figs.\ \ref{figure3} and
\ref{figure4} surrounded by the $\kappa^{\text{PB}}_{+}(\lambda)$ and
$\kappa^{\text{PB}}_{-}(\lambda)$ curves) is always contained inside
the larger region bounded by $\kappa^\text{NPB}_{-}(\lambda) \leqslant
\kappa \leqslant \kappa^\text{NPB}_{+}(\lambda)$, and the only
difference between the PB region and the NPB one is the dependence of
the scaling variable with $\kappa$ and $\lambda$, the furthest
constrained phase space for the RFG model is given by
\begin{equation}\label{eq:rfg_boundary_exact}
\max\left(\kappa^\text{lepton}_{\min},
 \kappa^\text{NPB}_{-}\right) \leqslant 
 \kappa \leqslant \min \left(\kappa^\text{lepton}_{\max}, 
 \kappa^\text{NPB}_{+}\right),
\end{equation}
provided that the maximum on the left-hand side of expression
\eqref{eq:rfg_boundary_exact} is always smaller than the minimum on
the right-hand side of the same expression in the range of
$\lambda$-values ranging from $\lambda=0$ to $\lambda=\lambda_{\max}$,
where $\lambda_{\max}$ is given in Eq.\ \eqref{eq:lambda_max} for a
fixed neutrino/antineutrino energy.

The curves $\kappa^\text{lepton}_{\max}(\lambda)$ and
$\kappa^\text{lepton}_{\min}(\lambda)$, given in
Eqs.\ \eqref{kappa_max_lepton} and \eqref{kappa_min_lepton}
respectively, are monotonically decreasing and increasing with
$\lambda$, respectively. Both curves reach the same value when
$\lambda=\lambda_{\max}$, i.e, when
$\kappa^\text{lepton}_{\max}(\lambda_{\max})=
\kappa^\text{lepton}_{\min}(\lambda_{\max})=\epsilon_{\nu}$.

\begin{figure*}[htb]
\centering
\includegraphics[width=\linewidth, height=7cm]{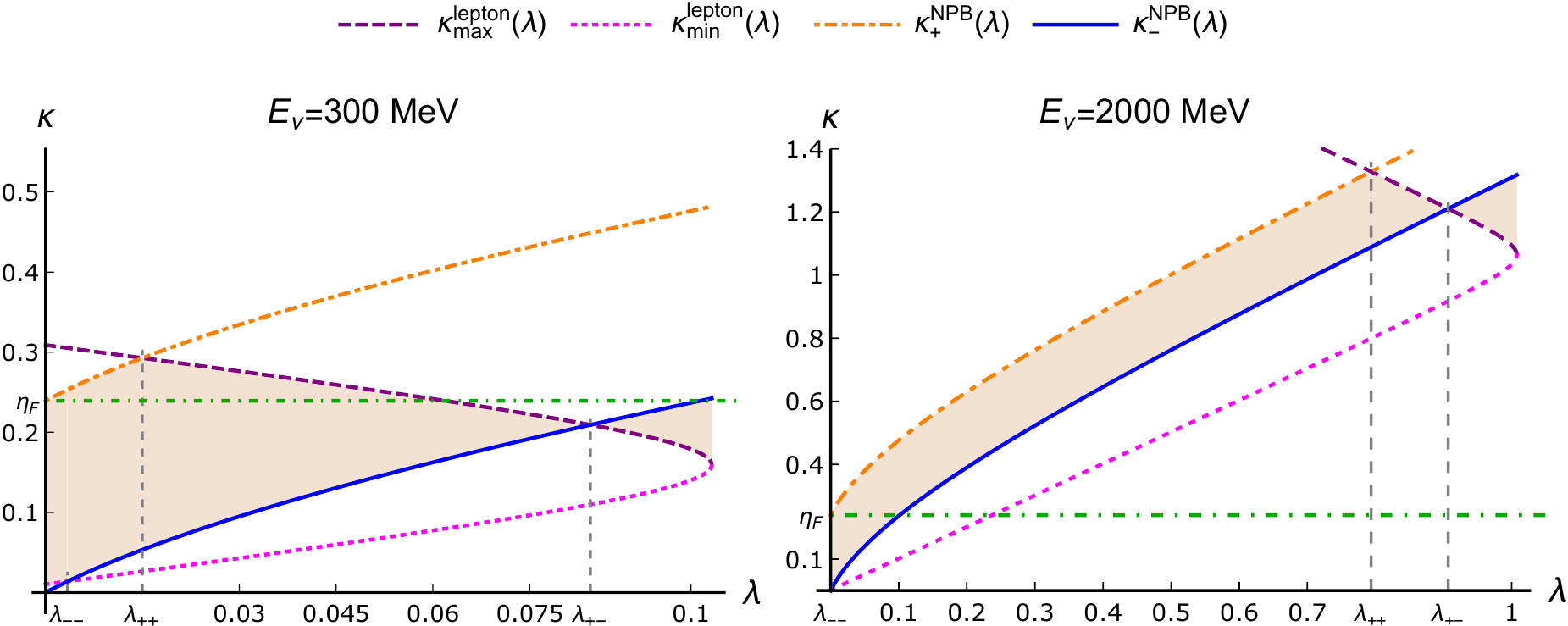}
\caption{Available phase space in $(\lambda,\kappa)$ variables in the
  RFG model for $E_{\nu}=300$ MeV (left panel) and for $E_{\nu}=2000$
  MeV (right panel), shown as the shaded regions for
  $\lambda\le\lambda_{+-}$. Also displayed are the different cut
  points between the curves constraining the lepton and nucleon
  kinematics, labeled as in Eqs.\ \eqref{lambda++solution} and
  \eqref{lambdapm-solution} of appendix\ \ref{appendix-a4}.  The value
  of $\eta_F$ has been taken as $0.239$, corresponding to a Fermi
  momentum of $k_F=225$ MeV/c. Also note that the curves
  $\kappa^\text{lepton}_{\max,\min}(\lambda)$ (dashed and dotted
  lines, respectively) are actually two different branches of the same
  curve.}\label{fig: phase-space-RFG}
\end{figure*}

In Fig.\ \ref{fig: phase-space-RFG} we show the phase space in the
$(\lambda,\kappa)$ variables for two different neutrino energies in
the RFG model. We have also shown the cutting points between the
different curves $\kappa^\text{lepton}_{\max,\min}(\lambda)$ and
$\kappa^\text{NPB}_{\pm}(\lambda)$, which constrain the lepton and
nuclear model kinematics in the RFG, respectively. Note that, because
we have shown the plots for the case of muon neutrinos, $m_{\mu}=106$
MeV/c${}^2$, for neutrino energies close to the muon mass the phase
space is mostly constrained by the lepton kinematics (left panel).
However, for higher neutrino energies (right panel), the available
phase space is almost entirely constrained by the nuclear model
kinematics in the RFG (thin solid and dot-dashed lines corresponding
to the limits of the RFG scaling function). In this latter case,
lepton kinematics plays a really minor role, except in the region of
the endpoint in $\lambda$, which corresponds to the largest energy
transfers to the nucleus (and consequently the least energy carried by
the muon), so one starts to see the effects of the muon mass as if one
were in the situation of the left panel.
 
 \section{SuSAM* model case}\label{susam_model}
\begin{figure}[htb]
\centering
 \includegraphics[width=\linewidth]{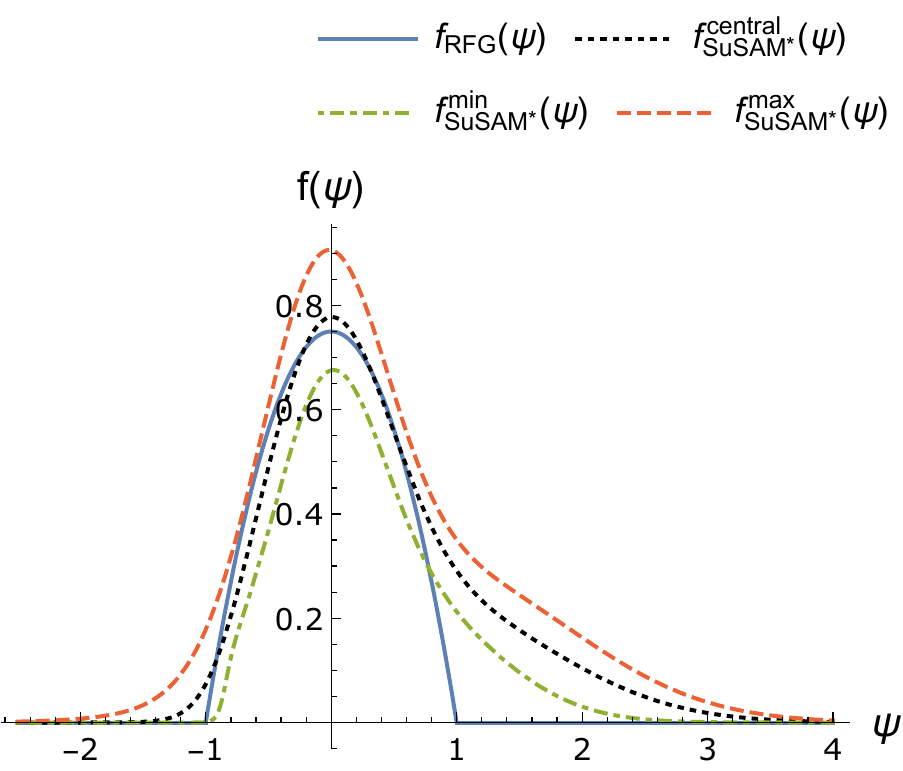}
\caption{Different scaling functions used in this article as a
  function of the scaling variable $\psi$. The well-known scaling
  function of the RFG model is shown in solid style, while the three
  SuSAM*-model scaling functions, extracted from a global fit to
  ``QE'' electron scattering data off nuclei in
  Ref.\ \cite{Amaro:2018xdi}, are shown as dotted, dot-dashed and
  dashed lines for the central, the lower and upper bounds,
  respectively. The scaling functions of the SuSAM* model plotted in
  this figure correspond to the parameters denoted as Band C in Table
  I of Ref.\ \cite{Amaro:2018xdi}.}
\label{figure7}
\end{figure}

The SuSAM* model is theoretically based in the Walecka (or
$\sigma-\omega$) model \cite{Walecka:1974qa, Serot:1984ey} for
relativistic nuclear matter.  The Walecka model was the first
relativistic, many-body, quantum-field theory model that exhibited
saturation in nuclear matter. The Relativistic Mean Field (RMF)
version of the model for nuclear matter has constant scalar and
time-like vector potentials, associated to the expectation values of
the scalar and time-like component of the vector fields. However, for
nuclear matter, the RMF version is exactly solvable and the dynamic
nucleon fields can be expanded as plane-wave solutions as in a free
theory, because the dynamics due to the scalar and vector potentials
is hidden in a shift of the nucleon mass and the energy. Thus, the
effect of the condensed value of the scalar field is to shift the mass
of the nucleon, reducing it. What we have done in previous works
\cite{Amaro:2015zja,*Amaro:2015zjaErratum, Amaro:2017pkd,
  Martinez-Consentino:2017ryk, RuizSimo:2018kdl, Amaro:2018xdi} is to
use this underlying well-founded theory to phenomenologically adjust
the relativistic effective mass for several nuclear species, assuming
that the ``QE" electron scattering data scale within an uncertainty
band, that has been also estimated; and that these ``QE" electron
scattering data can be selected from the whole inclusive data by means
of a density criterion.

In this model, the Pauli blocking is treated exactly as in the RFG,
i.e, by means of using Eq.\ \eqref{e0_definition}. This treatment can
be problematic and probably not excessively well-founded, but it is
the easiest way to incorporate it. For problems related to this
treatment of the Pauli blocking in the SuSAM* model, the reader is
referred to Sect.\ \ref{subsec:double_differential}, where a more
detailed discussion confronted with the results is given.

The binding energy in the SuSAM* model is simulated with the decrease
in the nucleon mass, i.e, by using the relativistic effective (shifted
by the condensed value of the scalar $\sigma$ field) nucleon mass,
although this mass (and the Fermi momentum) is fitted for each nucleus
for which we have inclusive $(e,e^\prime)$ data \cite{Amaro:2018xdi}
from where selecting ``QE" points using the density criterion
mentioned above.  This is not the first time such an attempt has been
done to describe QE electron scattering (see also
Refs.\ \cite{Wehrberger:1993zu,Rosenfelder:1980nd}), but as far as we
know, it is the first serious attempt to translate it to describe CCQE
neutrino scattering.

In Fig.\ \ref{figure7} we show the scaling functions of the two models
we discuss in this article. In solid line style the scaling function
of the RFG model is depicted, whose expression was given in
Eq.\ \eqref{rfg_scalfunc}. The other three scaling functions are those
of the SuSAM* model, in particular, those extracted in a global fit to
the world ``QE'' electron scattering data
\cite{Benhar:2006er,Benhar:2006wy} extracted out from the inclusive
data stored in the web site of Ref.\ \cite{Benhar:2006er}. These three
scaling functions (shown as central, min and max in
Fig.\ \ref{figure7}) were obtained in Ref.\ \cite{Amaro:2018xdi} after
a selection procedure based on the scaling hypothesis of the QE data,
and the lower and upper scaling functions correspond to the estimation
of the uncertainty or thickness of the super-scaling band where the
bulk of the QE data tend to accumulate.

The functional form of the scaling functions of the SuSAM* model
depicted in Fig.\ \ref{figure7} is:
\begin{equation}\label{scaling_function_susam}
f_{\text{SuSAM*}}(\psi)=\frac{a_3\; e^{-\frac{(\psi-a_1)^2}{2\,a^2_2}}
  +b_3\;
  e^{-\frac{(\psi-b_1)^2}{2\,b^2_2}}}{1+e^{-\frac{\psi-c_1}{c_2}}},
\end{equation}
where the parameters $a_i$, $b_i$ and $c_i$ can be found, for the
three different scaling functions of Fig.\ \ref{figure7}, in Table I
of Ref.\ \cite{Amaro:2018xdi}, corresponding to the set labeled as
Band C.  It is worth noting the asymmetry shown by the scaling
functions of the SuSAM* model, which have longer tails towards
positive values of the scaling variable $\psi$ than they have for
negative ones, in contrast with the symmetric RFG scaling function.

In sect.\ \ref{subsec: analytic-boundaries-RFG} we have discussed the
boundaries in the $(\lambda,\kappa)$-plane where the scaling variable
$\psi$ is between $-1$ and $+1$, and therefore, this comprises the
region where the RFG scaling function of Eq.\ \eqref{rfg_scalfunc} is
different from zero. Because of this, the five response functions
entering in the double differential (with respect to the final lepton
kinematic variables) CCQE neutrino/antineutrino-nucleus cross section
contribute only inside this boundary for the RFG model (see sections
II.A and II.B of Ref.\ \cite{RuizSimo:2018kdl}).

Now, to do the same for the SuSAM* scaling functions, the procedure
follows the lines sketched in
appendix\ \ref{difficult-form-boundaries}. In this appendix, we obtain
$\kappa^\text{NPB}_{\pm}(\lambda)$ for the RFG model by imposing
$\epsilon_0\equiv \kappa\sqrt{1+1/\tau}- \lambda=\epsilon_F$, which is
the equivalent condition to $\psi=\pm 1$.  All we have to do to extend
it for the SuSAM* model is to identify extreme values of the scaling
variable, namely $\psi_\text{extr}$, where we can safely affirm that
the SuSAM* scaling functions are negligible beyond these extreme
values, one on the left and the other on the right. Note that, given
the asymmetry of the SuSAM* scaling functions, these extreme values
are not going to be necessarily the same at the left and at the right
of the QE peak position.

Let us assume we are in the region where $\lambda>\tau$, and the sign
function appearing in Eq.\ \eqref{psi_def} is positive. If the
positive extreme value ($\psi_\text{extr}$) for the scaling variable
in the SuSAM* model is larger than $1$, this obviously means that
$\epsilon_0>\epsilon_F$. The limiting curve in the $(\lambda,\kappa)$
plane will be obtained when
$\psi\equiv\sqrt{\frac{\epsilon_0-1}{\epsilon_F-1}}=\psi_\text{extr}$,
where $\psi_\text{extr}$ has to be chosen properly as a large value
where the scaling function of the SuSAM* model can be totally
neglected beyond that value. This last equation is totally equivalent
to:
\begin{equation}\label{condition_e0_susam}
\epsilon_0\equiv \kappa\sqrt{1+\frac{1}{\tau}}-\lambda=1+ \left(
\epsilon_F - 1 \right) \psi^2_\text{extr}.
\end{equation}
The same equation would have been obtained for the case
$\lambda<\tau$, with the negative sign function in
Eq.\ \eqref{psi_def}, as it also happened in the RFG case. Note that
if we choose $\psi_\text{extr}=\pm1$, we recover the condition of the
RFG, $\epsilon_0=\epsilon_F$, as it should be.

To obtain the boundaries of the phase space in the $(\lambda,\kappa)$
plane for the SuSAM* model, note that Eq.\ \eqref{condition_e0_susam}
is the same as that for the RFG ($\epsilon_0=\epsilon_F$) with the
right-hand side replaced by $1+(\epsilon_F-1)\psi^2_\text{extr}$
instead of $\epsilon_F$. Consequently, we can take the
Eq.\ \eqref{upper_lower_boundary} of
appendix\ \ref{difficult-form-boundaries} and replace any appearance
of $\epsilon_F$ by the new $\epsilon^\prime_F\equiv
1+(\epsilon_F-1)\psi^2_\text{extr}$, where $\psi_\text{extr}$ does not
have necessarily to be equal for the $\kappa^\text{NPB}_{+}(\lambda)$
(corresponding to negative values of $\psi$) and for the
$\kappa^{\text{NPB}}_{-}(\lambda)$ (corresponding to positive values
of the scaling variable) functions, because of the asymmetry of the
scaling functions in the SuSAM* model.

The last important point that remains to be shown is that, for the
SuSAM* model, it is still true that $\kappa^\text{NPB}_{-}(0)=0$ and
that $\kappa^\text{NPB}_{+}(0)>\eta_F\equiv\sqrt{\epsilon^2_F-1}$
(these proofs are provided in appendix\ \ref{appendix-b}). This is
important because mainly the effect of choosing a wider scaling
function than that of the RFG is to broaden the available phase space
shown in Fig.\ \ref{fig: phase-space-RFG} between the thin dot-dashed
and solid curves, thus increasing the domain of integration in
$(\lambda,\kappa)$ space and obtaining a larger total cross section
for a fixed neutrino/antineutrino energy, $\sigma(E_{\nu})$.

The fact that $\kappa^\text{NPB}_{+}(0)>\eta_F$ is related to the high
momentum components, larger than the Fermi momentum, that real nuclei
have in its ground state. These high momentum components, mainly
produced by interaction and short-range correlations (SRC)
\cite{RuizSimo:2017tcb,Sick:1980ey,Ramos:1989hqs,Arrington:2011xs,
  Muther:1995zz,Giusti:1999sv,Stoitsov:1993zz,Alvioli:2012qa,
  Vanhalst:2012ur,VanCuyck:2016fab,Fomin:2017ydn,Amaro:1998rr,
  Mazziotta:2001qh,Weise:1972klw,RuizSimo:2016vsh,Wiringa:2013ala,
  Schiavilla:1987ziw}, are totally missing in the RFG model, but not
in the phenomenological scaling function of the SuSAM* model, which
has been obtained from a global fit to selected ``QE'' electron
scattering data from \emph{nuclei}.

It is also well known that the effects of SRC are mainly present in
the left tail of the scaling function
\cite{Berardo:2011fx,Tornow:1981tb, Day:2008zz}, i.e, for
$\psi_{\text{extr}}<-1$, which is precisely the left extreme $\psi$
value to be adequately chosen for the curve
$\kappa^\text{NPB}_{+}(\lambda)$, but they also show up in the right
tail of the scaling function. This connection between high momentum
components and scaling violations, and their effects in the total
integrated QE neutrino cross section are deferred for a forthcoming
study.

There are, in principle, other nuclear effects implicitly incorporated
in the SuSAM* scaling function, which has been obtained in the scaling
analysis of Ref.\ \cite{Amaro:2018xdi}, by fitting globally all the
inclusive $(e,e^\prime)$ scattering data of the nuclei present in the
database of \cite{Benhar:2006er}, but selecting only those ``QE''
points that scale within an uncertainty band, using a population
density criterion to keep or reject them. Therefore, we expect that,
besides SRC, other nuclear effects such as final-state interactions
(FSI), long range correlations (RPA), 1p-1h and 2p-2h MEC
contributions..., are also phenomenologically incorporated in the
scaling function of the SuSAM* model.

However, a recent scaling re-analysis of the inclusive $(e,e^\prime)$
scattering data off ${}^{12}$C has been carried out in
Ref.\ \cite{Martinez-Consentino:2021avr}, where the 2p-2h MEC have
been explicitly accounted for within the same model of RMF in nuclear
matter with relativistic effective mass and vector energy in which the
SuSAM* model is based. The authors have finally obtained essentially
the same scaling function and band than in the SuSAM* model used here,
even although the theoretical 2p-2h MEC contribution was subtracted
from the experimental data before carrying out the scaling
analysis. These findings at least seem to hint that the 2p-2h MEC
contributions present in the electron scattering data are not so
relevant to extract a QE scaling function.

\begin{figure*}[htb]
\centering
\includegraphics[width=\linewidth, height=9cm]{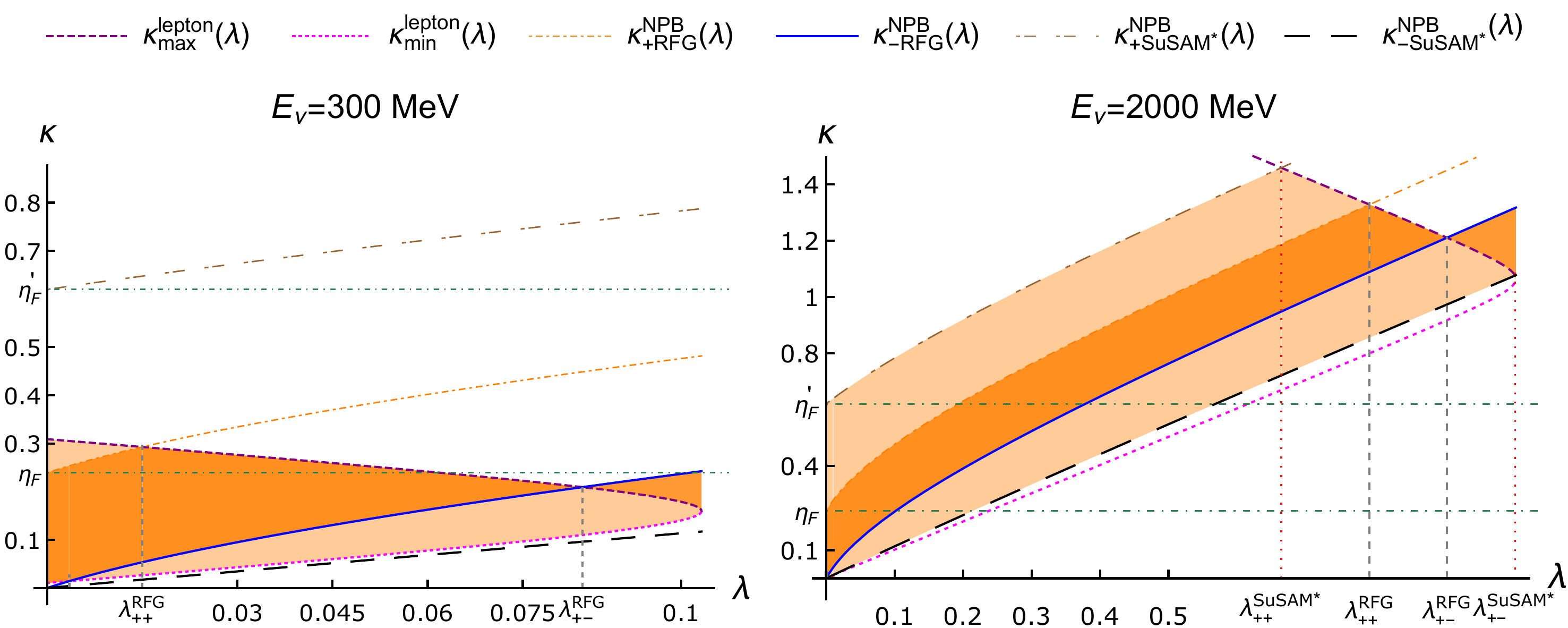}
\caption{Comparison of the available phase spaces in
  $(\lambda,\kappa)$ variables in the RFG (darker shade) and SuSAM*
  (paler shade) models for $E_{\nu}=300$ MeV (left panel) and for
  $E_{\nu}=2000$ MeV (right panel), shown as the shaded regions for
  $\lambda\le\lambda^{\text{model}}_{+-}$, in general.  Also displayed
  are the different cut points between the curves constraining the
  lepton and scaling model kinematics, labeled as in Fig.\ \ref{fig:
    phase-space-RFG} for the different models. Note that the available
  phase space gets much more enlarged in the SuSAM* model, due to the
  tails of the scaling function.}
  \label{fig: phase-space-RFG-susam}
\end{figure*}

In Fig.\ \ref{fig: phase-space-RFG-susam} we show the comparison
between the available phase space in the RFG (already shown in
Fig.\ \ref{fig: phase-space-RFG}) and SuSAM* scaling models, for the
same two neutrino energies as in Fig.\ \ref{fig: phase-space-RFG}.
The most remarkable difference is the enlargement of the phase space,
shown as the paler shade, in the SuSAM* model. This enlargement is
only attributable to the tails of the super-scaling function of the
SuSAM* model, which are absent in the RFG, because the curves
delimiting the boundaries from the lepton kinematics constraints are
the same, they do not depend at all on the scaling function.  The main
consequence of this enlargement of the phase space will be reflected
in a larger total integrated cross section. Of course, this increase
in the integrated cross section will depend on the values attained by
the double differential (with respect to $(\lambda,\kappa)$ variables)
CCQE cross section in the enlarged region. We can ensure that there is
going to be a clear increase, because in the regions outside the phase
space of the RFG, but close to its boundaries, the differential cross
section will still be substantial because the scaling function of the
SuSAM* is truly different from zero for scaling variables larger than
$1$ and lesser than $-1$, which corresponds to points lying in the
paler shaded regions. In any case, as expected, when the values of
$(\lambda,\kappa)$ are approaching the two-fold thin dashed curve
$\kappa^{\text{NPB}}_{+\text{SuSAM*}}(\lambda)$ and the long-dashed
thin line $\kappa^{\text{NPB}}_{-\text{SuSAM*}}(\lambda)$, their
contribution to the total cross section will be very small because in
these zones of the phase space the SuSAM* scaling function becomes
negligible.
 
It is also worth noting that the SuSAM* boundaries have been
calculated in Fig.\ \ref{fig: phase-space-RFG-susam} for
$\psi_\text{left}=-2.5$ and $\psi_\text{right}=6$ for the upper and
lower SuSAM* boundaries, corresponding to the two-fold dashed and
long-dashed thin curves, respectively. These extreme values of the
scaling variable for the left and right tails of the SuSAM*
super-scaling function have been chosen thinking in the values
attained by the central SuSAM* function (shown in Fig.\ \ref{figure7})
at them, which amount to roughly a factor $10^{-7}$ of the value of
this scaling function at the peak. Also, in Fig.\ \ref{fig:
  phase-space-RFG-susam} we show, in the $\kappa$-axis, the value
$\eta^\prime_F\approx0.62$, defined as
\begin{equation*}\label{etaprimaf}
\eta^\prime_F=\sqrt{\epsilon^{\prime\,2}_F-1}
\end{equation*}
with $\epsilon^\prime_F=1 + \left( \epsilon_F - 1 \right)
\psi^2_{\text{left}}$, as the horizontal thin dot-dashed line.

The analytical delimitation of the phase space boundaries in the
$(\lambda,\kappa)$ variables is important because when integrating
over the final lepton kinematics in order to obtain the total CCQE
integrated cross section as a function of the neutrino/antineutrino
energy, we can make the integration procedure as efficient as
possible, as we are evaluating the integrand only where it is
different from zero. This is particularly important when the neutrino
energy is really huge, $E_{\nu}\sim50 - 100$ GeV, because then the
contribution of the QE peak is concentrated at small values of the
energy transfer $\omega$ (if compared with the neutrino energy) and at
very forward angles. As for huge neutrino energies the allowed
interval in $\omega$ is also huge, it is convenient to constrain as
much as possible the angular interval (related to $\kappa$) where
truly integrating.
  
\section{Results}\label{sec:results}
In this section, we show the results for the CCQE double differential
$\frac{d^2\sigma}{dT_{\mu}\,d\cos\theta_\mu}$ and
$\frac{d^2\sigma}{d\kappa\,d\lambda}$ neutrino and antineutrino cross
sections for the RFG and SuSAM* models, as well as the fully
integrated total cross sections in both models.

\subsection{Double differential cross sections}
\label{subsec:double_differential}

\begin{figure*}[htb]
\centering
\begin{subfigure}[t]{0.5\linewidth}
\centering
\includegraphics[width=\linewidth, height=7.5cm]{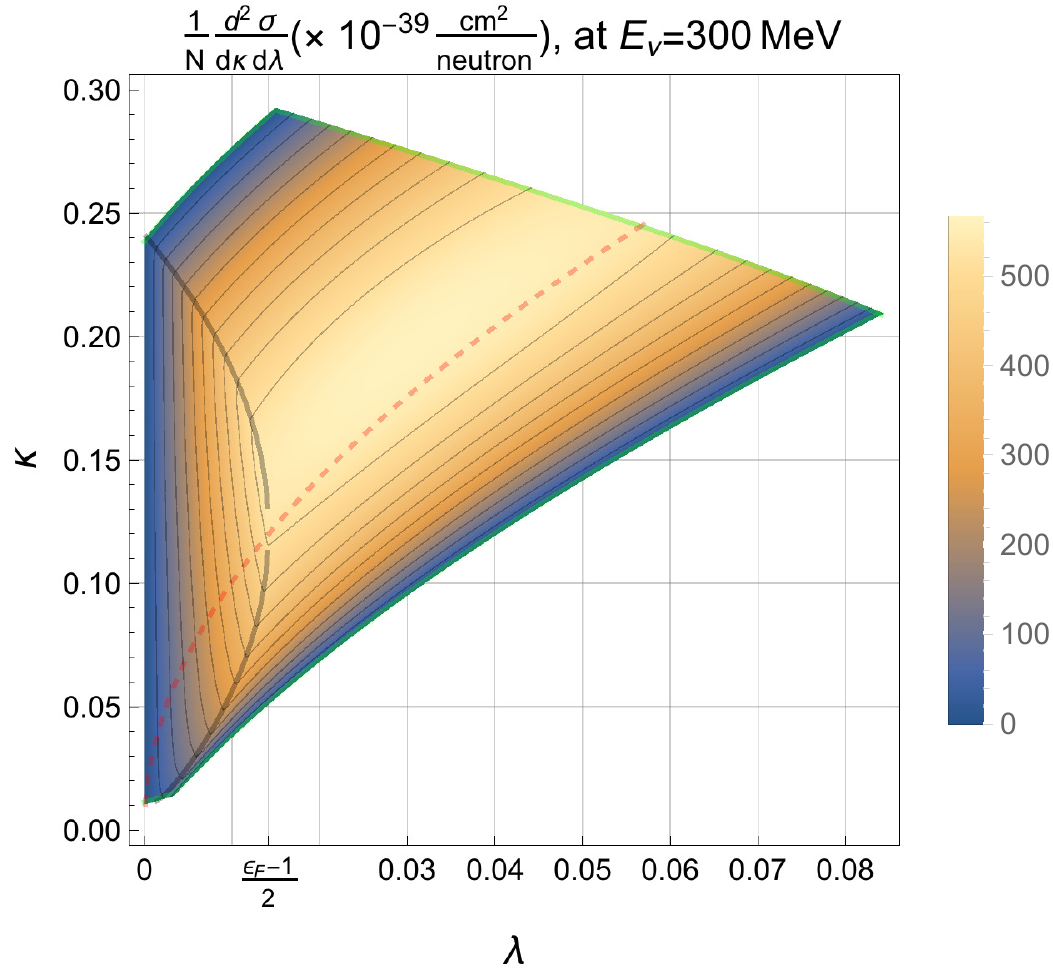}
\end{subfigure}
\begin{subfigure}[t]{0.49\linewidth}
\centering
\includegraphics[width=\linewidth, height=7.5cm]{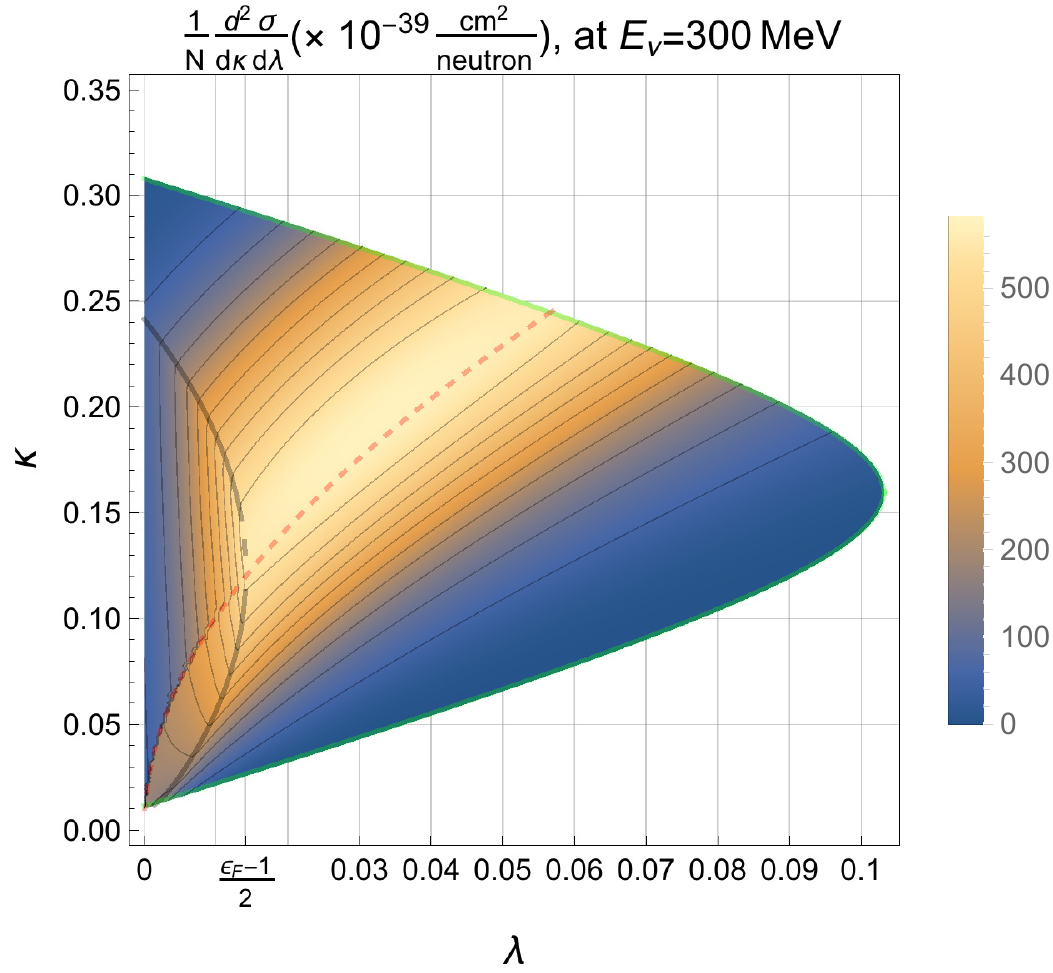}
\end{subfigure}
\caption{Comparison of the density plots for the $\nu_\mu$ CCQE double
  differential $\frac{d^2\sigma}{d\kappa\,d\lambda}$ cross section per
  neutron in ${}^{12}$C for the RFG (left panel) and SuSAM* (right
  panel) models at $E_{\nu}=300$ MeV.  Note that the available phase
  spaces in the different models are those shown in the left panel of
  Fig.\ \ref{fig: phase-space-RFG-susam}.  We show in short-dashed
  style the curve $\kappa_\text{QE}(\lambda)$, where the QE peak is
  placed; while in solid fashion we also display the boundary of the
  PB region, already shown in Figs.\ \ref{figure3} and
  \ \ref{figure4}.}\label{fig: xsect-RFG-susam-300MeV}
\end{figure*}

In Fig.\ \ref{fig: xsect-RFG-susam-300MeV} we show the double
differential CCQE $\frac{d^2\sigma}{d\kappa\,d\lambda}$ cross section
per neutron for the $(\nu_{\mu},\mu^{-})$ reaction off ${}^{12}$C, at
incident neutrino energies of $300$ MeV, for the two models discussed
in this work: RFG (left panel) and SuSAM* (right panel). The available
phase spaces in the two models at this neutrino energy are those
already depicted in the left panel of Fig.\ \ref{fig:
  phase-space-RFG-susam}.  Note that, although not exactly the same,
both scales in the two panels are very similar, as well as the values
reached by the cross section. In Fig.\ \ref{fig:
  xsect-RFG-susam-300MeV} we also show as the short-dashed line the
curve $\kappa=\kappa_\text{QE}(\lambda)$, where $\lambda=\tau$ and
$\psi=0$, i.e, the curve corresponding to the position of the QE peak,
which, as expected, runs over the region of largest cross section. In
solid style, it is also shown the boundary of the PB region. As
already mentioned in sect.\ \ref{subsec: analytic-boundaries-RFG}, in
the PB region the scaling variable $\psi$ only depends on $\lambda$
and not on $\kappa$. Because of this, the contour lines (curves with
the same value of the cross section) inside the PB region are almost
vertical lines, because the dependence on $\kappa$ mainly enters
through the lepton kinematic factors $V_K$ and the nuclear response
functions $U_K$ of Eq.\ \eqref{d2sigma_dtmu_dcosmu_v2} and it is very
mild at least for the RFG (left panel) model. Notice also that at the
boundaries of the PB region, the contour lines show a sudden change of
their direction. This is because at these boundaries the scaling
variable $\psi$ starts to sharply depend on $\kappa$ as well.

There is, nevertheless, a remarkable difference between the left and
right panels of Fig.\ \ref{fig: xsect-RFG-susam-300MeV} in the PB
region: in the SuSAM* model (right panel) the color gradient along
vertical lines of constant $\lambda$ changes abruptly when crossing
the QE curve, especially for small values of $\lambda$; however this
effect is totally absent in the left panel, corresponding to the RFG
model. The reason for this is because of the properties of the scaling
functions in the two models. In the RFG, the scaling function given by
Eq.\ \eqref{rfg_scalfunc} is an even function of $\psi$. This means
that for constant $\lambda$ there is no difference in being above or
below the QE curve inside the PB region (the only difference is the
sign of the scaling variable, but not the value of the scaling
function in the RFG model).  However, the situation is very different
in the SuSAM* model because its scaling function, given by
Eq.\ \eqref{scaling_function_susam}, also depends on $\psi$ and not
only on $\psi^2$. Hence, a simple change of sign in the scaling
variable can produce a large difference in the scaling function (see
dotted line of Fig.\ \ref{figure7}), thus inducing a sudden change in
the value of the cross section when passing from positive values of
the scaling variable (below the short-dashed curve) to negative ones
(above the same curve) in the PB region. Nonetheless, this effect
seems to be quite pronounced only for small values of $\lambda$ in the
PB region, and not so perceptible for $\lambda$ values closer to the
end point of the PB region, given by $\lambda_{-}$ in
Eq.\ \eqref{lambda_minus}.

The physical reason for this sharp discontinuity in the value of the
differential cross section when crossing the
$\kappa_{\text{QE}}(\lambda)$ curve inside the PB region can be surely
related to the treatment of the Pauli blocking effect in the SuSAM*
model. As discussed already in Sect.\ \ref{susam_model}, the treatment
of the Pauli blocking in the SuSAM* model is exactly the same as that
of the RFG, which for sure is not the ideal one, although being the
simplest one. For instance, in Refs.\ \cite{Gonzalez-Jimenez:2014eqa,
  Megias:2014kiaWithErratum} for the SuSA and SuSAv2 models, a totally
different and surely more well-founded approach has been used to
incorporate the Pauli blocking, the so-called ``mirror" scaling
function subtraction. At the present stage of the SuSAM* model we have
treated the Pauli blocking in the simplest way, and although we are
aware of this limitation (that could be amended in future refinements
of the model), it is not the purpose of this article to discuss these
drawbacks in detail.  The fairest option we can take is to warn the
reader of this issue and of other possible ways of incorporating the
Pauli blocking when using ``by hand" phenomenological scaling
functions that do not derive from a well-known momentum distribution,
as it is also the case of the SuSAM* model.

\begin{figure*}[htb]
\centering
\begin{subfigure}[t]{0.5\linewidth}
\centering
\includegraphics[width=\linewidth, height=7.5cm]{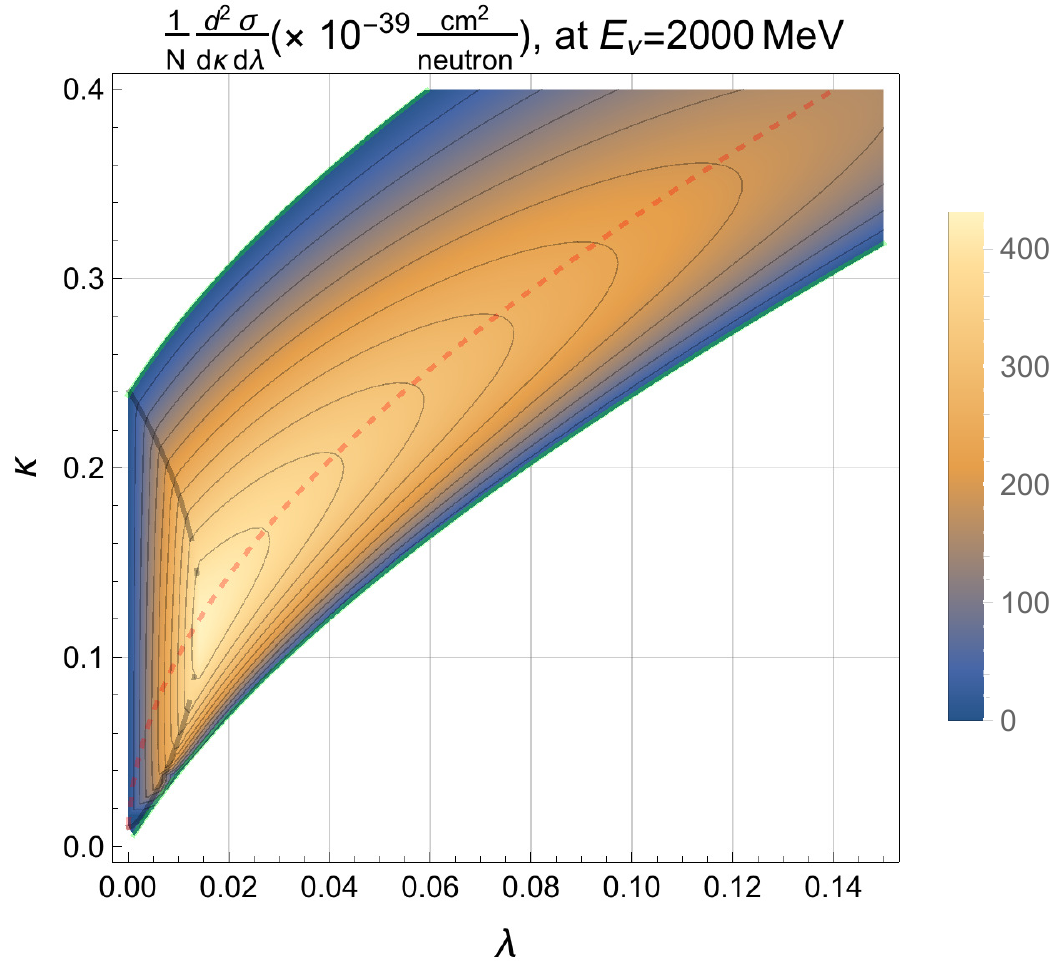}
\end{subfigure}
\begin{subfigure}[t]{0.49\linewidth}
\centering
\includegraphics[width=\linewidth, height=7.5cm]{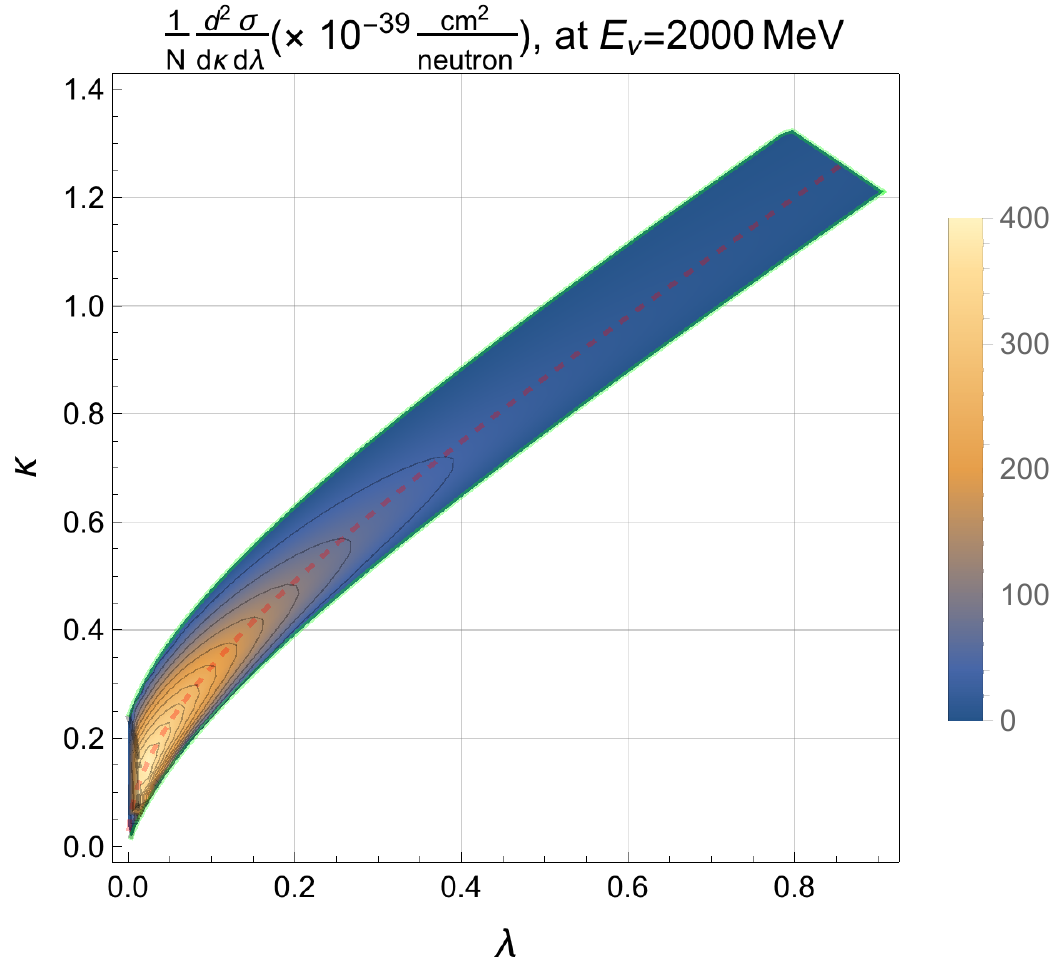}
\end{subfigure}
\caption{Density plots for the $\nu_\mu$ double differential CCQE
  $\frac{d^2\sigma}{d\kappa\,d\lambda}$ cross section per neutron in
  ${}^{12}$C in the RFG model at $E_{\nu}=2000$ MeV.  The left panel
  highlights the PB region, while the right one shows the full phase
  space region corresponding to those already shown in the right
  panels of Figs.\ \ref{fig: phase-space-RFG} and \ref{fig:
    phase-space-RFG-susam}.  Lines have the same meaning as in
  Fig.\ \ref{fig: xsect-RFG-susam-300MeV}.}
  \label{fig: xsect-RFG-2000MeV}
\end{figure*}

\begin{figure*}[htb]
\centering
\begin{subfigure}[t]{0.5\linewidth}
\centering
\includegraphics[width=\linewidth, height=7.5cm]{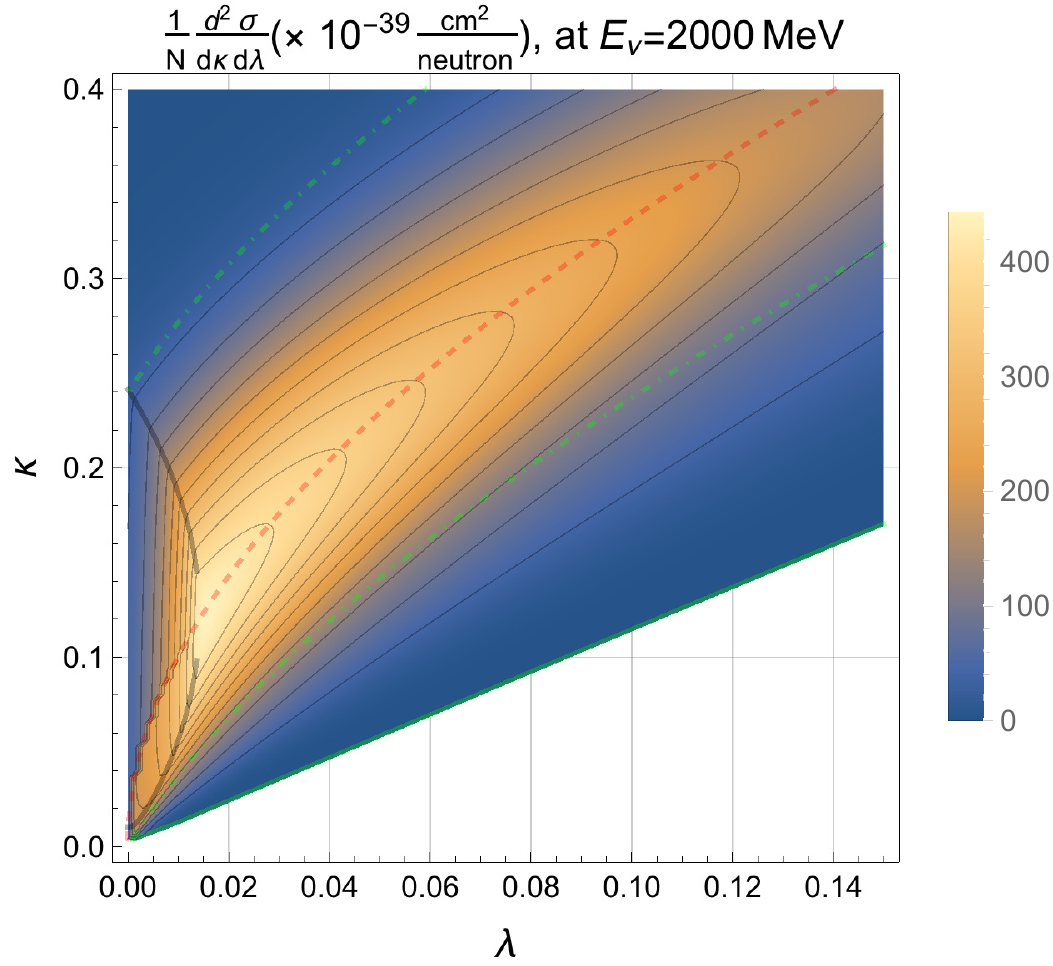}
\end{subfigure}
\begin{subfigure}[t]{0.49\linewidth}
\centering
\includegraphics[width=\linewidth, height=7.5cm]{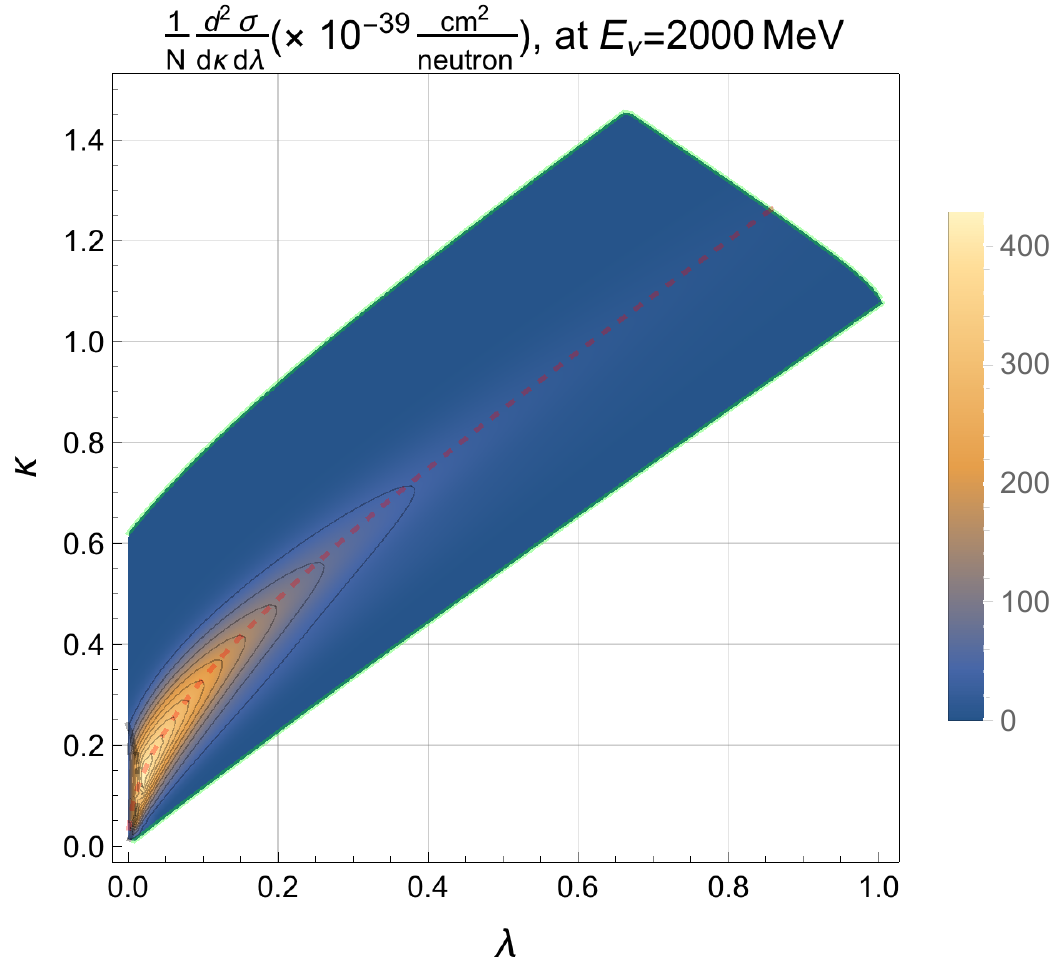}
\end{subfigure}
\caption{Same as Fig.\ \ref{fig: xsect-RFG-2000MeV} but for the SuSAM*
  model. Short-dashed and solid lines have the same meaning as in
  Figs.\ \ref{fig: xsect-RFG-susam-300MeV} and \ref{fig:
    xsect-RFG-2000MeV}. Notice, however, in the left panel, the
  dot-dashed lines that correspond to the upper and lower boundaries
  of the RFG model, i.e, the same boundaries shown in the left panel
  of Fig.\ \ref{fig: xsect-RFG-2000MeV}. }\label{fig:
  xsect-susam-2000MeV}
\end{figure*}

In Figs.\ \ref{fig: xsect-RFG-2000MeV} and \ref{fig:
  xsect-susam-2000MeV} we show the density and contour plots of the
double differential CCQE $\frac{d^2\sigma}{d\kappa\,d\lambda}$ cross
section per neutron for $\nu_\mu$ reactions on ${}^{12}$C at a fixed
neutrino energy of $2000$ MeV. Figure \ref{fig: xsect-RFG-2000MeV}
corresponds to the RFG model, while Fig.\ \ref{fig:
  xsect-susam-2000MeV} shows the results for the SuSAM* one. Left
panels highlight the region of small values of energy transfers
$\lambda$, i.e, showing clearly the PB region, while right panels in
both figures show the full phase space.  At this neutrino energy, the
boundary of the phase space is basically delimited by the curves
obtained from the scaling model conditions (limited by imposing the
condition that the scaling function is zero or negligible), and not
from the lepton kinematics, as it happened in Fig.\ \ref{fig:
  xsect-RFG-susam-300MeV} for smaller neutrino energy. Besides that,
the sharp boundaries of the phase space for the RFG model, shown in
Fig.\ \ref{fig: xsect-RFG-2000MeV}, are due to the sharp way in which
the RFG scaling function goes to zero at $\psi=\pm1$. However, in
Fig.\ \ref{fig: xsect-susam-2000MeV}, the phase space extends further
than for the RFG case just because the SuSAM* scaling function has
tails beyond $\psi=\pm1$. Actually, the phase space of the SuSAM*
model would extend even further than what is shown in Fig. \ref{fig:
  xsect-susam-2000MeV}, but with negligible values of the cross
section, already visible in the own figure.

In general, the values of the cross section in both models at
$E_{\nu}=2000$ MeV are very similar in the same regions of the
$(\lambda,\kappa)$ phase space. In Fig.\ \ref{fig:
  xsect-susam-2000MeV}, there seems to be a non negligible cross
section in the SuSAM* model in regions of the phase space below and
close to the lower boundary of the RFG model, according to its color
legend.  These additional contributions will have a large impact in
the total integrated CCQE cross section shown later on in
Fig.\ \ref{fig: total_xsect_nu_nubar} at $E_{\nu}=2000$ MeV.

Nonetheless, the most important feature of Figs.\ \ref{fig:
  xsect-RFG-2000MeV} and \ref{fig: xsect-susam-2000MeV}, if compared
with Fig.\ \ref{fig: xsect-RFG-susam-300MeV} for $E_{\nu}=300$ MeV of
incident neutrino energy, is that the maximum of the double
differential cross section $\frac{d^2\sigma}{d\kappa\,d\lambda}$
depends very little on the neutrino energy. Indeed, for $E_{\nu}=300$
MeV the maximum of the cross section is around $550\times10^{-39}$
cm$^2$/neutron, while for $E_{\nu}=2000$ MeV this maximum is around
$450\times10^{-39}$ cm$^2$/neutron. This remarkable feature makes this
double differential cross section especially well-suited to be used in
MC generators to select the kinematics of the final lepton events for
fixed neutrino energy. This is especially relevant for the generators
that use the acceptance-rejection method to select the events, because
using this method it is necessary to normalize the double differential
cross section to its maximum value.  \emph{And if this maximum value
depends very weakly with the neutrino energy, one can efficiently set
a fixed maximum suitable for all the neutrino energies.} We will see
that this efficiency would not be so attainable if one uses the double
differential cross section
$\frac{d^2\sigma}{dT_{\mu}\,d\cos\theta_{\mu}}$, given in
Eq.\ \eqref{d2sigma_dtmu_dcosmu_v2}, instead of
$\frac{d^2\sigma}{d\kappa\,d\lambda}$, just because the maximum of the
former depends very strongly on the neutrino energy.

\begin{figure*}[htb]
\centering
\begin{subfigure}[t]{0.5\linewidth}
\centering
\includegraphics[width=\linewidth, height=7.5cm]{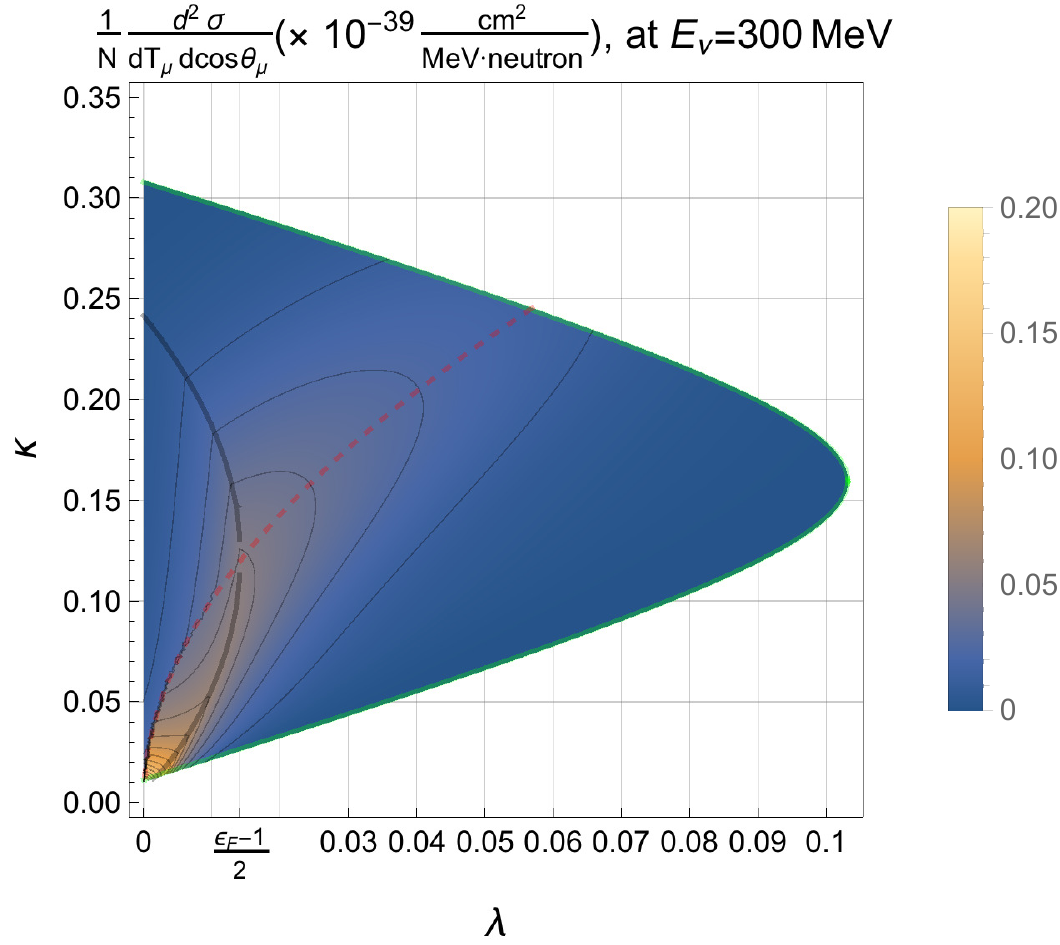}
\end{subfigure}
\begin{subfigure}[t]{0.49\linewidth}
\centering
\includegraphics[width=\linewidth, height=7.5cm]{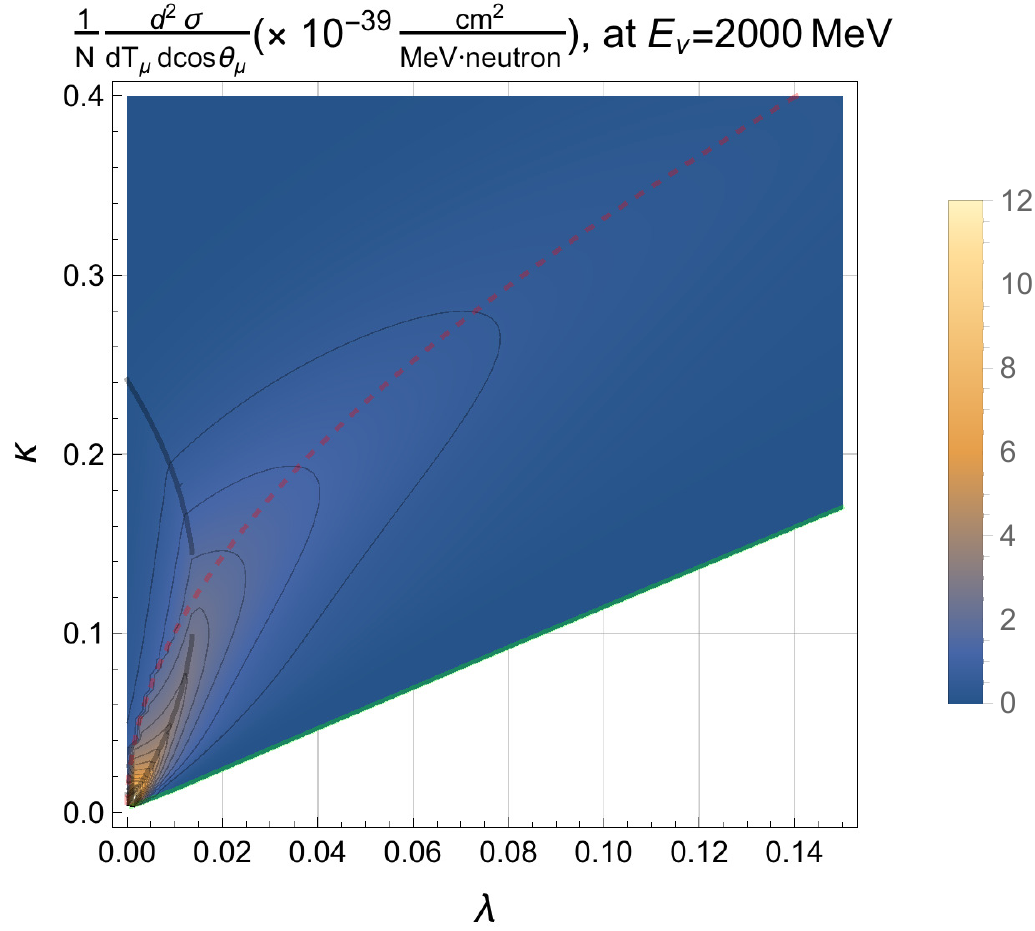}
\end{subfigure}
\begin{subfigure}[t]{0.5\linewidth}
\centering
\includegraphics[width=\linewidth, height=7.5cm]{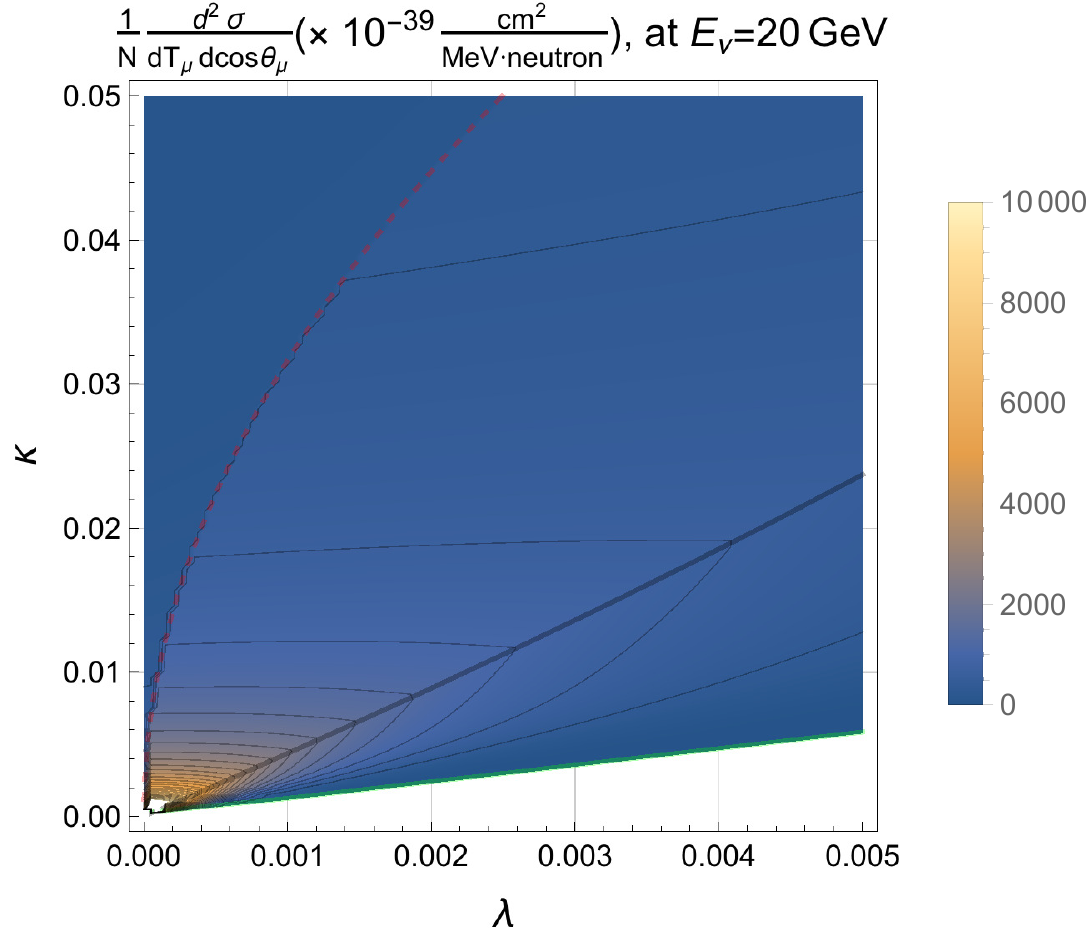}
\end{subfigure}
\caption{Comparison of the density plots for the CCQE
  $\nu_{\mu}$-double differential cross section
  $\frac{1}{N}\frac{d^2\sigma}{dT_{\mu}\,d\cos\theta_{\mu}}$ off
  ${}^{12}$C in the SuSAM* model for three different neutrino
  energies, highlighting the PB region, where the bulk of the cross
  section is concentrated.  In the top left panel we show the density
  plot for $E_{\nu}=300$ MeV; in the top right panel we display that
  for $E_{\nu}=2000$ MeV; while in the bottom panel the plot for
  $E_{\nu}=20$ GeV is shown as well. Curves on the plot have the same
  meaning as they had in Figs.\ \ref{fig:
    xsect-RFG-susam-300MeV}--\ref{fig: xsect-susam-2000MeV}. The white
  hole in the bottom left corner of the bottom panel means that in
  that region the cross section is reaching values larger than the
  maximum shown in its scale.}
  \label{fig: xsectTmucosmu-susam}
\end{figure*}

Indeed, if we inspect Fig.\ \ref{fig: xsectTmucosmu-susam}, where the
CCQE $\nu_{\mu}$-double differential cross section
$\frac{d^2\sigma}{dT_{\mu}\,d\cos\theta_{\mu}}$ per neutron has been
plotted for three different neutrino energies in the SuSAM* model, we
can conclude two main things: First, the height of the peak of this
cross section is strongly growing with the neutrino energy, as it can
be seen from the values taken in the graduated color scale. Second,
the larger the neutrino energy is, the more concentrated the bulk of
the cross section is in a smaller region of the phase space, although
this last conclusion can get overshadowed by the differences in the
figures' scales. Moreover, the gradient of the cross section grows
strongly with the neutrino energy for this differential cross section
(larger variations of the cross section in a smaller region of the
phase space, which makes the contour lines of constant cross section
to appear closer and closer as the neutrino energy increases).  Note,
in particular, that this behavior is very striking in the bottom panel
of Fig.\ \ref{fig: xsectTmucosmu-susam}, i.e, for $E_{\nu}=20$ GeV.
In this latter panel, in the bottom left corner, there is a white hole
which means that there, the double differential cross section is much
larger than the maximum value shown in its scale.

These conclusions should be compared with those of the
$\frac{d^2\sigma}{d\kappa\,d\lambda}$ cross section per neutron shown
in Figs.\ \ref{fig: xsect-RFG-susam-300MeV}--\ref{fig:
  xsect-susam-2000MeV}, where they were the opposite, i.e, the peak of
the cross section was almost flat with the neutrino energy (we have
also checked that this statement is also true for $E_{\nu}=20$ GeV,
although not shown in any figure), and the variation of the cross
section over the phase space is much softer. These two special
features make the double differential cross section
$\frac{d^2\sigma}{d\kappa\,d\lambda}$ much more suitable to generate
the final lepton events in any MC generator, specially those which use
the acceptance-rejection method.  Indeed, the event generation
consists in the following steps:
\begin{itemize}
	\item randomly select a point in kinematic phase space,
          e.g. $\lambda_0$ and $\kappa_0$;
	\item randomly choose uniformly distributed variable $t$ in
          range (0,1);
	\item accept event if
          $\frac{d^2\sigma}{d\kappa\,d\lambda}|_{\lambda=\lambda_0,\kappa=\kappa_0}>t\max\left({\frac{d^2\sigma}{d\kappa\,d\lambda}}\right)$
          and reject otherwise.
\end{itemize} 

For this method to be efficient, the differential cross section has to
be as flat as possible for all neutrino energies because of two main
reasons: first, it allows maximum search algorithm to be more
efficient (seeking the maximum more accurately in a shorter time);
second, the fewer attempts to select kinematic variables that are
rejected, the faster the events are generated.

\begin{figure*}[htb]
\centering
\begin{subfigure}[t]{0.5\linewidth}
\centering
\includegraphics[width=\linewidth, height=7.5cm]{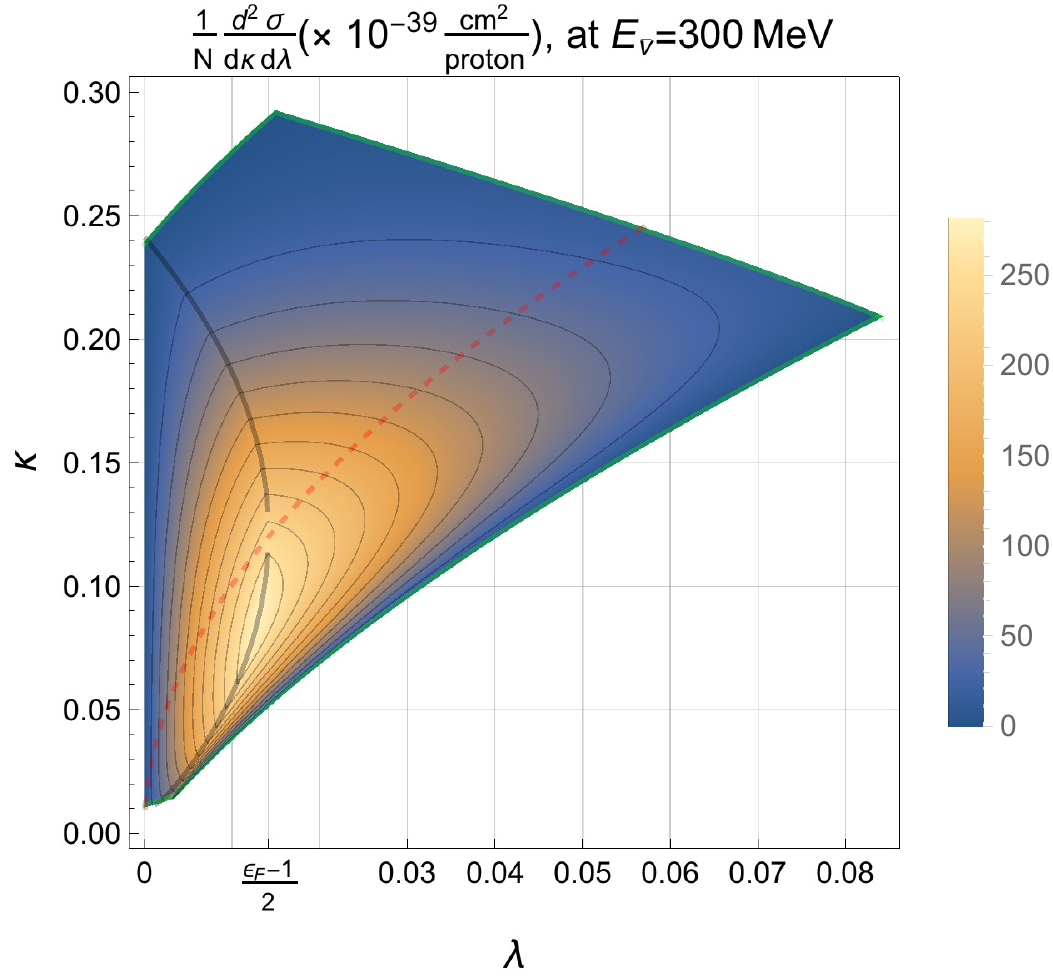}
\end{subfigure}
\begin{subfigure}[t]{0.49\linewidth}
\centering
\includegraphics[width=\linewidth, height=7.5cm]{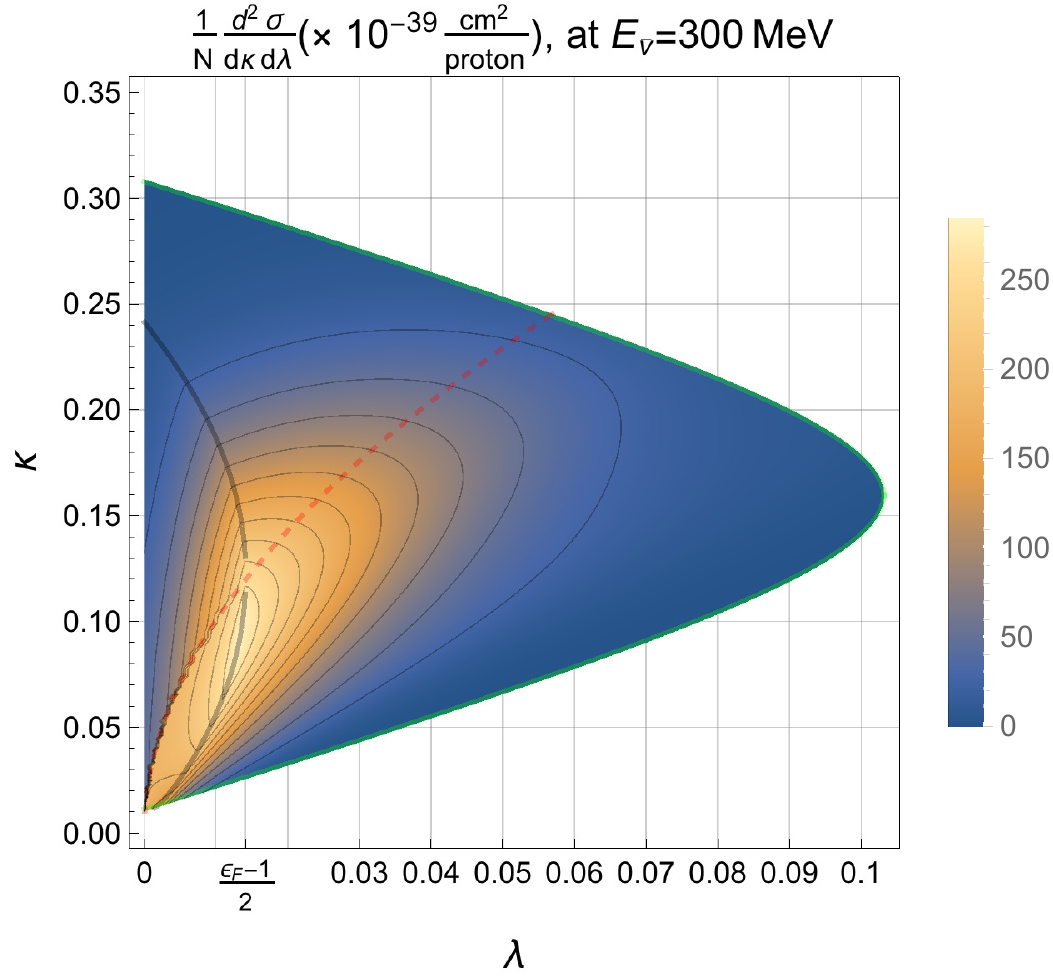}
\end{subfigure}
\caption{Comparison of the density and contour plots for the
  $\bar{\nu}_\mu$ CCQE double differential
  $\frac{d^2\sigma}{d\kappa\,d\lambda}$ cross section per proton in
  ${}^{12}$C for the RFG (left panel) and SuSAM* (right panel) models
  at $E_{\bar{\nu}}=300$ MeV.  Lines have the same meaning as in
  Fig.\ \ref{fig: xsect-RFG-susam-300MeV}. Note that now the cross
  section is roughly half than that for the neutrino case, but also
  much smaller along other regions of the whole phase space because of
  the minus sign in Eq.\ \eqref{d2sigma_dtmu_dcosmu_v2}, which applies
  for CCQE antineutrino scattering. Notice as well about the
  difference this minus sign makes in the contour lines of constant
  double differential cross section.}\label{fig:
  xsect-RFG-susam-300MeV-antinu}
\end{figure*}

\begin{figure*}[htb]
\centering
\begin{subfigure}[t]{0.5\linewidth}
\centering
\includegraphics[width=\linewidth, height=7.5cm]{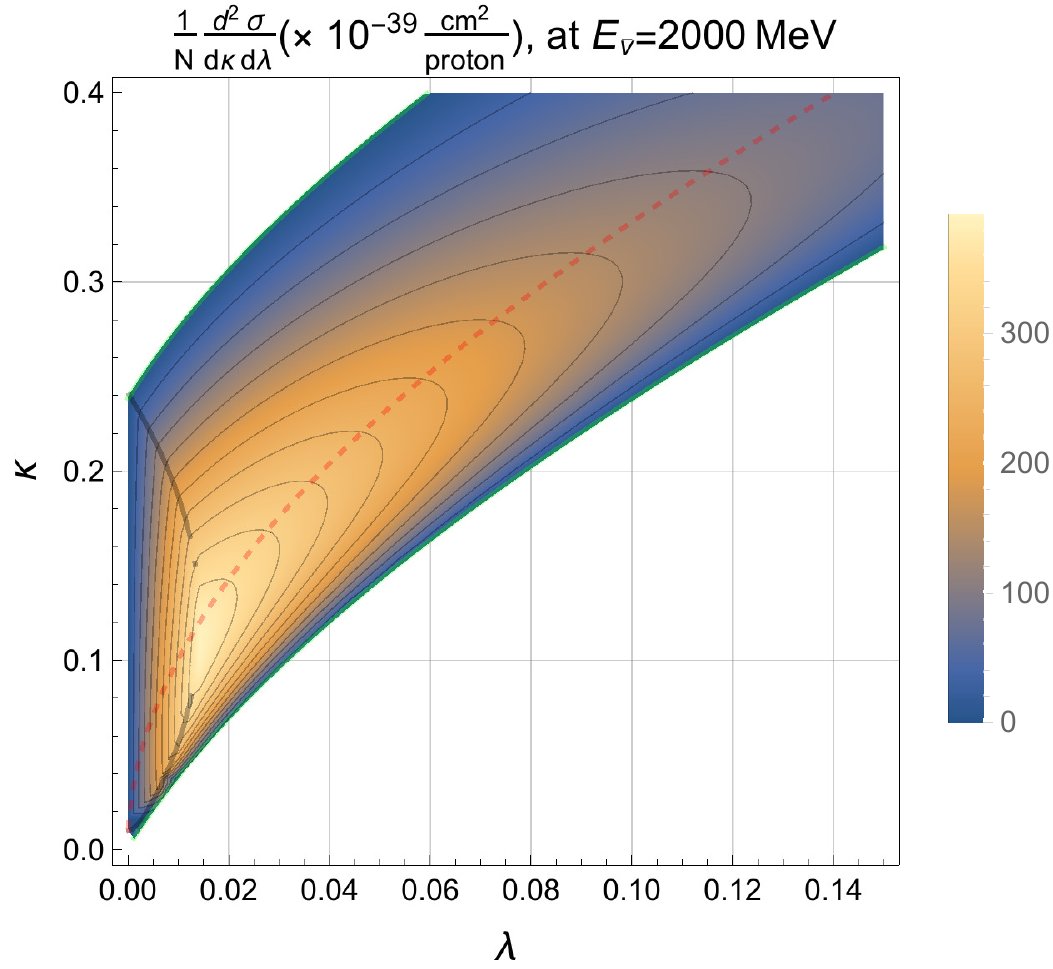}
\end{subfigure}
\begin{subfigure}[t]{0.49\linewidth}
\centering
\includegraphics[width=\linewidth, height=7.5cm]{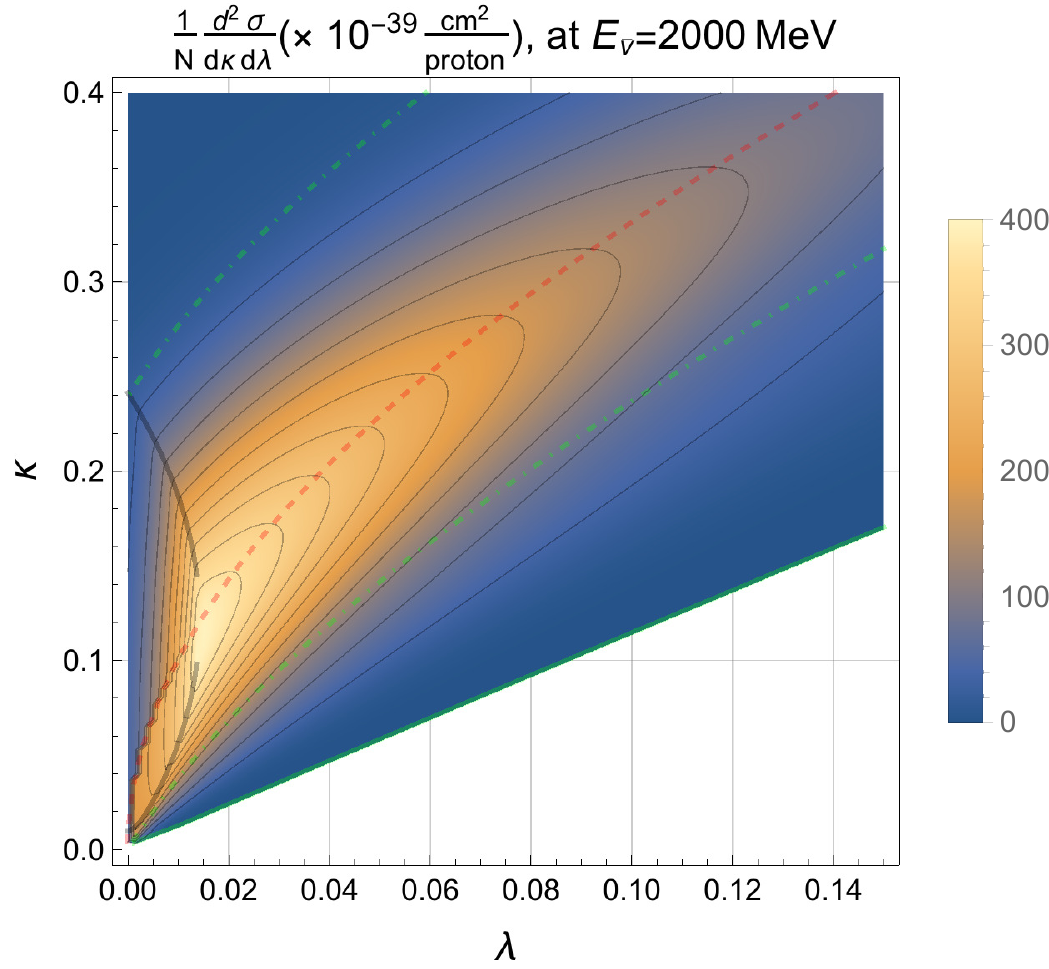}
\end{subfigure}
\caption{Same as Fig.\ \ref{fig: xsect-RFG-susam-300MeV-antinu} but
  for $E_{\bar{\nu}}=2000$ MeV. The left panel (RFG model) should be
  compared with the left panel of Fig.\ \ref{fig: xsect-RFG-2000MeV},
  while the right one (SuSAM* model) should be compared with the left
  one of Fig.\ \ref{fig: xsect-susam-2000MeV}.  These plots highlight
  the PB region and do not show the full phase space.  Note again that
  the double differential cross section is smaller for CCQE
  antineutrino scattering than it is for neutrino case. The dot-dashed
  lines in the right panel are the boundaries of the RFG model shown
  in the left panel.}
\label{fig: xsect-RFG-susam-2000MeV-antinu}
\end{figure*}

In Figs.\ \ref{fig: xsect-RFG-susam-300MeV-antinu} and \ref{fig:
  xsect-RFG-susam-2000MeV-antinu} we show the CCQE
$\bar{\nu}_{\mu}$-induced double differential cross section
$\frac{d^2\sigma}{d\kappa\,d\lambda}$ per proton for two different
antineutrino energies, respectively. In the left panels we display the
density plots for the RFG model, while in the right ones we show those
for the SuSAM* model. The main conclusion that can be drawn from these
figures if compared with the corresponding ones for the neutrino case
is that, as expected, the antineutrino cross sections are smaller than
their neutrino counterparts. This is especially clear in
Fig.\ \ref{fig: xsect-RFG-susam-300MeV-antinu}, if compared with
Fig.\ \ref{fig: xsect-RFG-susam-300MeV}, because the values in the
scales of the figure for antineutrinos are roughly half of the values
shown in Fig.\ \ref{fig: xsect-RFG-susam-300MeV}, and the regions
where the maximum values are reached in Fig.\ \ref{fig:
  xsect-RFG-susam-300MeV-antinu} are clearly smaller in size than
those of Fig.\ \ref{fig: xsect-RFG-susam-300MeV}, despite the fact
that the available phase space is exactly the same.

The comparison can be less clear for the case of antineutrinos of
$E_{\bar{\nu}}=2000$ MeV (Fig.\ \ref{fig:
  xsect-RFG-susam-2000MeV-antinu}) if one compares the corresponding
model with the left panels of Figs.\ \ref{fig: xsect-RFG-2000MeV} and
\ref{fig: xsect-susam-2000MeV}, because in this case the color scales
reach similar values, although a bit smaller for
antineutrinos. However, one can notice that the number of contour
lines of constant cross section that enter completely inside the shown
phase space (this is the same area of phase space shown in the left
panels of Figs.\ \ref{fig: xsect-RFG-2000MeV} and \ref{fig:
  xsect-susam-2000MeV}) is larger in Fig.\ \ref{fig:
  xsect-RFG-susam-2000MeV-antinu} (8 contour lines out of 10) than it
was for the neutrino case (6 contour lines out of 10). This means
that, even although the color scales could be considered similar, the
contour lines for the antineutrino case appear more concentrated in
the \emph{same} region of phase space than their neutrino
counterparts.  Thus, we can conclude that larger cross sections extend
far beyond the \emph{same} phase space shown in Fig.\ \ref{fig:
  xsect-RFG-susam-2000MeV-antinu} for the neutrino case than for the
antineutrino one, yielding a larger CCQE total cross section for
neutrinos than for antineutrinos.

\subsection{Total integrated cross section}\label{subsec:total_xsect}
In this section we discuss the integrated CCQE cross sections both for
neutrinos and antineutrinos off ${}^{12}$C, when integration over the
$(\lambda,\kappa)$ phase space is carried out.

First of all, we want to point out a thorough description of how the
$(\lambda,\kappa)$ phase space behaves as the neutrino energy
increases. Notice that the curves $\kappa^\text{NPB}_{\pm}(\lambda)$,
either those described by Eq.\ \eqref{eq: npb_pm_kappa_curves}, or by
Eq.\ \eqref{upper_lower_boundary} (which are actually the same
expressions, as explained in
appendix\ \ref{difficult-form-boundaries}), do not depend at all on
the reduced neutrino energy $\epsilon_{\nu}$.  Therefore these
boundaries are always the same irrespective of the values taken by the
neutrino energy.  The dependence on the neutrino energy is in the
curves $\kappa^\text{lepton}_{\max,\min}(\lambda)$ given by
Eqs.\ \eqref{kappa_max_lepton} and \eqref{kappa_min_lepton}.

\begin{figure*}[htb]
\centering
\includegraphics[width=\linewidth, height=15cm]{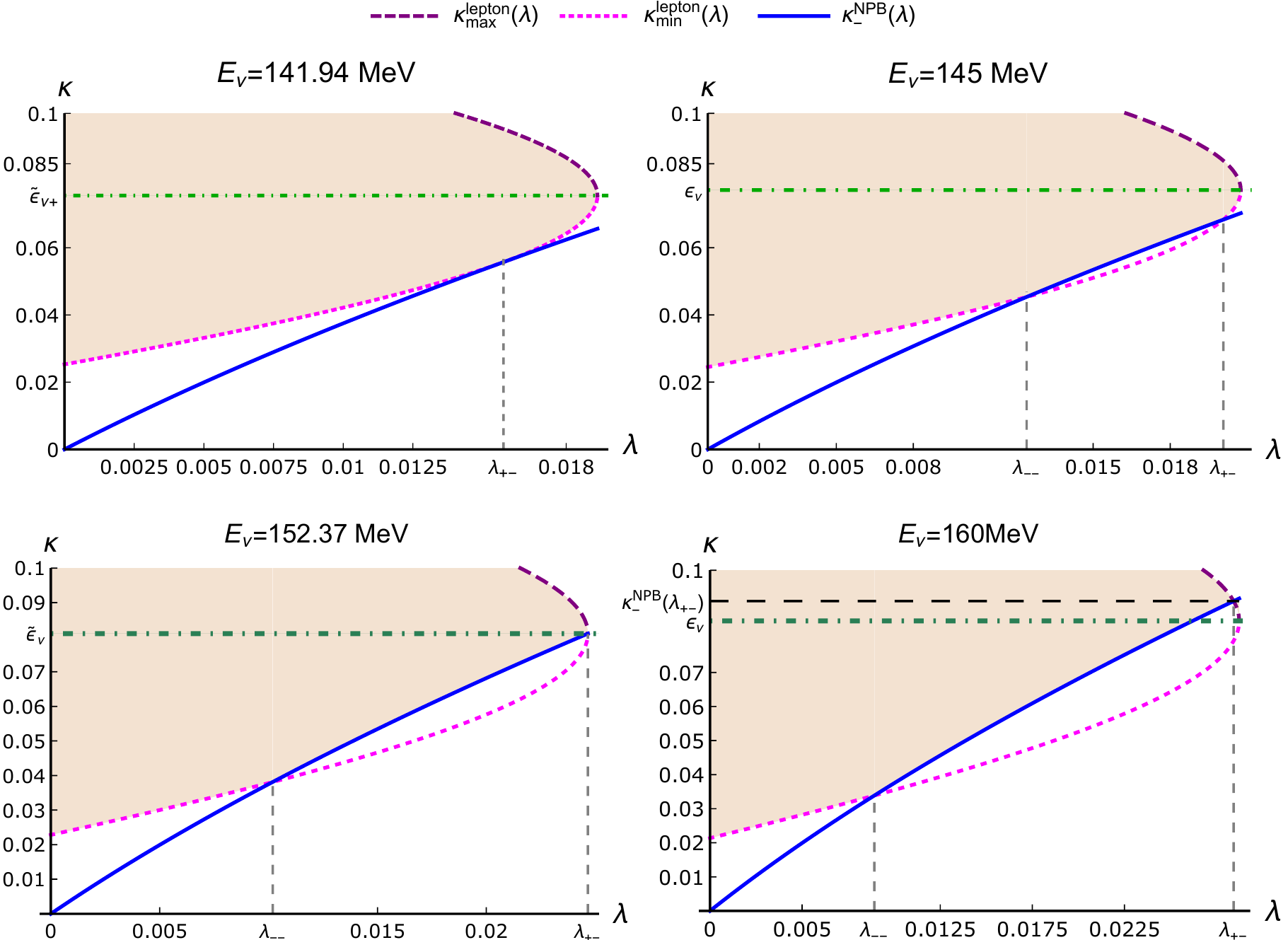}
\caption{Plot of the phase space for the RFG model at four different
  neutrino energies, where different situations arise.  In the top
  left panel, the reduced neutrino energy is given by
  $\tilde{\epsilon}_{\nu+}$ in Eq.\ \eqref{lambdapmm_unique}, and the
  cut between the curves $\kappa^\text{lepton}_{\min}(\lambda)$ and
  $\kappa^{\text{NPB}}_{-}(\lambda)$ is sole and tangent.  However,
  when the neutrino energy increases a bit (top right panel), the two
  different solutions $\lambda_{\pm-}$ given by
  Eq.\ \eqref{lambdapm-solution} appear, first as cuts between the
  curves $\kappa^\text{lepton}_{\min}(\lambda)$ and
  $\kappa^{\text{NPB}}_{-}(\lambda)$.  For higher neutrino energies,
  as that shown in the bottom left panel, the $\lambda_{+-}$ cut given
  by Eq.\ \eqref{lambdapm-solution} occurs exactly at
  $\lambda=\lambda_{\max}$, and this happens for the reduced neutrino
  energy $\tilde{\epsilon}_{\nu}$ given in
  Eq.\ \eqref{epsilon_nu_corner}.  Finally, in the bottom right panel
  we show the situation for a bit larger neutrino energy.  In this
  case, the $\lambda_{+-}$ solution given by
  Eq.\ \eqref{lambdapm-solution} corresponds to a cut point between
  the curves $\kappa^{\text{NPB}}_{-}(\lambda)$ and
  $\kappa^{\text{lepton}}_{\max}(\lambda)$.}\label{fig: cuts_lambdapm}
\end{figure*}

At low neutrino energies, the phase space is completely bounded by the
final lepton kinematics, i.e, by the curves
$\kappa^\text{lepton}_{\max,\min}(\lambda)$ solely. This is the case,
for instance, of the right panels of Figs.\ \ref{fig:
  xsect-RFG-susam-300MeV} and \ref{fig:
  xsect-RFG-susam-300MeV-antinu}. In this case, there are no cutting
points between the two lepton kinematic branches and between the upper
and lower $\kappa^\text{NPB}_{\pm}(\lambda)$ curves.  This same effect
would occur in the left panels of Figs.\ \ref{fig:
  xsect-RFG-susam-300MeV} and \ref{fig:
  xsect-RFG-susam-300MeV-antinu}, corresponding to the RFG model, but
at a neutrino energy lower than 300 MeV, because in the RFG the
$\kappa^\text{NPB}_{\pm}(\lambda)$ curves are squeezed with respect to
those of the SuSAM* model. This can be seen, for instance, in the top
left panel of Fig.\ \ref{fig: cuts_lambdapm} for the RFG.

As the neutrino energy increases, the cutting points between the
curves $\kappa^\text{lepton}_{\max,\min}(\lambda)$ and
$\kappa^\text{NPB}_{\pm}(\lambda)$ start to appear. One of these cuts
is discussed in appendix\ \ref{appendix-a4}, labelled as
$\lambda_{++}$, given in Eq.\ \eqref{lambda++solution} and shown in
Figs.\ \ref{fig: phase-space-RFG} and \ref{fig:
  phase-space-RFG-susam}.  The other two additional cuts that can
occur are those given by Eq.\ \eqref{lambdapm-solution}, labelled as
$\lambda_{\pm-}$, and shown in Figs.\ \ref{fig: phase-space-RFG} and
\ref{fig: phase-space-RFG-susam} as well.  However, there is a reduced
neutrino energy $\epsilon_{\nu}$ for which
$\lambda_{+-}=\lambda_{--}$. This happens when the radicand of
Eq.\ \eqref{lambdapm-solution} is zero. Thus, we can find the
$\epsilon_{\nu}$ value for this to happen by equating the radicand of
Eq.\ \eqref{lambdapm-solution} to zero and solving the second degree
equation for $\epsilon_{\nu}$. The result is
\begin{equation}\label{lambdapmm_unique}
\tilde{\epsilon}_{\nu\pm}=
\frac{\widetilde{m}_{\mu}(\widetilde{m}_{\mu}\pm1)}{\epsilon^\prime_F-\eta^\prime_F},
\end{equation}
where $\epsilon^\prime_F\equiv1+(\epsilon_F-1)\psi^2_\text{right}$ and
$\eta^\prime_F=\sqrt{\epsilon^{\prime\,2}_F-1}$. Of course, if one
wants to recover the results of the RFG, one substitutes
$\psi_\text{right}=1$.

In principle, the $\tilde{\epsilon}_{\nu-}$ solution can be ruled out
for electron and muon neutrinos because it is negative
\footnote{The case of $\tau$ neutrinos should be taken with care and studied
in depth because both solutions could be positive if the nucleon mass were the
relativistic effective mass of the Walecka model 
\protect\cite{Walecka:1974qa,Serot:1984ey,Wehrberger:1993zu,Rosenfelder:1980nd},
which is the underlying theoretical model in which the SuSAM* is based on.},
but $\tilde{\epsilon}_{\nu+}$ is positive and must be considered. At this
reduced neutrino energy $\tilde{\epsilon}_{\nu+}$, there is a single
and tangent cut between the curves
$\kappa^\text{lepton}_{\min}(\lambda)$ and
$\kappa^\text{NPB}_{-}(\lambda)$, as it can be seen in the top left panel
of Fig.\ \ref{fig: cuts_lambdapm}.

If the neutrino energy continues increasing, the two $\lambda$-cuts
given by Eq.\ \eqref{lambdapm-solution} are different, but still both
cuts occur between the $\kappa^\text{lepton}_{\min}(\lambda)$ and
$\kappa^\text{NPB}_{-}(\lambda)$ curves (as it can be observed in the
top right panel of Fig.\ \ref{fig: cuts_lambdapm}), until a higher
neutrino energy ($\tilde{\epsilon}_{\nu}$) is reached, at which
$\lambda_{+-}=\lambda_{\max}$, with $\lambda_{\max}$ given by
Eq.\ \eqref{eq:lambda_max}. In this range of values for
$\epsilon_{\nu}\in\left[\tilde{\epsilon}_{\nu+},
  \tilde{\epsilon}_{\nu}\right]$, the range of integration in
$\lambda$ still runs from $\lambda\in\left[0,\lambda_{\max}\right]$.

To find $\tilde{\epsilon}_{\nu}$ one could equate
$\lambda_{+-}=\lambda_{\max}$ and try to solve it for
$\epsilon_{\nu}$, but this is very difficult because the equation
turns out to be a third degree equation in
$\epsilon_{\nu}$. Nonetheless, there is a very easy way to obtain this
value of $\tilde{\epsilon}_{\nu}$: we can equate
$\kappa^\text{lepton}_{\max}(\lambda_{\max})=
\kappa^\text{NPB}_{-}(\lambda_{\max})$ and solve it for
$\epsilon_{\nu}$. Given that
$\kappa^\text{lepton}_{\max}(\lambda_{\max})=\epsilon_{\nu}$, we can
take Eq.\ \eqref{eq: npb_pm_kappa_curves} for
$\kappa^\text{NPB}_{-}(\lambda_{\max})$ and solve the equation for
$\epsilon_{\nu}$. The result is straightforward
\begin{equation}\label{epsilon_nu_corner}
\tilde{\epsilon}_{\nu}=
\frac{\widetilde{m}_{\mu}(\epsilon^\prime_F-\widetilde{m}_{\mu})}{\epsilon^\prime_F-\eta^\prime_F-2\widetilde{m}_{\mu}}.
\end{equation}
This is the situation shown in the bottom left panel of
Fig.\ \ref{fig: cuts_lambdapm}.  And now, we can ensure that for
$\epsilon_{\nu} > \tilde{\epsilon}_{\nu}$, the $\lambda_{+-}$ cut
given by Eq.\ \eqref{lambdapm-solution} is lesser than
$\lambda_{\max}$, but now it is a cut between the curve
$\kappa^\text{NPB}_{-}(\lambda)$ and the upper branch of the lepton
kinematics boundary, $\kappa^\text{lepton}_{\max}(\lambda)$. This can
be observed in the bottom right panel of Fig.\ \ref{fig:
  cuts_lambdapm}.

Now, the integration range in the $\lambda$ variable is further
constrained to be $\lambda\in\left[0,\lambda_{+-} \right]$ where
$\lambda_{+-}<\lambda_{\max}$ (only valid when $\epsilon_{\nu} >
\tilde{\epsilon}_{\nu}$).  Thus we can integrate the double
differential cross section $\frac{d^2\sigma}{d\kappa\,d\lambda}$ in
the region of the phase space where it is truly different from zero,
thus making the integration algorithm the most efficient as
possible. The integrated total CCQE cross section can now be written
\begin{equation*}
\sigma(E_{\nu})=\int^{\lambda_u}_{0} d\lambda 
\int^{\kappa_u(\lambda)}_{\kappa_d(\lambda)} d\kappa\,
\frac{d^2\sigma}{d\kappa\,d\lambda}(E_{\nu}),
\end{equation*}
where
\begin{eqnarray}
\lambda_u&=&\left\{
\begin{array}{c}
\lambda_{\max} \qquad \text{if $\epsilon_{\nu} \leqslant 
\tilde{\epsilon}_{\nu}$}, \\
\lambda_{+-} \qquad \text{if $\epsilon_{\nu} > \tilde{\epsilon}_{\nu}$;}
\end{array}
\right. \nonumber \\
\kappa_d(\lambda)&=&\max\left( \kappa^\text{NPB}_{-}(\lambda),
\kappa^\text{lepton}_{\min}(\lambda) \right), \nonumber \\
\kappa_u(\lambda)&=&\min\left( \kappa^\text{NPB}_{+}(\lambda),
\kappa^\text{lepton}_{\max}(\lambda) \right).  \nonumber
\end{eqnarray}

\begin{figure*}[htb]
\centering
\includegraphics[width=0.495\linewidth]{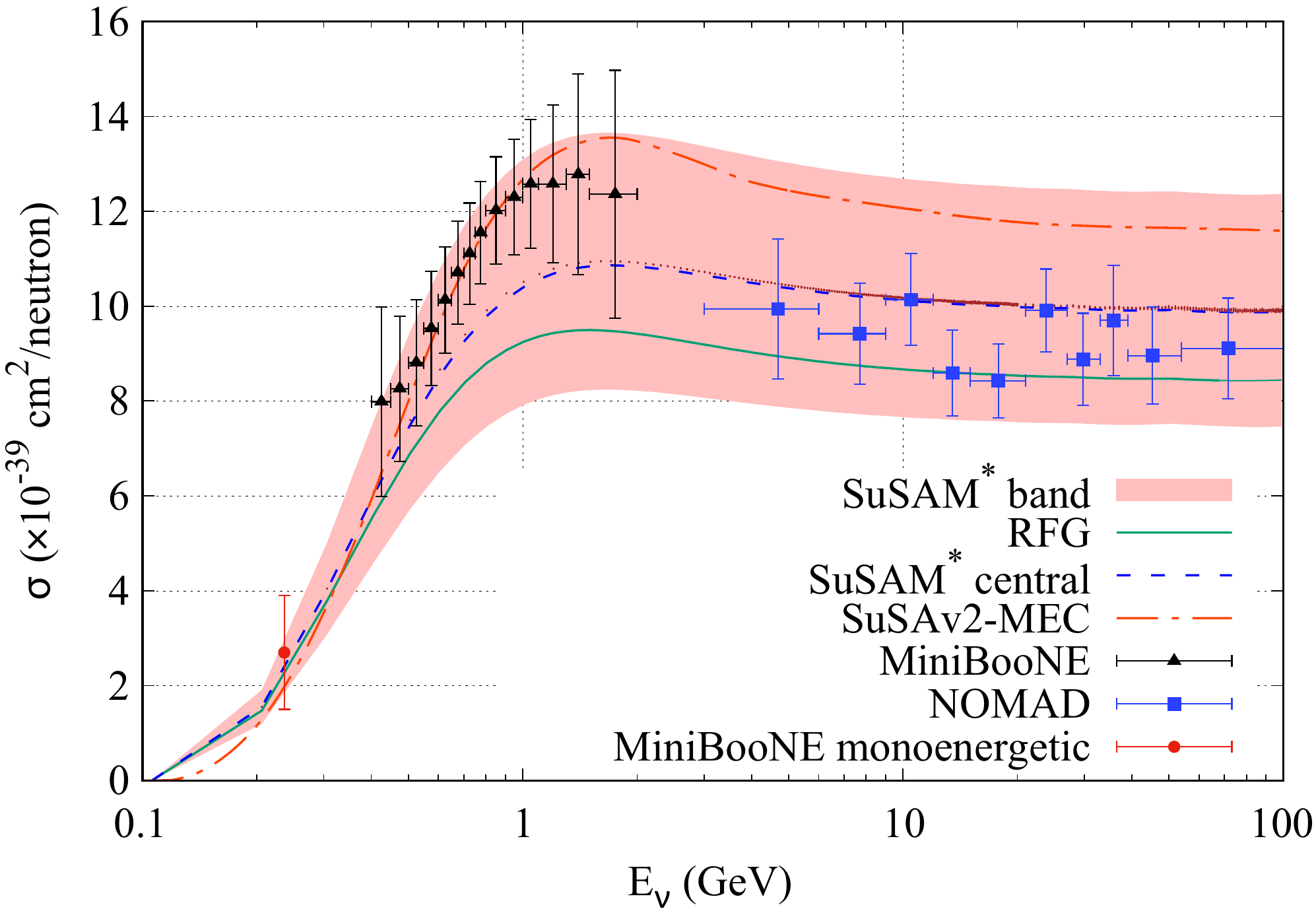}
\includegraphics[width=0.495\linewidth]{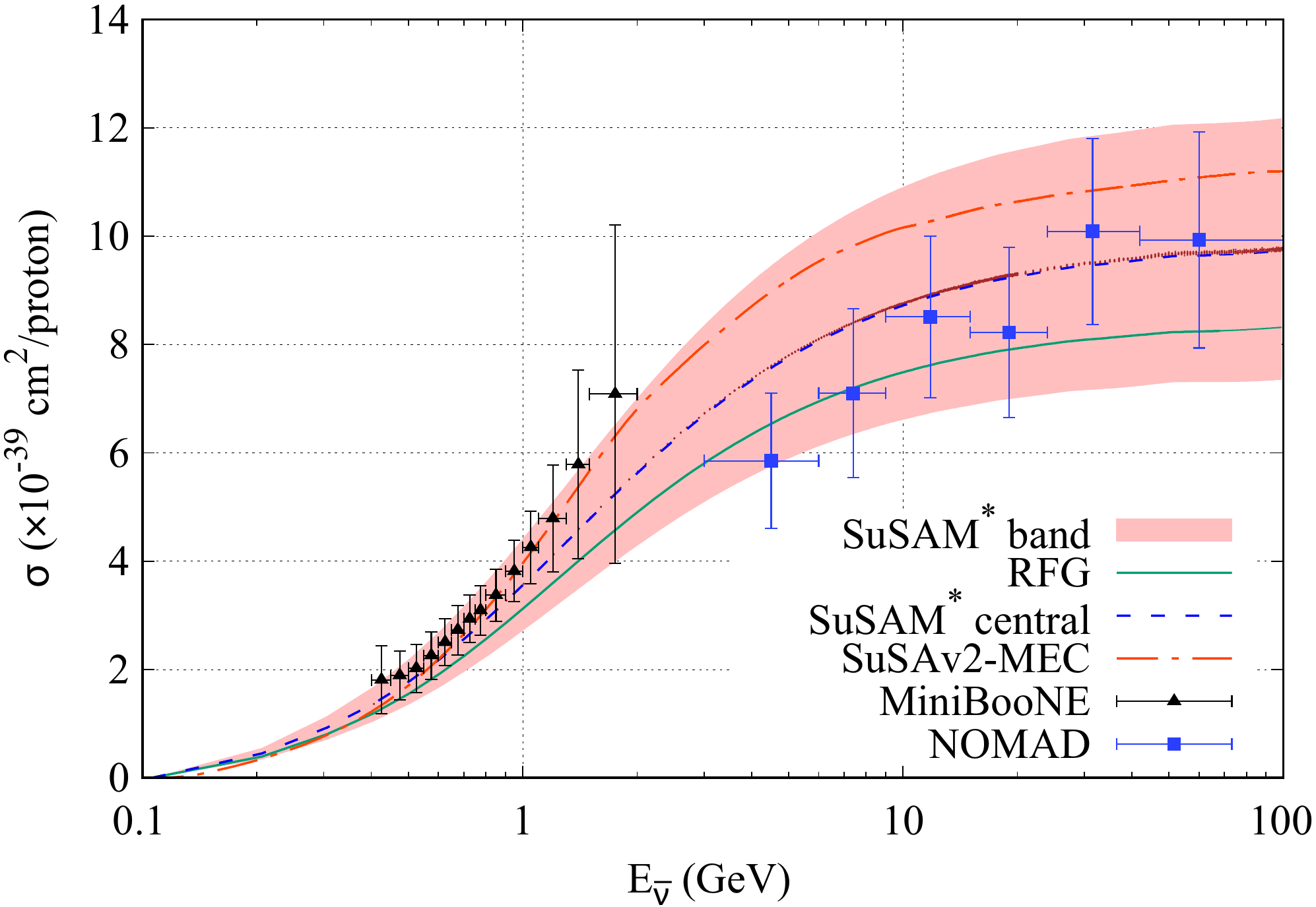}
\caption{Plot of the total CCQE cross section $\sigma(E_{\nu})$
  (normalized per interacting nucleon) as a function of the
  neutrino/antineutrino energy $E_{\nu(\bar{\nu})}$ for the two models
  discussed in this work: RFG (solid line) and SuSAM* (short-dashed
  line). In the left panel the neutrino total cross section per
  neutron off ${}^{12}$C is displayed along with the experimental
  measurements of MiniBooNE \cite{Aguilar-Arevalo:2010zc,
    Aguilar-Arevalo:2018ylq} and NOMAD \cite{Lyubushkin:2008pe}.  In
  the right panel, we show the same for the antineutrino total cross
  section per proton, compared with the measurements of MiniBooNE
  \cite{Aguilar-Arevalo:2013dva} and NOMAD \cite{Lyubushkin:2008pe}
  collaborations. For both models, RFG and SuSAM*, the nucleon
  relativistic effective mass $m^*_N=0.83\,m_N$ has been taken, as
  well as a Fermi momentum of $k_F=212$ MeV/c, accordingly to the
  global fit to ``QE" electron scattering data performed in
  Ref.\ \cite{Amaro:2018xdi}. Additionally, in dot-dashed style, it is
  also shown the SuSAv2-MEC model prediction, which has been taken
  from Ref.\ \cite{Megias:2016fjk}. The Monte Carlo band (shown as
  points surrounding the SuSAM* central line) is also displayed (see
  main text for an explanation about how it has been obtained).}
\label{fig: total_xsect_nu_nubar}
\end{figure*}

In Fig.\ \ref{fig: total_xsect_nu_nubar} we show the results for the
total CCQE integrated cross section off ${}^{12}$C for the two models
discussed in this work: RFG (solid line) and SuSAM* (short-dashed
line). The left panel is for muon neutrino scattering, while the right
one corresponds to muon antineutrino.  We have displayed the
uncertainty band of the SuSAM* model, taken as the area between the
predictions for the total cross sections obtained by taking the
$f^{\max,\min}_{\text{SuSAM*}}(\psi)$ scaling functions depicted in
Fig.\ \ref{figure7}, instead of taking the
$f^{\text{central}}_{\text{SuSAM*}}(\psi)$ scaling function of the
same figure, which is the one we have used throughout this article.
To compare with another important and relevant scaling model, already
incorporated in GENIE\ \cite{Dolan:2019bxf}, the SuSAv2-MEC model of
Ref.\ \cite{Megias:2016fjk}, we have plotted the curve of this model
in Fig.\ \ref{fig: total_xsect_nu_nubar} in dot-dashed style as well.

\begin{figure}[htb]
 \centering
 \includegraphics[bb = 60 70 760 530, scale=0.325]{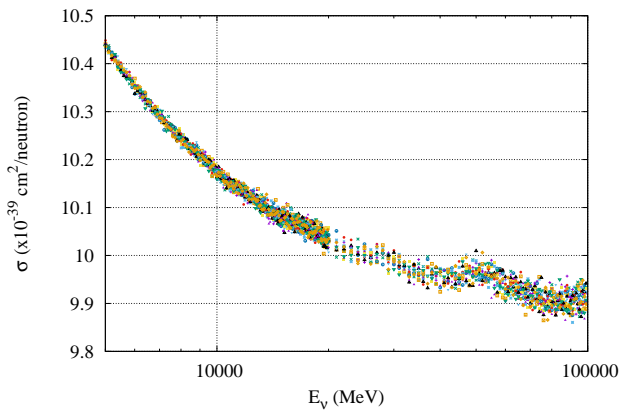}
 \caption{Plot of the Monte Carlo band obtained as explained in the
   main text.  Notice that the spreading of this band occurs in a
   scale much thinner than that displayed in Fig.\ \ref{fig:
     total_xsect_nu_nubar}.}
 \label{fig: Monte Carlo band}
\end{figure}

In Fig.\ \ref{fig: total_xsect_nu_nubar} we also display, in the form
of a very narrow band of points surrounding the SuSAM* central curve,
a Monte Carlo band that has been obtained by choosing an uniformly
distributed value for the scaling function between the minimum and the
maximum scaling functions of the SuSAM* model (figure\ \ref{figure7})
for each kinematic point $\psi(\lambda,\kappa)$ in the integration
procedure for each neutrino/antineutrino energy. Although it is almost
imperceptible in the scale of Fig.\ \ref{fig: total_xsect_nu_nubar},
for each neutrino/antineutrino energy, there are 12 dots forming this
extremely narrow band. Only for the higher and higher energies the
spread of these points starts to become appreciable. This can be seen
in Fig.\ \ref{fig: Monte Carlo band}, where a zoom view of this band
is shown, exhibiting that the scale of fluctuations of this Monte
Carlo band is much thinner than that shown in Fig.\ \ref{fig:
  total_xsect_nu_nubar}, where it is seen almost as a single curve
without spreading.

An argument in favor of choosing an uniform distribution to draw a
value for the scaling function for each kinematic point being
integrated is that, in the super-scaling analysis performed in
Ref.\ \cite{Amaro:2018xdi}, the super-scaling band was found almost
equally populated between the minimum and maximum scaling functions of
the SuSAM* model. In fact, that band was obtained by using a density
criterion, i.e, by keeping those points with at least more than a
fixed number of neighboring points inside a circle of given radius.

The main conclusion that can be drawn by comparing the curves of the
three different scaling models (RFG, SuSAM* and SuSAv2-MEC) is that
all of them lie inside the uncertainty band of the SuSAM* model. It is
true that this band is very large, but not so large if compared with
the experimental uncertainties, which it is even truer for the
antineutrino total cross section (right panel of Fig.\ \ref{fig:
  total_xsect_nu_nubar}), where one can see that the theoretical
uncertainty of the SuSAM* model is of the same order as the error bars
of the experimental points. In fact, the central prediction of the
SuSAM* model is in between those of the RFG and the SuSAv2-MEC, and as
observed in Fig.\ \ref{fig: total_xsect_nu_nubar}, it passes closer to
both sets of the experimental data shown in the figure.  These are the
results of MiniBooNE
\cite{Aguilar-Arevalo:2010zc,Aguilar-Arevalo:2013dva,
  Aguilar-Arevalo:2018ylq} (for intermediate neutrino energies and for
the new technique based on kaon decay at rest
\cite{Aguilar-Arevalo:2018ylq}) and NOMAD \cite{Lyubushkin:2008pe}
(for the high neutrino energy range) experiments.  It is worth another
remark: the SuSAM* predictions (band and central curve) are solely
based on the super-scaling properties of the selected ``QE" electron
scattering data out of the total inclusive ($e,e^\prime$) data, in the
global fit carried out in Ref.\ \cite{Amaro:2018xdi}, and no CCQE
neutrino scattering parameter has been fitted at all. The same can be
said for the SuSAv2-MEC model, which is based on another scaling
function \cite{Megias:2016lke} and with the contribution of the weak
charged meson-exchange currents (MEC) calculated in
Ref.\ \cite{Simo:2016ikv}. The SuSAv2-MEC model describes the
MiniBooNE data very well, but systematically overestimates the cross
section at the NOMAD energies. In fact, this could point to a conflict
between the MEC contribution used in this model and the NOMAD data.

It is not the purpose of this work to discuss the discrepancies
between both sets of data shown in Fig.\ \ref{fig:
  total_xsect_nu_nubar}, because the experimental collaborations
recognize in their works \cite{Aguilar-Arevalo:2013dva} that the
experiments use different detector technologies and assume different
topologies in defining CCQE events. And, in addition, the
neutrino/antineutrino energy drawn in the abscissa axes of
Fig.\ \ref{fig: total_xsect_nu_nubar} is the true neutrino energy,
while in the experiments the energy is the reconstructed one (except
for the kaon decay at rest technique), which assumes an educated guess
to obtain it from the measured final lepton kinematic variables via an
unfolding procedure.  In fact, the problems related to the
reconstruction of the neutrino energy have been addressed in a series
of articles
\cite{Butkevich:2008ef,Martini:2012fa,Nieves:2012yz,Martini:2012uc,
  Leitner:2010kp,Ankowski:2014yfa,Mosel:2013fxa,
  DeRomeri:2016qwo,Lu:2015hea,Ankowski:2015jya,
  Munteanu:2019llq,Furmanski:2016wqo}.
  
The total integrated CCQE cross sections shown in Fig.\ \ref{fig:
  total_xsect_nu_nubar} have been obtained with the set of parameters
($k_F=212$ MeV/c and $M^*=m^*_N/m_N=0.83$) for ${}^{12}$C obtained in
the global fit to ``QE" electron scattering data of
Ref.\ \cite{Amaro:2018xdi}, and given in Table II of the same
reference. We did not try to adjust these and any other parameters of
the model (e.g., axial mass of the nucleon) to the neutrino data.
This is important to be stressed, because in the previous figures and
formulae of this work, we have utterly used the values of $k_F=225$
MeV/c and $M^*=1$ for the Fermi momentum and the relativistic
effective mass, respectively. We used these values in the calculations
of the previous figures because we did not want to bother the reader
with additional complications related to the underlying Walecka model
\cite{Walecka:1974qa, Serot:1984ey} (see also
Refs.~\cite{Wehrberger:1993zu,Rosenfelder:1980nd}) in which the SuSAM*
approach is based. It is also worth warning the reader that in this
work we have used the usual dipole axial-vector form factor with an
charged-current axial-vector mass of $M_A=1.008$ GeV, and the set of
vector form factors taken from the Galster parametrization given in
Ref.\ \cite{Galster:1971kv}. This value of $M_A$ was obtained in a
global fit to all available self-consistent data on CCQE
(anti)neutrino scattering on nuclei, within the Smith-Monitz RFG
model\ \cite{Smith:1972xhWithErratum} and so-called running
axial-vector mass of nucleon\ \cite{Kakorin:2020atz}.  Since $M_A$ is
an effective model-dependent parameter, we plan to adjust it from a
global fit within the SuSAM* model.

It is worth noting that all the formulae appearing in this work can be
translated to the real SuSAM* model by just changing the value of the
Fermi momentum and the free nucleon mass $m_N\rightarrow m^*_N$, where
$m^*_N$ is the value of the relativistic effective mass. These changes
affect the values of $\eta_F$, $\epsilon_F$, $\epsilon^\prime_F$,
$\eta^\prime_F \ldots$, but the form of the equations obtained in this
work remains the same. Of course, what also does not change at all is
the form of the scaling functions shown in Fig.\ \ref{figure7}. What
changes is the value of the scaling variable $\psi$ for a given
kinematics ($\omega$,$q$), but not the form of the scaling functions.

\begin{figure*}[htb]
\centering
\includegraphics[width=0.495\linewidth]{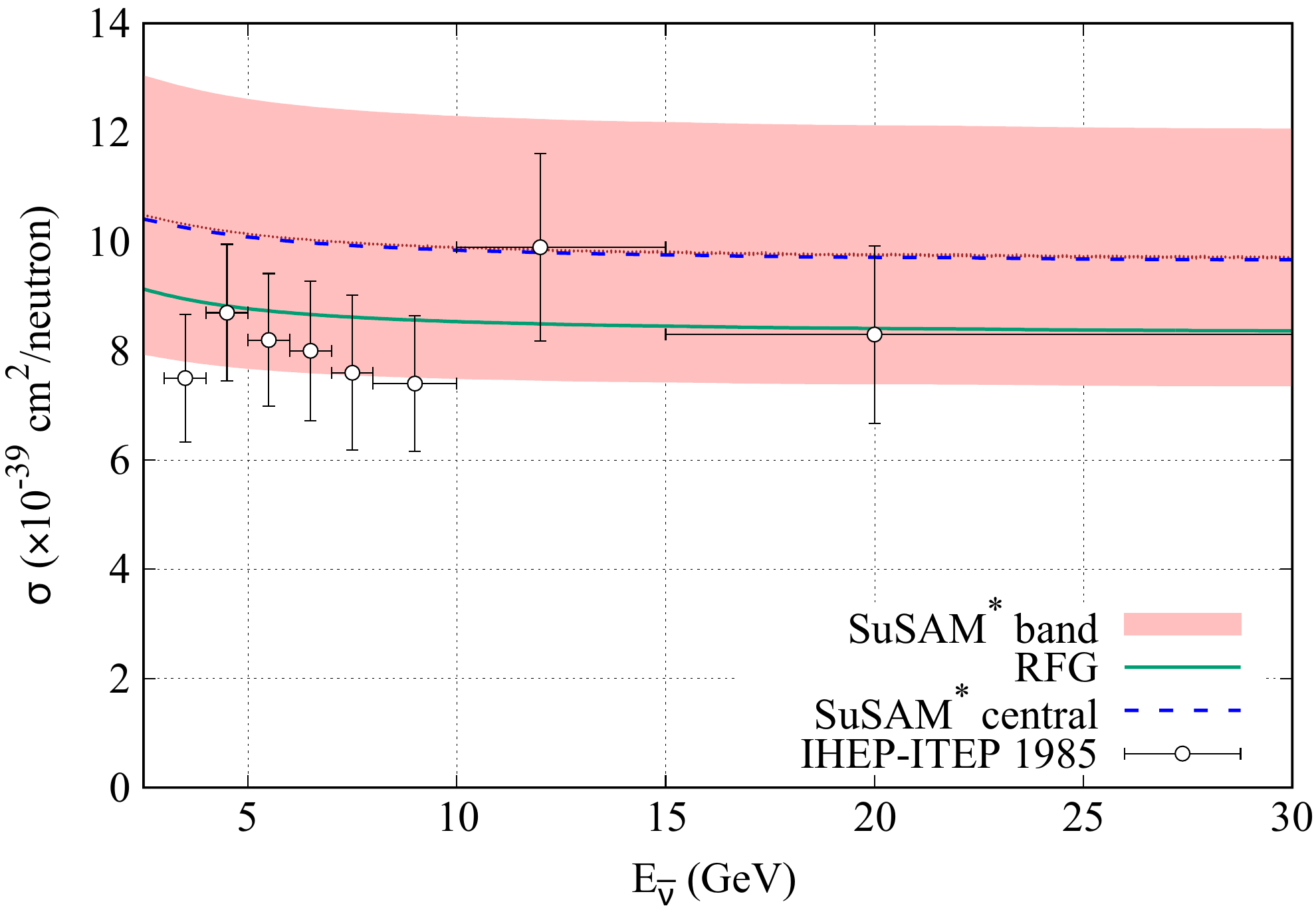}
\includegraphics[width=0.495\linewidth]{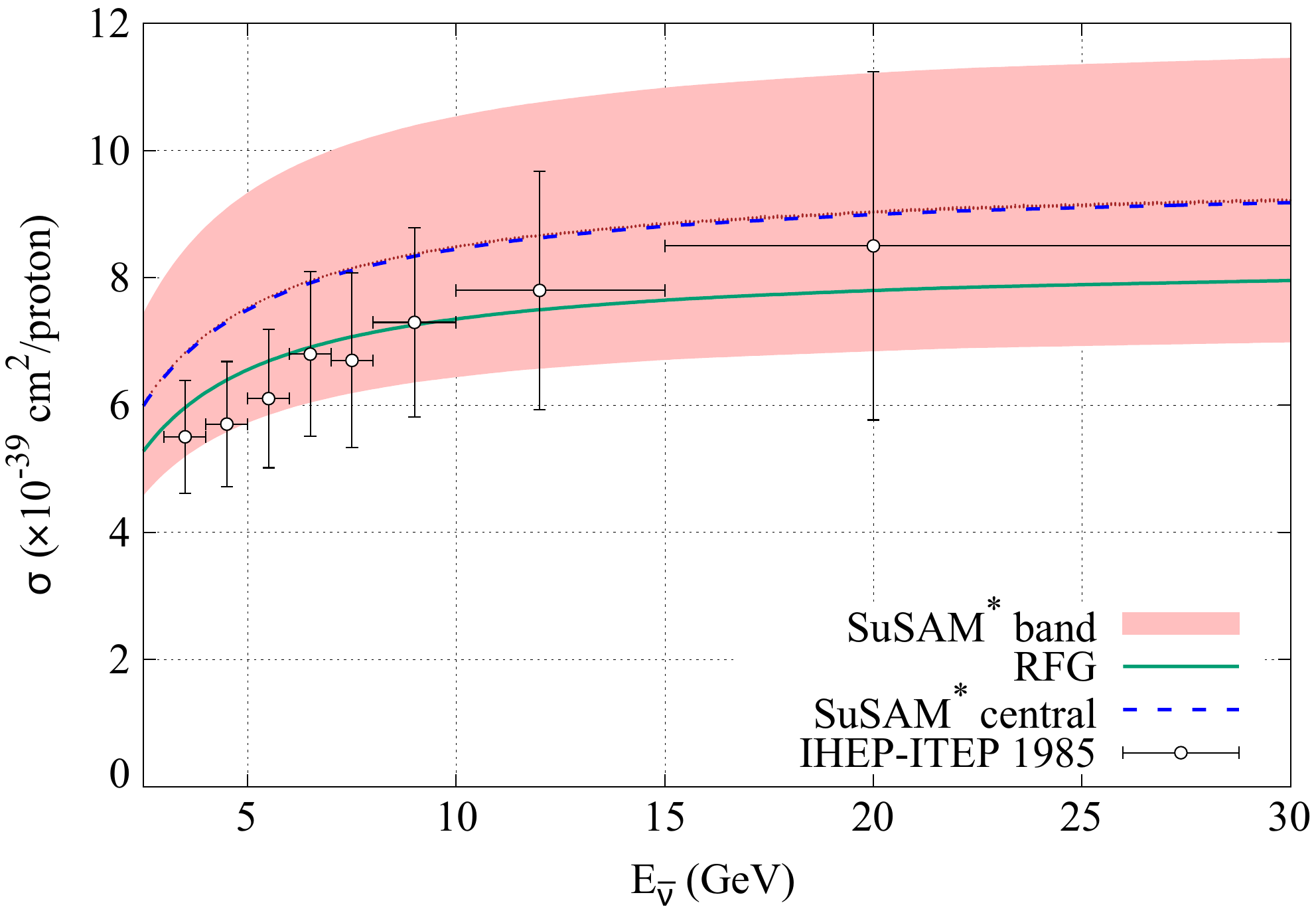}
\caption{Same as figure\ \ref{fig: total_xsect_nu_nubar} but for muon
  (anti)neutrino, (right)left panel, CCQE scattering off ${}^{27}$Al.
  The experimental points are taken from the experiment of
  Ref.\ \cite{Belikov:1983kg}. The values taken for the theoretical
  calculations of the RFG and SuSAM* models are $k_F=233$ MeV/c for
  the Fermi momentum and $m^*_N=0.80m_N$ for the relativistic
  effective mass (see ``Global fit'' parameters in Table II of
  Ref.\ \cite{Amaro:2018xdi}).}
\label{fig: total_xsect_nu_nubar_Al}
\end{figure*}

\begin{figure*}[htb]
\centering
\includegraphics[width=0.495\linewidth]{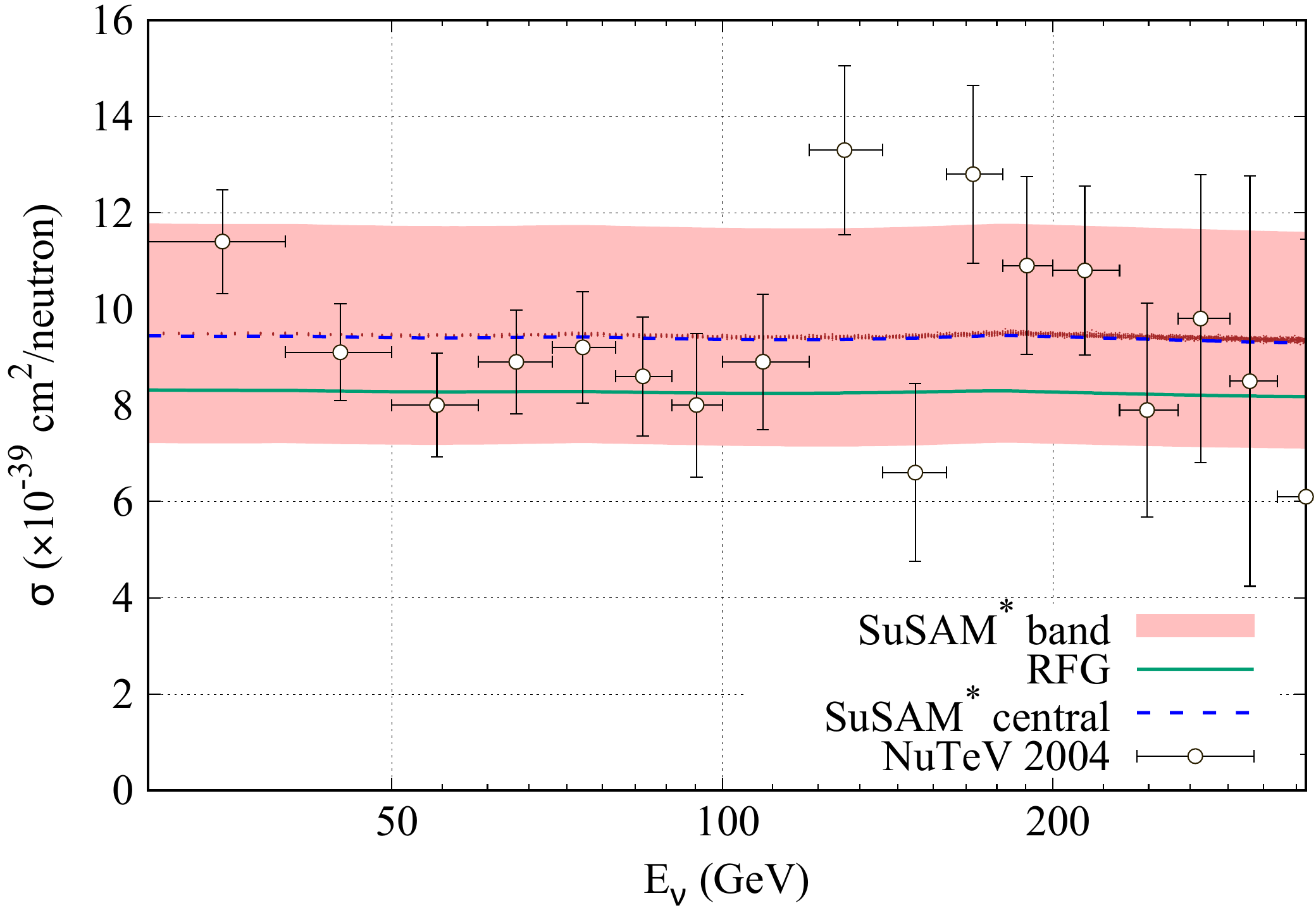}
\includegraphics[width=0.495\linewidth]{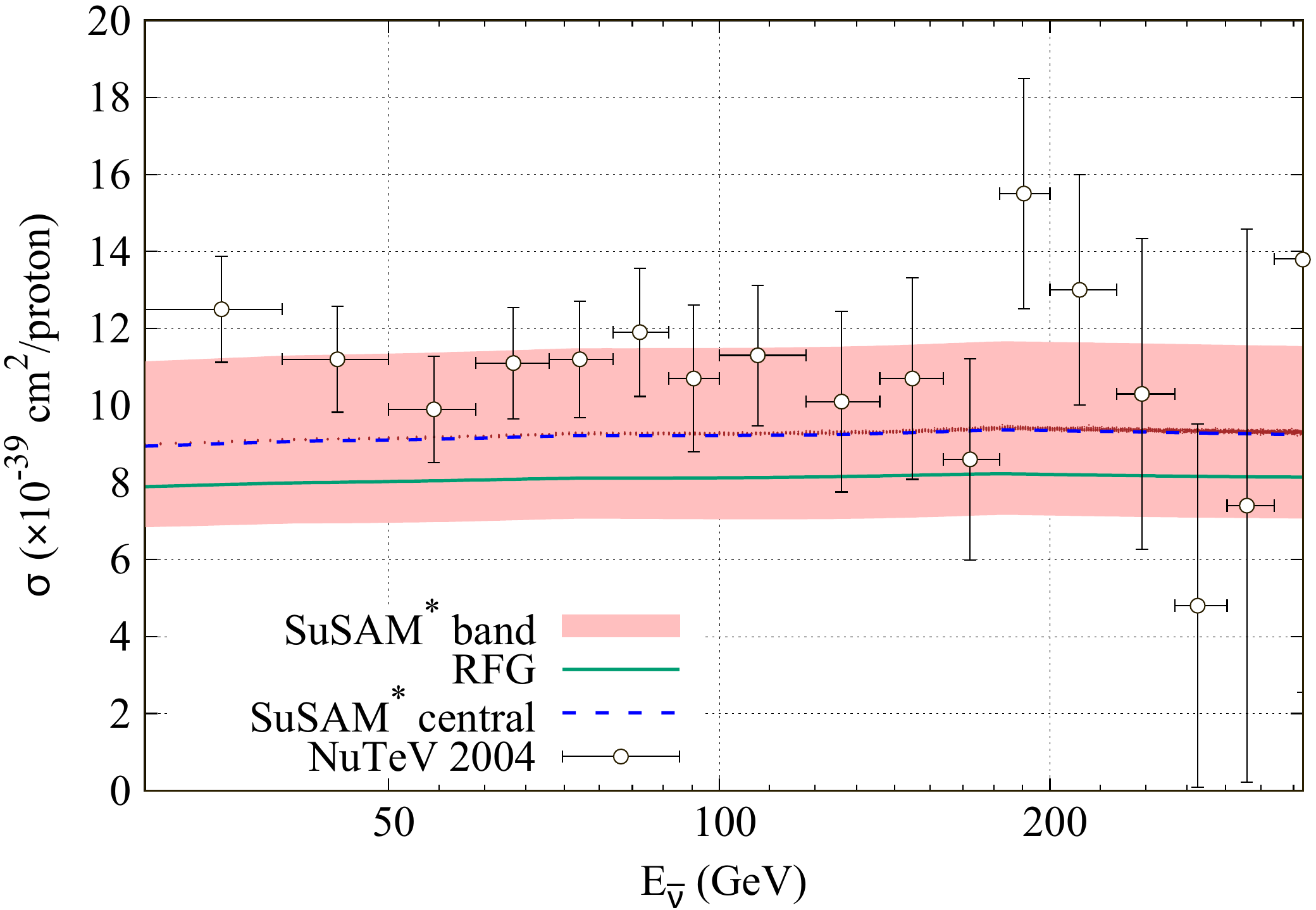}
\caption{Same as figures\ \ref{fig: total_xsect_nu_nubar} and
  \ref{fig: total_xsect_nu_nubar_Al} but for muon (anti)neutrino,
  (right)left panel, CCQE scattering off ${}^{56}$Fe.  The
  experimental points are taken from Ref.\ \cite{Suwonjandee:2004aw}.
  The values taken for the theoretical calculations of the RFG and
  SuSAM* models are $k_F=240$ MeV/c for the Fermi momentum and
  $m^*_N=0.72m_N$ for the relativistic effective mass (see ``Global
  fit'' parameters in Table II of Ref.\ \cite{Amaro:2018xdi}).}
\label{fig: total_xsect_nu_nubar_Fe}
\end{figure*}

What can also be stressed from the inspection of Fig.\ \ref{fig:
  total_xsect_nu_nubar} is the effect of nuclear correlations in the
integrated cross section. The RFG model does not contain nuclear
correlations, not either high momentum components in its nuclear
ground state. However, the SuSAM* model does contain them
phenomenologically, because its scaling function has been fitted to a
selected sample of ``QE'' electron scattering data extracted from the
inclusive ($e,e^\prime$) reaction data from a large list of different
nuclear targets (see Ref.\ \cite{Amaro:2018xdi}). Therefore, the
SuSAM* contains high momentum components in its nuclear model,
although phenomenologically. In fact, the tails of the SuSAM* scaling
function, that extend beyond $\psi=\pm1$, partially account for these
high momentum components, producing the enlarging of the available
phase space if compared with the RFG (see in particular the right
panel of Fig.\ \ref{fig: phase-space-RFG-susam}, where the cut of the
upper boundary of the SuSAM* model with the $\kappa$-axis occurs at
$\kappa=\eta^\prime_F$, which can be considered as playing the role of
an effective higher Fermi momentum). One can also compare the left
panels of Figs.\ \ref{fig: xsect-RFG-2000MeV} and \ref{fig:
  xsect-susam-2000MeV}, or the left and right panels of
Fig.\ \ref{fig: xsect-RFG-susam-2000MeV-antinu} for CCQE antineutrino
scattering, where one can observe that for the SuSAM* model, beyond
the boundaries of the RFG denoted by the dot-dashed lines, there is
still a significant region of the phase space with a non-negligible
contribution to the double differential cross section. This is mainly
responsible of the enhancement observed in the total cross section in
both panels of Fig.\ \ref{fig: total_xsect_nu_nubar} for the SuSAM*
model with respect to the RFG.

The bands in Fig.\ \ref{fig: total_xsect_nu_nubar} have been obtained
by integrating the differential cross sections calculated with the
minimum and maximum scaling functions fitted to the QE electron data
and shown in Fig.\ \ref{figure7}.  But the width of these bands do not
necessarily correspond to the theoretical error, but an upper bound of
it.  The determination of the theoretical error in the total cross
section is not the objective of this work.  Nonetheless, we have
examined a possible way to estimate this error statistically, shown in
Fig.\ \ref{fig: Monte Carlo band}, corresponding to the cross section
of neutrinos between 5000 and 100000 MeV.

For a fixed energy of the neutrino we perform the numerical
integration over $q$ and $\omega$ using random values for the scaling
function $f(\psi)$ within the band. That is, we choose $f(\psi)=x$,
where $x$ is a random value between $f_{\rm min}(\psi)$ and $f_{\rm
  max}(\psi)$ with an uniform probability distribution. A value of $x$
is sampled for each kinematics $(q,\omega)$ inside the integral. In
this way, we estimate the statistical error as if we randomly chose
experimental points within the quasi-elastic band. Repeating the
calculation of the cross section many times, we obtain a series of
points that form the band shown in Fig.\ \ref{fig: Monte Carlo
  band}. We see that now the dispersion of the points, which is the
statistical error calculated, is much smaller than the width of the
bands of Fig.\ \ref{fig: total_xsect_nu_nubar}.

Finally, in Figs.\ \ref{fig: total_xsect_nu_nubar_Al} and \ref{fig:
  total_xsect_nu_nubar_Fe} we show similar results as those of
Fig.\ \ref{fig: total_xsect_nu_nubar} for older
experiments\ \cite{Belikov:1983kg, Suwonjandee:2004aw} using
${}^{27}$Al and ${}^{56}$Fe targets, respectively. The comparison of
the SuSAM* band with the experimental results of
Ref.\ \cite{Belikov:1983kg}, shown in Fig.\ \ref{fig:
  total_xsect_nu_nubar_Al} for muon neutrinos (left panel) and muon
antineutrinos (right panel), seems to indicate that these data are
closer to the lower bounds of the band, accumulating along the RFG
curve. Nonetheless, the error bars are large enough, especially for
the antineutrino induced CCQE reactions, to conclude the same than in
Fig.\ \ref{fig: total_xsect_nu_nubar}, i.e, that the size of the
vertical error bars is similar to the theoretical uncertainty derived
from the SuSAM* model for the total CCQE cross sections as a function
of the neutrino/antineutrino energy,
$\sigma(E_{\nu(\bar{\nu})})$. Similar conclusions can be drawn by
inspecting Fig.\ \ref{fig: total_xsect_nu_nubar_Fe}, except that for
CCQE muon antineutrino scattering (right panel), the experimental
measurements have a trend to accumulate along the upper bounds of the
SuSAM* band, but with very large uncertainties.

\section{Conclusions}\label{conclusions}
In this work we have thoroughly analyzed the analytical boundaries of
the phase space for the CCQE double differential cross section
$\frac{d^2\sigma}{d\kappa\,d\lambda}$ within the scaling formalism for
the RFG model, where these boundaries can be more easily obtained.
This allows to perform the integration of this double differential
cross section only in the region where it is truly different from
zero, thus making the integration algorithm as efficient as possible.
We have also easily extended the formalism to accommodate the SuSAM*
model as well, taking into account the tails of the scaling function.

We have analyzed these double differential cross sections for CCQE
muon neutrino and antineutrino scattering off ${}^{12}$C at several
neutrino/antineutrino energies as a benchmark.  Our results show that
the $\frac{d^2\sigma}{d\kappa\,d\lambda}$ cross section has very good
properties to be implemented in the MC neutrino event generators,
basically because of two main reasons: it is quite flat regardless of
the neutrino energy; and it has a significant contribution in a larger
region of the available phase space if compared with the usual
$\frac{d^2\sigma}{dT_{\mu}\,d\cos\theta_{\mu}}$ cross section,
especially at very high neutrino energies
$E_{\nu(\bar{\nu})}\gtrsim10-20$ GeV.  We think these features of the
$\frac{d^2\sigma}{d\kappa\,d\lambda}$ cross section make it especially
well-suited to be used to generate final lepton events in any MC
generator that uses the acceptance-rejection method.

We have used the analytical boundaries obtained in this work to
integrate the double differential cross section in order to obtain the
CCQE total integrated $\sigma(E_{\nu(\bar{\nu})})$ cross section for
the two models studied in this work: RFG and SuSAM*. The effect of the
tails of the phenomenological SuSAM* scaling function, that partially
account for nuclear correlations in the model, are directly
responsible of an enhancement of about a 17--18$\%$ for intermediate
and high neutrino energies. The same conclusion can be drawn for CCQE
antineutrino scattering with roughly the same enhancement in
percentage.

Finally, we have also compared the fully integrated CCQE total cross
section in the RFG and SuSAM* models with past measurements carried
out by several experiments using targets of ${}^{12}$C, ${}^{27}$Al
and ${}^{56}$Fe. The main conclusion here is that all these
measurements lie inside the uncertainty band of the SuSAM* model,
being these uncertainties of roughly the same size as the experimental
error bars. Nonetheless, we interpret the large uncertainty band of
the SuSAM* model as lower and upper bounds for the true theoretical
error in the total CCQE neutrino/antineutrino cross sections, based
solely on our scaling analysis of electron scattering
data\ \cite{Amaro:2018xdi}. The reader, by no means, should have the
impression that this uncertainty band reflects the true error, just
that the true error must be inside the band. This last statement, at
first sight, can seem futile; but the experimental errors are also
large, and even with that, different sets of data can become
incompatible with the others.

Future works can be done based on the findings of this study.  In
particular, we are working on the study of how nuclear correlations
can be approximately and phenomenologically incorporated in the RFG
model.  \\

\section{Acknowledgements}

This work has been partially supported by the former Spanish
Ministerio de Economia y Competitividad and ERDF (European Regional
Development Fund) under contract FIS2017-85053-C2-1P, by the Junta de
Andaluc\'ia grant No.\ FQM225, by contract PID2020-114767GB-I00 funded
by MCIN/ AEI /10.13039/501100011033 and by the Russian Science
Foundation grant No.\ 18-12-00271.

\appendix

\section{Analytical formulae for the boundaries in the RFG model}
\label{appendix-a}
 The RFG model requires that $\epsilon_0\le\epsilon_F$, otherwise the
 scaling variable would take values $\left|\psi \right|>1$.  The
 physical constraint of $\tau\ge0$ implies that we have to search only
 in the region where $\kappa\ge\lambda$.  Finally, it can be shown
 that if $\kappa\ge\eta_F$ (or $q\ge2k_F$, which corresponds to the
 NPB region) then $\epsilon_0=\kappa
 \sqrt{1+1/\tau}-\lambda\ge\epsilon_F -2\lambda$ for all positive
 values of the energy transfer $\lambda$ \ \footnote{Note that
 $\epsilon_0=\kappa \sqrt{1+1/\tau}-\lambda$ can be also hold in the
 region where $\kappa<\eta_F$ if $\kappa
 \sqrt{1+1/\tau}>\epsilon_F-\lambda$ (see Eq.~\eqref{e0_cond1} and
 below).},\ \footnote{This statement can be rigorously and
 mathematically proved, but the easiest way to convince any reader of
 it is to have a look at Fig. 1 of Ref.\ \cite{Alberico:1988bv}}.  In
 the NPB region, typically $q\gtrsim500$ MeV/c, and the above
 condition can be rewritten as $\kappa\gtrsim1/4$.

As discussed in Sect.\ \ref{subsec: analytic-boundaries-RFG}, the
boundaries of the RFG scaling variable ($-1$, $+1$) are reached when
$\epsilon_0=\epsilon_F$ as it follows from
Eq.~\eqref{psi_def}. Solving the equation $\epsilon_0=\kappa
\sqrt{1+1/\tau}-\lambda=\epsilon_F$ in the NPB region (corresponding
to $\kappa\ge\eta_F$) we get two different curves in the
$(\lambda,\kappa)$ plane. One of them,
$\kappa^\text{NPB}_{+}(\lambda)$, is always greater than
$\kappa_{\text{QE}}(\lambda)$ and corresponds to $\psi=-1$, just
because $\kappa^\text{NPB}_{+}(\lambda) >\sqrt{\lambda(\lambda+1)}$
and this implies $\lambda<\tau$:
  \begin{align}
   \kappa^\text{NPB}_{+}(\lambda)
   >&\ \kappa_\text{QE}(\lambda)\equiv 
   \sqrt{\lambda(\lambda+1)}\ge 0\nonumber\\
   \Longleftrightarrow&\ 
   (\kappa^\text{NPB}_{+}(\lambda))^2-\lambda^2>\lambda
   \nonumber\\
   \Longleftrightarrow&\ 
   \tau^\text{NPB}_{+}(\lambda)
   > \lambda \nonumber\\
   \Rightarrow&\ \text{sign}(\lambda-
   \tau^\text{NPB}_{+}(\lambda))=
   -1.\label{kappaplus_sign_func}
  \end{align}
Therefore, along the curve $\kappa^\text{NPB}_{+}(\lambda)$ the
scaling variable is always equal to $-1$.  Analogously, there is
another curve, solution of $\epsilon_0= \epsilon_F$, called
$\kappa^{\text{NPB}}_{-}(\lambda)$, which is always lesser than
$\kappa_{\text{QE}}(\lambda)$, and where (by similar arguments as
those proven in Eq.\ \eqref{kappaplus_sign_func}) the scaling variable
is always equal to $+1$.

The expressions of these two curves,
$\kappa^\text{NPB}_{\pm}(\lambda)$, are given below in two different
ways in the appendices\ \ref{subsec:easy} and
\ \ref{difficult-form-boundaries}.

\subsection{Obtaining
$\kappa^\text{NPB}_{\pm}(\lambda)$}\label{subsec:easy}

One of the easiest ways to obtain the limiting curves
$\kappa^{\text{NPB}}_{\pm}(\lambda)$ can be found in Eq.\ (A.2) of
appendix A of Ref.\ \cite{RuizSimo:2017hlc} (see also Eqs.~(C7)--(C9)
of Ref.\ \cite{Amaro:2019zos}).  This latter equation allows to find
the lowest and highest $\omega$ limits for fixed $q$.  These two
limits can be found from the equation $\epsilon_0=\epsilon_F$:
\[
\omega_{\pm}=E_{k_F \pm q}-E_F=\sqrt{(k_F \pm q)^2+m^2_N}-\sqrt{k^2_F+m^2_N}.  
\]
Dividing on both sides of the above equations by $2m_N$, and writing
everything in terms of the dimensionless variables given in
Eqs.\ \eqref{lambda_def} and \eqref{etafermi_def}, we obtain the
boundaries in $\lambda_{\pm}$ for fixed $\kappa$:
\begin{equation}\label{lambda_pm_boundaries}
\lambda_{\pm}(\kappa)=\frac12 \sqrt{(\eta_F\pm2\kappa)^2+1}-
\frac12\epsilon_F. 
\end{equation}
The problem with the boundaries given in \eqref{lambda_pm_boundaries}
is that they are given as curves $\lambda=\lambda(\kappa)$, whereas we
want them in the form of curves $\kappa=\kappa(\lambda)$.  Thus, we
have to find the inverse functions. Then, writing
$\lambda_{+}(\kappa)\equiv \frac12 \sqrt{(\eta_F+2\kappa)^2+1}-
\frac12\epsilon_F = \lambda$, and solving for $\kappa$, we obtain the
lower bound $\kappa^\text{NPB}_{-}(\lambda)$, which is given by
\begin{eqnarray}\label{kappa_minus_lambda_func}
\kappa^\text{NPB}_{-}(\lambda)&=&\frac12 
\sqrt{(\epsilon_F+2\lambda)^2-1} - \frac{\eta_F}{2}.
\end{eqnarray}

\begin{figure}[htb]
\centering
\includegraphics[width=\linewidth]{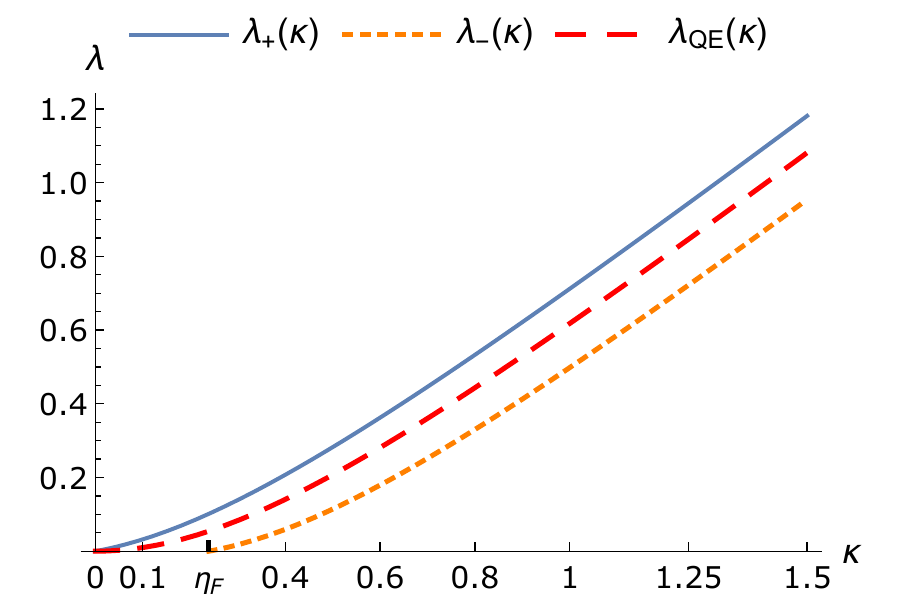}
\caption{Plot of the two limiting curves $\lambda_{\pm}(\kappa)$ as a
  function of $\kappa$ in the RFG model in the NPB region, i.e, for
  $\kappa\ge\eta_F$ (notice that for $\kappa<\eta_F$ the
  $\lambda_{-}(\kappa)$ curve reaches negative values, which are
  forbidden; this is because it is entering in the PB region).  In
  this figure, we have taken $\eta_F=0.239$.  The dashed curve
  $\lambda_\text{QE}(\kappa)= -\frac12+\frac12 \sqrt{1+4\kappa^2}$
  corresponds to the inverse of $\kappa_\text{QE}(\lambda)=\kappa$.}
\label{figure1b}
\end{figure}

In Fig.\ \ref{figure1b} one can inspect that the inversion of
$\lambda_{-}(\kappa)\equiv\frac12
\sqrt{(\eta_F-2\kappa)^2+1}-\frac12\epsilon_F=\lambda$ needs a bit of
care because it must be solved for $\kappa$ in the region where
$\lambda\ge0$.
\begin{eqnarray}
&&\lambda_{-}(\kappa)=\lambda \Longleftrightarrow
(\eta_F-2\kappa)^2=(2\lambda+\epsilon_F)^2-1 \nonumber\\
&&\Longleftrightarrow \left| \eta_F-2\kappa \right|=
\sqrt{(\epsilon_F+2\lambda)^2-1} \label{lambdam_step1}
\end{eqnarray}
if $\kappa\ge\eta_F$ then $\left| \eta_F-2\kappa
\right|=2\kappa-\eta_F$ and thus Eq.\ \eqref{lambdam_step1} becomes
\begin{equation}\label{kappa_plus_lambda_func}
\kappa^\text{NPB}_{+}(\lambda)=\frac12\sqrt{(\epsilon_F+2\lambda)^2-1}
+ \frac{\eta_F}{2}.
\end{equation}
Eqs.\ \eqref{kappa_minus_lambda_func} and
\eqref{kappa_plus_lambda_func} are plotted as the thick solid and
short-dashed lines of Fig.\ \ref{figure1}.

\subsection{Alternative form of obtaining
$\kappa^\text{NPB}_{\pm}(\lambda)$}\label{difficult-form-boundaries}
In the NPB region, it is true that
$\epsilon_0=\kappa\sqrt{1+1/\tau}-\lambda$. The maximum value is
$\epsilon_0=\epsilon_F$ and this last equation defines two curves in
the $(\lambda,\kappa)$ plane. Taking the square and using
$\tau=\kappa^2-\lambda^2$, we obtain the following bi-quadratic
equation in $\kappa$:
\begin{equation*}
\kappa^4-\left(2\lambda^2+\phi\right)\kappa^2 + 
\lambda^2\left(\lambda^2+\phi+1\right)=0,
\end{equation*}
where $\phi=\epsilon_F(\epsilon_F+2\lambda)-1$. Now, making a change
in a variable $t\equiv \kappa^2$ we arrive to a quadratic equation
\begin{equation}\label{t2_polynomial}
t^2 - \left(2\lambda^2+\phi\right) t + \lambda^2\left(\lambda^2+\phi+1\right)=0,
\end{equation}
whose two roots are
\begin{equation}\label{tpm_lambda_equation}
t_{\pm}(\lambda)=\frac{2\lambda^2+\phi
\pm\sqrt{\phi^2-4\lambda^2}}{2}
\end{equation}
It can be easily shown that the discriminant of
Eq.~\eqref{tpm_lambda_equation} is always positive for $\lambda\ge0$,
because it can be written as
\begin{equation}\label{discriminant}
\phi^2-4\lambda^2=
\eta^2_F\left(4\lambda^2+4\lambda\epsilon_F+\eta^2_F\right)>0\quad
\text{if} \quad \lambda\ge0.
\end{equation}
This ensures that the roots $t_{\pm}(\lambda)$ are real. With this, we
can write Eq.\ \eqref{t2_polynomial} as
\begin{equation}\label{kappa2_factorized}
\left[\kappa^2-t_{+}(\lambda)\right]\left[\kappa^2-t_{-}(\lambda)\right]=0.
\end{equation}
It is also easy to demonstrate that both roots $t_{\pm}(\lambda)$,
besides being real, are also positive. To this end we first write the
negative of the coefficient of $t$ in Eq.\ \eqref{t2_polynomial} as
\begin{equation}\label{b_coefficient}
2\lambda^2+\phi=
2\lambda^2+2\lambda\epsilon_F+\eta^2_F > 0 \quad \text{if}\quad \lambda\ge0.
\end{equation}
With this, it is obvious that $t_{+}(\lambda)$ is positive for
$\lambda\ge0$. To demonstrate the same for $t_{-}(\lambda)$, it is
enough to prove that the square of Eq.\ \eqref{b_coefficient} is
greater than the discriminant given in Eq.\ \eqref{discriminant}. This
comes from
\begin{eqnarray}
&&\left( 
2\lambda^2+2\lambda\epsilon_F+\eta^2_F
\right)^2\ge \eta^2_F\left(4\lambda^2+4\lambda\epsilon_F+
\eta^2_F\right) \Longleftrightarrow \nonumber\\
&&4\lambda^2\left(\lambda^2 + 2\epsilon_F\lambda + \epsilon^2_F\right)\ge0,
\nonumber
\end{eqnarray}
which is true if $\lambda\ge0$. With these proofs we can be sure that
the four roots of $\kappa$ in Eq.\ \eqref{kappa2_factorized} are all
real as well.  This means that the boundary in the
$(\lambda,\kappa)$-plane where $-1\le \psi \le 1$ is bounded by the
curves
\begin{equation}\label{upper_lower_boundary}
\kappa^{\text{NPB}}_{\pm}(\lambda)=\sqrt{\frac{\left(2\lambda^2+\phi
\right)\pm\sqrt{\phi^2-4\lambda^2}}{2}},
\end{equation}
in the NPB region, i.e, for $\kappa\ge\eta_F$.

It could seem that these two curves given by
Eq.\ \eqref{upper_lower_boundary} are totally different from those
obtained in appendix\ \ref{subsec:easy} and given in
Eqs.\ \eqref{kappa_minus_lambda_func} and
\eqref{kappa_plus_lambda_func}, but they are actually the same, and
already plotted in Fig.\ \ref{figure1}. One way of proving this is by
raising to the square the $\kappa^\text{NPB}_{\pm}(\lambda)$ functions
obtained in appendix\ \ref{subsec:easy}, given by
Eqs.\ \eqref{kappa_minus_lambda_func} and
\eqref{kappa_plus_lambda_func}; then an easy but lengthy algebra
manipulation can demonstrate that the square of
Eq.\ \eqref{kappa_plus_lambda_func} is equal to $t_{+}(\lambda)$ and
that the square of Eq.\ \eqref{kappa_minus_lambda_func} is also equal
to $t_{-}(\lambda)$, both jointly given in
Eq.\ \eqref{tpm_lambda_equation}.

\subsection{Obtaining $\kappa^\text{PB}_{\pm}(\lambda)$}
\label{PB-region}

Up to now we have been discussing the boundaries in the
$(\lambda,\kappa)$-plane of the NPB region, where the following
identity holds true
\begin{equation}\label{e0_cond1}
 \epsilon_0\equiv\max\left( 
 \kappa
 \sqrt{1+\frac{1}{\tau}}
 -\lambda,\epsilon_F-
 2\lambda
 \right)=\kappa
 \sqrt{1+\frac{1}{\tau}}
 -\lambda.
\end{equation}
As it was stated above, Eq.\ \eqref{e0_cond1} always holds when
$\kappa\ge\eta_F$, but not necessarily when $\kappa<\eta_F$.  We will
see below that for $\lambda\le\kappa<\eta_F$ (since $\tau\ge0$), there
are some regions in the $(\lambda,\kappa)$-plane where $\epsilon_0$
can be equal to the second argument of the maximum function appearing
in Eq.\ \eqref{e0_cond1}, while there are other regions where
$\epsilon_0$ is still equal to the first argument of the maximum
function.

In order to delimit these boundaries note that
\begin{equation}\label{PB_cond1}
 \kappa
 \sqrt{1+\frac{1}{\tau}}
 -\lambda\le \epsilon_F-2\lambda
 \Longleftrightarrow
 \kappa
 \sqrt{1+\frac{1}{\tau}}\le
 \epsilon_F-\lambda.
 \end{equation}
The latter inequality defines some region in the
$(\lambda,\kappa)$-plane. As both sides of the inequality given in
Eq.\ \eqref{PB_cond1} are positive (because we are looking for
solutions where $\lambda < \eta_F < \epsilon_F$), taking the square,
substituting $\tau=\kappa^2-\lambda^2$, and rearranging terms we
finish with another bi-quadratic inequality for $\kappa$:
\begin{equation}\label{biquadratic2}
 \kappa^4 -\rho\kappa^2 + \left(
 \lambda\epsilon_F-\lambda^2
 \right)^2\le 0,
\end{equation}
where $\rho=2\lambda^2-2\lambda\epsilon_F+\eta^2_F$.  Performing the
usual trick of solving the inequality by making the change of variable
$u\equiv \kappa^2$, we obtain that the equality holds for
\begin{equation}\label{upm_sols1}
 u_{\pm}(\lambda)=
 \frac{\rho\pm 
 \sqrt{\rho^2-4\left(\lambda
 \epsilon_F-\lambda^2
 \right)^2}}{2}.
\end{equation}
For Eq.\ \eqref{upm_sols1} to have real solutions, the discriminant
must be positive at least in the region of $\lambda$-values where we
are seeking a solution.

It is not difficult to write the condition for the discriminant to be
positive as
\begin{eqnarray}
 &&\rho^2-4\left(\lambda
 \epsilon_F-\lambda^2
 \right)^2\ge0 \Longleftrightarrow
 \nonumber\\
 &&\left[\rho + 2 
 \left(\lambda\epsilon_F-
 \lambda^2\right)
 \right]\nonumber\left[\rho - 2 
 \left(\lambda\epsilon_F-
 \lambda^2\right)
 \right]\ge0 \Longleftrightarrow\\
&& 4\lambda^2-
 4\lambda\epsilon_F + \eta^2_F
 \ge0.\label{discrim_cond1}
\end{eqnarray}
The last equality has two roots for $\lambda$. They are
\begin{equation}
\label{discrim_roots1}
 \lambda_{\pm}=
 \frac{\epsilon_F\pm 1}{2}>0.
\end{equation}
Hence, the last inequality of Eq.\ \eqref{discrim_cond1} can be
written as
\begin{equation}
\label{discrim_roots}
 4(\lambda-\lambda_{+})
 (\lambda-\lambda_{-})\ge 0.
\end{equation}
The only meaningful solution to Eq.\ \eqref{discrim_roots} is that
$\lambda\le\lambda_{-} <\lambda_{+}$ \footnote{The other possibility,
i.e, that $\lambda\ge\lambda_{+} >\lambda_{-}$ can be ruled out
because then $\lambda\ge (\epsilon_F+1)/2>1>\eta_F\equiv k_F/m_N$
(even with effective nucleon masses as in the SuSAM* model, we will
always have that the Fermi momentum is smaller than the nucleon mass,
regardless of this mass being the free nucleon mass or the
relativistic effective one), and we are seeking solutions in the
region where $\lambda< \eta_F$.}. The next step is to see if
$\lambda_{-}$ is lesser than $\eta_F$ or not, because if so then the
interval in $\lambda$ where to have real roots for $u_{\pm}(\lambda)$
is further constrained compared to the interval defined by $0 \le
\lambda < \eta_F$. Clearly,
\begin{align}
 \nonumber
 \lambda_{-}\equiv&\
 \frac{\epsilon_F-1}{2}=
 \frac{\sqrt{1+\eta^2_F}-1}{2}\\
 =&\ \frac{\eta^2_F}{2\left(
 \sqrt{1+\eta^2_F}+1
 \right)}<\frac{\eta^2_F}{4}
 <\eta_F, \label{lambda_minus}
\end{align}
where the last inequalities hold because $0<\eta_F<1$, which implies
that $\eta^2_F < \eta_F$.  So we can conclude that the discriminant of
Eq.\ \eqref{upm_sols1} is positive and $u_{\pm}(\lambda)$ are real
roots of Eq.\ \eqref{biquadratic2} for $0\le \lambda \le \lambda_{-} <
\eta_F$.  The next step is wondering about the sign and magnitude of
the coefficient of $\kappa^2$ in Eq.\ \eqref{biquadratic2} in the
region of $\lambda$-values between 0 and $\lambda_{-}$.  The reason
for this is because depending upon its sign and magnitude, the roots
$u_{\pm}(\lambda)$ can be negative and we want them to be positive
because
$u_{\pm}(\lambda)=\left[\kappa^{\text{PB}}_{\pm}(\lambda)\right]^2$
should be the square of real roots of Eq.\ \eqref{biquadratic2}.  To
this end, we set out the following inequality and seek for their
solutions:
\begin{equation}
\label{b_coefficient_PB_reg}
\rho\ge 0.
\end{equation}
The equality $\rho=0$ has two positive roots for $\lambda$:
\begin{equation*}
 \lambda^\prime_{\pm}=
 \frac{\epsilon_F\pm
 \sqrt{2-\epsilon^2_F}}{2}=
 \frac{\sqrt{1+\eta^2_F}\pm
 \sqrt{1-\eta^2_F}}{2}>0.
\end{equation*}
We have to compare them with $\lambda_{-}$, given in
Eq.\ \eqref{discrim_roots1}.  The reason for this is because if any of
the two new roots $\lambda^\prime_{\pm}$ is lesser than $\lambda_{-}$,
then the interval in $\lambda$ where to seek the boundary of the PB
region can be, again, further constrained from the last condition
$0\le\lambda\le\lambda_{-}$. It is easy to see that
$\lambda^\prime_{+}$ is clearly greater than $\lambda_{-}$:
\begin{equation*}
 \lambda^\prime_{+}=
 \frac{\epsilon_F+
 \sqrt{1-\eta^2_F}}{2} >
 \frac{\epsilon_F}{2} >
 \frac{\epsilon_F-1}{2}\equiv
 \lambda_{-}
\end{equation*}
On the other hand, it is also straightforward to see that
$\lambda^\prime_{-}$ is greater than $\lambda_{-}$ as well:
\begin{align*}
 \lambda^\prime_{-}\equiv&\
 \frac{\sqrt{1+\eta^2_F}-
 \sqrt{1-\eta^2_F}}{2} \\
 =&\ \frac{\eta^2_F}{
 \sqrt{1+\eta^2_F}+
 \sqrt{1-\eta^2_F}}
 >\frac{\eta^2_F}{4}>
 \lambda_{-}, 
\end{align*}
where in the second step we have multiplied and divided by
$\sqrt{1+\eta^2_F}+ \sqrt{1-\eta^2_F}$.

Finally, it is worth noting that the inequality
\eqref{b_coefficient_PB_reg} can be rewritten as
\begin{equation}
\label{b_coefficient_PB_reg_root_form}
 2(\lambda-\lambda^\prime_{+})
 (\lambda-\lambda^\prime_{-})
 \ge0,
\end{equation}
which is absolutely fulfilled if $\lambda\le\lambda_{-}<
\lambda^\prime_{-}< \lambda^\prime_{+}$, because then both parentheses
in Eq.\ \eqref{b_coefficient_PB_reg_root_form} are negative and their
product is positive. Having found the most restrictive region in the
$\lambda$ variable where Eqs.\ \eqref{discrim_cond1} and
\eqref{b_coefficient_PB_reg} are simultaneously fulfilled, we can
assert that the roots $u_{\pm}(\lambda)$ given in
Eq.\ \eqref{upm_sols1} are both real and
\emph{positive} \footnote{This last feature can be stated because the
discriminant of Eq.\ \eqref{upm_sols1} is positive and lesser than
$\rho^2$.  Thus the square root of the discriminant is also lesser
than $\rho$, and then $u_{-}(\lambda)$ is necessarily positive in the
region where $0\le\lambda\le \lambda_{-}$.}. Thus, we can rewrite
Eq.\ \eqref{biquadratic2} as
 \begin{equation*}
\left[\kappa^2-u_{+}(\lambda)
 \right]
 \left[\kappa^2-u_{-}(\lambda)
 \right]\le0 
 \end{equation*}
From the above inequality, it is obvious that the only solution is
\begin{equation*}
\sqrt{u_{-}(\lambda)}\le \kappa \le \sqrt{u_{+}(\lambda)},
\end{equation*}
where we have taken the square roots because the solutions
$u_{\pm}(\lambda)$ and $\kappa$ are all positive.  Therefore, we can
conclude from all this discussion that the region where PB makes
$\epsilon_0$ to be equal to the second argument,
$\epsilon_F-2\lambda$, of the maximum function displayed in
Eq.\ \eqref{e0_cond1}, corresponds to the region
$\kappa^\text{PB}_{-}(\lambda)\le \kappa \le
\kappa^\text{PB}_{+}(\lambda)$ in the region where
$0\le\lambda\le\lambda_{-}$ with
\begin{equation*}
 \kappa^\text{PB}_{\pm}(\lambda)=
 \sqrt{\frac{\rho \pm  
 \sqrt{\rho^2-4\left(\lambda
 \epsilon_F-\lambda^2
 \right)^2}}{2}}
\end{equation*}

It is also easy to find the values of
$\kappa^\text{PB}_{\pm}(\lambda)$ for $\lambda=0$ and
$\lambda=\lambda_{-}$. They are
\begin{align*}
\kappa^\text{PB}_{\pm}(\lambda_{-})&=\frac{\eta_F}{2}\\
\kappa^\text{PB}_{-}(0)=&\ 0, \quad
\kappa^\text{PB}_{+}(0)=\eta_F.
\end{align*}
The first two above equations can be easily found by noticing that,
for $\lambda=\lambda_{-}=(\epsilon_F-1)/2$, the discriminant of
$u_{\pm}(\lambda)$ is exactly zero (see Eq.\ \eqref{discrim_roots}),
and then there is no difference between
$\kappa^{\text{PB}}_{+}(\lambda_{-})$ and
$\kappa^\text{PB}_{-}(\lambda_{-})$. It is also easy to notice that
$\kappa_{\text{QE}}(\lambda_{-})=\eta_F/2$:
\begin{equation*}
\kappa_\text{QE}(\lambda_{-})=\sqrt{\lambda_{-}
(\lambda_{-}+1)}=\sqrt{\frac{\epsilon^2_F-1}{4}}=\frac{\eta_F}{2}.
\end{equation*}
This means that for $\lambda=\lambda_{-}$,
$\epsilon_0=\epsilon_F-2\lambda_{-}=1$ and then the scaling variable
at the point $(\lambda,\kappa)=(\lambda_{-},\eta_F/2)$ is exactly 0
(see definition given in Eq.\ \eqref{psi_def}).

\subsection{Obtaining $\kappa^{\text{lepton}}_{\text{max,min}}(\lambda)$}
\label{appendix-a4}

As both $\kappa^\text{NPB}_{\pm}(\lambda)$ curves are increasing
functions of $\lambda$ (see for example Figs.\ \ref{figure1},
\ref{figure3} or \ref{figure4}), it is interesting to look for the
cutting points between $\kappa^{\text{lepton}}_{\max}(\lambda)$ and
$\kappa^\text{NPB}_{\pm}(\lambda)$, or between
$\kappa^\text{lepton}_{\min}(\lambda)$ and
$\kappa^\text{NPB}_{\pm}(\lambda)$, if any. These cutting points will
help us to constrain and to understand the form of the available
phase-space in the RFG model for a fixed neutrino/antineutrino reduced
energy $\epsilon_{\nu}$.

We can start by looking for the $\lambda$ value where
$\kappa^{\text{NPB}}_{+}(\lambda)=\kappa^\text{lepton}_{\max}(\lambda)$.
This value can give us the $\lambda$ point where the minimum function
appearing in expression\ \eqref{eq:rfg_boundary_exact} changes from
selecting one curve to the other. In this case, to obtain the solution
for $\lambda$, it is better to use the expression for
$\kappa^\text{NPB}_{+}(\lambda)$ given in
Eq.\ \eqref{kappa_plus_lambda_func} rather than that given in
Eq.\ \eqref{upper_lower_boundary}, although both are equivalent, just
because the first one is much simpler to manipulate. To obtain the
solution it is necessary to square twice the equation, and we finish
with the following second degree equation for $\lambda$ after a
lengthy algebra manipulation:
\begin{equation*}
a\lambda^2+b\lambda+c=0
\end{equation*}
with 
\begin{gather*}
a\equiv1+4\epsilon_{\nu} \left(\epsilon_F + \eta_F \right), \\
b\equiv2\epsilon_F \left(\widetilde{m}^2_{\mu} + \epsilon_{\nu} \eta_F \right)
+2 \epsilon_{\nu} \left[   
\eta^2_F + 2 \widetilde{m}^2_{\mu} - 2 \epsilon_{\nu} \left(
\epsilon_F + \eta_F \right)\right], \\
c\equiv \widetilde{m}^2_{\mu} \left( 
\eta^2_F - 2 \epsilon_{\nu} \eta_F + \widetilde{m}^2_{\mu}
\right).
\end{gather*}
The above equation has two roots. Only the solution with the positive
square root is positive for some values of $\epsilon_{\nu}$. The other
solution is always negative and we discard it. The relevant solution,
which we call $\lambda_{++}$, is given by
\begin{align}
&\lambda_{++}= \frac{\zeta_+ \left(  2 \epsilon_{\nu} - \eta_F \right)
- \widetilde{m}^2_{\mu} \left( \epsilon_F + 2 \epsilon_{\nu}  \right)
}{\left(1+4\zeta_+\right)} \nonumber \\
& + \frac{\left| 2\epsilon_{\nu} - \eta_F  \right|
\sqrt{\widetilde{m}^4_{\mu} - 
\widetilde{m}^2_{\mu} + \zeta_+
\left(\zeta_+-2\widetilde{m}^2_{\mu} \right)}}
{\left(1+4\zeta_+\right)}, \label{lambda++solution}
\end{align} 
where we define $\zeta_+=\epsilon_{\nu}\left(\epsilon_F +
\eta_F\right)$.  In the above equation \eqref{lambda++solution} there
is a value for the reduced neutrino energy $\epsilon_{\nu}$ for which
$\lambda_{++}=0$. This value can be found by equating the numerator of
\eqref{lambda++solution} to zero and solving for
$\epsilon_{\nu}$. This value of $\epsilon_{\nu}$ is precisely that for
which the cut point between $\kappa^\text{NPB}_{+}(\lambda)$ and
$\kappa^\text{lepton}_{\max}(\lambda)$ occurs at $\lambda=0$. Again, a
lengthy and tedious algebraic manipulation leaves us with another
second degree equation in the variable $\epsilon_{\nu}$:
\begin{equation*}
a\,\epsilon_{\nu}^2+b\,\epsilon_{\nu}+c=0
\end{equation*}
with
\begin{gather*}
a\equiv8\eta_F\left(\eta_F+\epsilon_F\right), \\
b\equiv 2 \left(\eta_F - 2 \left( \eta_F + \epsilon_F \right)\left(  \eta^2_F + \widetilde{m}^2_{\mu} \right)\right), \\
c\equiv -\eta^2_F - \widetilde{m}^2_{\mu}.
\end{gather*}
Only the solution with the positive square root is again positive,
while the other solution is always negative and we discard it. The
meaningful solution is
\footnote{Another way to arrive to the same solution would have been
to solve $\kappa^\text{NPB}_{+}(0)= \kappa^{\text{lepton}}_{\max}(0)$
for $\epsilon_{\nu}$. As $\kappa^{\text{NPB}}_{+}(0)=\eta_F$ (see
equation \eqref{kappa_plus_lambda_func}), and 
$\kappa^{\text{lepton}}_{\max}(0)=\epsilon_{\nu}+\sqrt{\epsilon^2_{\nu} -
\widetilde{m}^2_{\mu}}$ (see definition given in
\eqref{kappa_max_lepton}), the solution $\epsilon_{\nu_{+}}$ would
have been obtained in a much simpler way.}
\begin{equation}\label{epsilon_nu_+}
\epsilon_{\nu_{+}}= \frac{\eta^2_F + 
\widetilde{m}^2_{\mu}}{2 \eta_F} > 0.
\end{equation}
Notice that the solution given in the above Eq.\ \eqref{epsilon_nu_+}
depends both on the reduced final lepton mass and on a nuclear
property, namely the Fermi momentum (in units of the nucleon mass).

Also note that, given the behavior of the curves
$\kappa^{\text{lepton}}_{\max}(\lambda)$ (which is a monotonically
decreasing function of $\lambda$, as already mentioned in
Sect.\ \ref{lepton-kinem-bound}), and $\kappa^\text{NPB}_{+}(\lambda)$
(which is monotonically increasing), for $\widetilde{m}_{\mu}
\leqslant \epsilon_{\nu} \leqslant \epsilon_{\nu_{+}}$ the upper limit
of the phase space of the QE double differential cross section with
respect to final lepton variables in the RFG model is bounded only by
the curve $\kappa^\text{lepton}_{\max}(\lambda)$.  Or said in other
words, if $\widetilde{m}_{\mu} \leqslant \epsilon_{\nu} \leqslant
\epsilon_{\nu_{+}}$, then the minimum function of the right-hand side
of inequality\ \eqref{eq:rfg_boundary_exact} is always the curve
$\kappa^\text{lepton}_{\max}(\lambda)$ for all the allowed $\lambda$
values.

On the other hand, the presence of the reduced lepton mass in
Eq.\ \eqref{epsilon_nu_+} means that the necessary ranges of neutrino
energies to allow the lepton kinematic constraints to determine by
themselves the upper boundary of the phase space, depend a lot on the
kind of neutrino flavor for charged current processes. For instance,
for tau neutrinos and for typical values of Fermi momenta ($\eta_F
\simeq 0.24$), the neutrino energy
$E_{\nu_{+}}=2m_N\,\epsilon_{\nu_{+}} \simeq 3.72$ GeV, which is
already a quite large neutrino energy for the intermediate neutrino
energy range. However, for muon neutrinos, $E_{\nu_{+}}\simeq 240$
MeV, which is a quite low neutrino energy. Of course, these values are
completely related to the threshold neutrino energies to produce a
$\tau$ lepton or a muon in charged-current elastic scattering with
nucleons, respectively.

Now we can look for the cut point between
$\kappa^\text{lepton}_{\max}(\lambda)$ and
$\kappa^\text{NPB}_{-}(\lambda)$, which will occur for a $\lambda$
value larger \footnote{The reason for this statement is because
$\kappa^\text{lepton}_{\max}(\lambda)$ is monotonically decreasing
with $\lambda$ and $\kappa^\text{NPB}_{-}(\lambda)$ is a monotonically
increasing function of $\lambda$, but smaller than
$\kappa^\text{NPB}_{+}(\lambda)$.} than $\lambda_{++}$, given in
Eq.\ \eqref{lambda++solution}. Again, after a lengthy calculation,
solving $\kappa^\text{NPB}_{-}(\lambda)=
\kappa^\text{lepton}_{\max}(\lambda)$ for $\lambda$, we find two
roots:
\begin{align}
\lambda_{\pm-}&= 
\frac{\zeta_- \left(  2 \epsilon_{\nu} + \eta_F \right)
- \widetilde{m}^2_{\mu} \left( \epsilon_F + 
2 \epsilon_{\nu}  \right)}{\left(1+4\zeta_-\right)} \nonumber \\
& \pm \frac{\left( 2\epsilon_{\nu} + \eta_F  \right)
\sqrt{\widetilde{m}^4_{\mu} - 
\widetilde{m}^2_{\mu} + \zeta_-\left( \zeta_-
 - 2  \widetilde{m}^2_{\mu} \right)}}{\left(1+4\zeta_-\right)},
 \label{lambdapm-solution}
\end{align}
where we define $\zeta_-=\epsilon_{\nu}\left(\epsilon_F -
\eta_F\right)$.  Both roots are physical (not complex numbers) for
some reduced neutrino energies which depend on the model scaling
function (see Fig.\ \ref{fig: cuts_lambdapm}).  Note that the
expression for the first root, $\lambda_{+-}$, corresponds to the
solution $\lambda_{++}$ given in Eq.\ \eqref{lambda++solution} if one
makes the replacement $\eta_F \mapsto -\eta_F$, which makes sense
because the only difference between $\kappa^\text{NPB}_{+}(\lambda)$
(given in Eq.\ \eqref{kappa_plus_lambda_func}), and
$\kappa^{\text{NPB}}_{-}(\lambda)$
(Eq.\ \eqref{kappa_minus_lambda_func}) is the sign of $\eta_F$.  The
other root, $\lambda_{--}$, when it is physical, always corresponds to
the cutting point between the curves $\kappa^\text{NPB}_{-}(\lambda)$
and $\kappa^{\text{lepton}}_{\min}(\lambda)$.  This latter solution
corresponds to the $\lambda$ point where the maximum function
appearing on the left-hand side of the inequality
\eqref{eq:rfg_boundary_exact} changes from one of its arguments to the
other. It could seem striking at first glance that this solution
appears when we have not used at all the
$\kappa^\text{lepton}_{\min}(\lambda)$ curve to obtain it, but (as it
can be seen from Fig.\ \ref{fig: phase-space-RFG}) the curves
$\kappa^\text{lepton}_{\max,\min}(\lambda)$ form actually two
different branches of the same unique curve, namely, $\left(\kappa -
\epsilon_{\nu} \right)^2= \left( \epsilon_{\nu} - \lambda \right)^2 -
\widetilde{m}^2_{\mu}$.

\section{Extension of the formulae to the SuSAM* model}
\label{appendix-b}

As discussed in Sects.\ \ref{susam_model} and
\ref{subsec:total_xsect}, all the above analytical formulae obtained
for the RFG model in the appendix\ \ref{appendix-a} can be used for
the SuSAM* model by just replacing
$\epsilon_F\rightarrow\epsilon^\prime_F$ and
$\eta_F\rightarrow\eta^\prime_F\equiv\sqrt{\epsilon^{\prime\,2}_F-1}$.

In this appendix\ \ref{appendix-b} we demonstrate that
$\kappa^{\text{NPB}}_{-}(0)=0$ and $\kappa^{\text{NPB}}_{+}(0)>\eta_F$
for the SuSAM* model, as stated in
Sect.\ \ref{susam_model}\ \footnote{For the demonstrations provided
here, we drop the label NPB from all the expressions in order to
shorten the already cumbersome notation}.

It is straightforward to prove that $\kappa^{2}_{-}(0)= t_{-}(0)=0$,
either from Eqs.\ \eqref{tpm_lambda_equation} or
\eqref{upper_lower_boundary}, irrespective of the values taken by
$\epsilon_F$ or $\epsilon^\prime_F$, provided that both are greater
than $1$, as it is the case. For the second demonstration we have,
from Eq.\ \eqref{tpm_lambda_equation},
\begin{align}
\kappa^2_{+}(0)=&\ t_{+}(0)=\frac{\epsilon^{\prime\,2}_F-1+
  \sqrt{\left(\epsilon^{\prime\,2}_F-1\right)^2}}{2} \nonumber \\
=&\ \epsilon^{\prime\,2}_F-1\nonumber \\ 
=&\ \left( 1 + \left( \epsilon_F
- 1 \right) \psi^2_\text{left}\right)^2-1, \nonumber
\end{align}
where $\psi_\text{left}$ is the negative value of the scaling
variable, lesser than $-1$, that one has to take to ensure that the
scaling function is negligible beyond that value. Finally, as
$\psi^2_{\text{left}}>1$, it is also true that
\begin{align}
&\ 1+\left(\epsilon_F-1 \right)\psi^2_\text{left} > \epsilon_F > 1
  \nonumber\\ 
&\ \Longrightarrow \kappa^2_{+}(0)= \left( 1 + \left(
  \epsilon_F - 1 \right) \psi^2_\text{left}\right)^2-1 > \epsilon^2_F
  -1 \equiv \eta^2_F \nonumber\\
&\ \Longleftrightarrow \kappa_{+}(0)>\eta_F, \nonumber
\end{align}
where in the first step we have multiplied the inequality
$\psi^2_{\text{left}}>1$ on both sides by $\left(\epsilon_F - 1
\right)$ without changing the direction of the inequality because
$\epsilon_F > 1$.

\bibliography{references}

\begin{thebibliography}{126}%
\makeatletter
\providecommand \@ifxundefined [1]{%
 \@ifx{#1\undefined}
}%
\providecommand \@ifnum [1]{%
 \ifnum #1\expandafter \@firstoftwo
 \else \expandafter \@secondoftwo
 \fi
}%
\providecommand \@ifx [1]{%
 \ifx #1\expandafter \@firstoftwo
 \else \expandafter \@secondoftwo
 \fi
}%
\providecommand \natexlab [1]{#1}%
\providecommand \enquote  [1]{``#1''}%
\providecommand \bibnamefont  [1]{#1}%
\providecommand \bibfnamefont [1]{#1}%
\providecommand \citenamefont [1]{#1}%
\providecommand \href@noop [0]{\@secondoftwo}%
\providecommand \href [0]{\begingroup \@sanitize@url \@href}%
\providecommand \@href[1]{\@@startlink{#1}\@@href}%
\providecommand \@@href[1]{\endgroup#1\@@endlink}%
\providecommand \@sanitize@url [0]{\catcode `\\12\catcode `\$12\catcode
  `\&12\catcode `\#12\catcode `\^12\catcode `\_12\catcode `\%12\relax}%
\providecommand \@@startlink[1]{}%
\providecommand \@@endlink[0]{}%
\providecommand \url  [0]{\begingroup\@sanitize@url \@url }%
\providecommand \@url [1]{\endgroup\@href {#1}{\urlprefix }}%
\providecommand \urlprefix  [0]{URL }%
\providecommand \Eprint [0]{\href }%
\providecommand \doibase [0]{https://doi.org/}%
\providecommand \selectlanguage [0]{\@gobble}%
\providecommand \bibinfo  [0]{\@secondoftwo}%
\providecommand \bibfield  [0]{\@secondoftwo}%
\providecommand \translation [1]{[#1]}%
\providecommand \BibitemOpen [0]{}%
\providecommand \bibitemStop [0]{}%
\providecommand \bibitemNoStop [0]{.\EOS\space}%
\providecommand \EOS [0]{\spacefactor3000\relax}%
\providecommand \BibitemShut  [1]{\csname bibitem#1\endcsname}%
\let\auto@bib@innerbib\@empty
\bibitem [{\citenamefont {Alvarez-Ruso}\ \emph {et~al.}(2014)\citenamefont
  {Alvarez-Ruso}, \citenamefont {Hayato},\ and\ \citenamefont
  {Nieves}}]{Alvarez-Ruso:2014bla}%
  \BibitemOpen
  \bibfield  {author} {\bibinfo {author} {\bibfnamefont {L.}~\bibnamefont
  {Alvarez-Ruso}}, \bibinfo {author} {\bibfnamefont {Y.}~\bibnamefont
  {Hayato}},\ and\ \bibinfo {author} {\bibfnamefont {J.~M.}\ \bibnamefont
  {Nieves}},\ }\bibfield  {title} {\bibinfo {title} {{Progress and open
  questions in the physics of neutrino cross sections at intermediate
  energies}},\ }\href {https://doi.org/10.1088/1367-2630/16/7/075015}
  {\bibfield  {journal} {\bibinfo  {journal} {New J.\ Phys.}\ }\textbf
  {\bibinfo {volume} {16}},\ \bibinfo {pages} {075015} (\bibinfo {year}
  {2014})},\ \Eprint {https://arxiv.org/abs/1403.2673} {arXiv:1403.2673
  [hep-ph]} \BibitemShut {NoStop}%
\bibitem [{\citenamefont {Balasi}\ \emph {et~al.}(2015)\citenamefont {Balasi},
  \citenamefont {Langanke},\ and\ \citenamefont
  {Mart\'{i}nez-Pinedo}}]{Balasi:2015dba}%
  \BibitemOpen
  \bibfield  {author} {\bibinfo {author} {\bibfnamefont {K.~G.}\ \bibnamefont
  {Balasi}}, \bibinfo {author} {\bibfnamefont {K.}~\bibnamefont {Langanke}},\
  and\ \bibinfo {author} {\bibfnamefont {G.}~\bibnamefont
  {Mart\'{i}nez-Pinedo}},\ }\bibfield  {title} {\bibinfo {title}
  {{Neutrino-nucleus reactions and their role for supernova dynamics and
  nucleosynthesis}},\ }\href {https://doi.org/10.1016/j.ppnp.2015.08.001}
  {\bibfield  {journal} {\bibinfo  {journal} {Prog.\ Part.\ Nucl.\ Phys.}\
  }\textbf {\bibinfo {volume} {85}},\ \bibinfo {pages} {33} (\bibinfo {year}
  {2015})},\ \Eprint {https://arxiv.org/abs/1503.08095} {arXiv:1503.08095
  [nucl-th]} \BibitemShut {NoStop}%
\bibitem [{\citenamefont {Mosel}(2016)}]{Mosel:2016cwa}%
  \BibitemOpen
  \bibfield  {author} {\bibinfo {author} {\bibfnamefont {U.}~\bibnamefont
  {Mosel}},\ }\bibfield  {title} {\bibinfo {title} {{Neutrino interactions with
  nucleons and nuclei: importance for long-baseline experiments}},\ }\href
  {https://doi.org/10.1146/annurev-nucl-102115-044720} {\bibfield  {journal}
  {\bibinfo  {journal} {Ann.\ Rev.\ Nucl.\ Part.\ Sci.}\ }\textbf {\bibinfo
  {volume} {66}},\ \bibinfo {pages} {171} (\bibinfo {year} {2016})},\ \Eprint
  {https://arxiv.org/abs/1602.00696} {arXiv:1602.00696 [nucl-th]} \BibitemShut
  {NoStop}%
\bibitem [{\citenamefont {Katori}\ and\ \citenamefont
  {Martini}(2018)}]{Katori:2016yel}%
  \BibitemOpen
  \bibfield  {author} {\bibinfo {author} {\bibfnamefont {T.}~\bibnamefont
  {Katori}}\ and\ \bibinfo {author} {\bibfnamefont {M.}~\bibnamefont
  {Martini}},\ }\bibfield  {title} {\bibinfo {title} {{Neutrino--nucleus cross
  sections for oscillation experiments}},\ }\href
  {https://doi.org/10.1088/1361-6471/aa8bf7} {\bibfield  {journal} {\bibinfo
  {journal} {J.\ Phys.\ G}\ }\textbf {\bibinfo {volume} {45}},\ \bibinfo
  {pages} {013001} (\bibinfo {year} {2018})},\ \Eprint
  {https://arxiv.org/abs/1611.07770} {arXiv:1611.07770 [hep-ph]} \BibitemShut
  {NoStop}%
\bibitem [{\citenamefont {Alvarez-Ruso}\ \emph {et~al.}(2018)\citenamefont
  {Alvarez-Ruso} \emph {et~al.}}]{Alvarez-Ruso:2017oui}%
  \BibitemOpen
  \bibfield  {author} {\bibinfo {author} {\bibfnamefont {L.}~\bibnamefont
  {Alvarez-Ruso}} \emph {et~al.} (\bibinfo {collaboration} {NuSTEC
  Collaboration}),\ }\bibfield  {title} {\bibinfo {title} {{NuSTEC White Paper:
  Status and challenges of neutrino-nucleus scattering}},\ }\href
  {https://doi.org/10.1016/j.ppnp.2018.01.006} {\bibfield  {journal} {\bibinfo
  {journal} {Prog.\ Part.\ Nucl.\ Phys.}\ }\textbf {\bibinfo {volume} {100}},\
  \bibinfo {pages} {1} (\bibinfo {year} {2018})},\ \Eprint
  {https://arxiv.org/abs/1706.03621} {arXiv:1706.03621 [hep-ph]} \BibitemShut
  {NoStop}%
\bibitem [{\citenamefont {Benhar}\ \emph {et~al.}(2017)\citenamefont {Benhar},
  \citenamefont {Huber}, \citenamefont {Mariani},\ and\ \citenamefont
  {Meloni}}]{Benhar:2015wva}%
  \BibitemOpen
  \bibfield  {author} {\bibinfo {author} {\bibfnamefont {O.}~\bibnamefont
  {Benhar}}, \bibinfo {author} {\bibfnamefont {P.}~\bibnamefont {Huber}},
  \bibinfo {author} {\bibfnamefont {C.}~\bibnamefont {Mariani}},\ and\ \bibinfo
  {author} {\bibfnamefont {D.}~\bibnamefont {Meloni}},\ }\bibfield  {title}
  {\bibinfo {title} {{Neutrino-nucleus interactions and the determination of
  oscillation parameters}},\ }\href
  {https://doi.org/10.1016/j.physrep.2017.07.004} {\bibfield  {journal}
  {\bibinfo  {journal} {Phys.\ Rept.}\ }\textbf {\bibinfo {volume} {700}},\
  \bibinfo {pages} {1} (\bibinfo {year} {2017})},\ \Eprint
  {https://arxiv.org/abs/1501.06448} {arXiv:1501.06448 [nucl-th]} \BibitemShut
  {NoStop}%
\bibitem [{\citenamefont {Giusti}\ and\ \citenamefont
  {Ivanov}(2020)}]{Giusti:2019cup}%
  \BibitemOpen
  \bibfield  {author} {\bibinfo {author} {\bibfnamefont {C.}~\bibnamefont
  {Giusti}}\ and\ \bibinfo {author} {\bibfnamefont {M.~V.}\ \bibnamefont
  {Ivanov}},\ }\bibfield  {title} {\bibinfo {title} {{Neutral current
  neutrino-nucleus scattering. Theory}},\ }\href
  {https://doi.org/10.1088/1361-6471/ab5251} {\bibfield  {journal} {\bibinfo
  {journal} {J.\ Phys.\ G}\ }\textbf {\bibinfo {volume} {47}},\ \bibinfo
  {pages} {024001} (\bibinfo {year} {2020})},\ \Eprint
  {https://arxiv.org/abs/1908.08603} {arXiv:1908.08603 [hep-ph]} \BibitemShut
  {NoStop}%
\bibitem [{\citenamefont {Amaro}\ \emph {et~al.}(2020)\citenamefont {Amaro},
  \citenamefont {Barbaro}, \citenamefont {Caballero}, \citenamefont
  {Gonz\'{a}lez-Jim\'{e}nez}, \citenamefont {Megias},\ and\ \citenamefont
  {Ruiz~Simo}}]{Amaro:2019zos}%
  \BibitemOpen
  \bibfield  {author} {\bibinfo {author} {\bibfnamefont {J.~E.}\ \bibnamefont
  {Amaro}}, \bibinfo {author} {\bibfnamefont {M.~B.}\ \bibnamefont {Barbaro}},
  \bibinfo {author} {\bibfnamefont {J.~A.}\ \bibnamefont {Caballero}}, \bibinfo
  {author} {\bibfnamefont {R.}~\bibnamefont {Gonz\'{a}lez-Jim\'{e}nez}},
  \bibinfo {author} {\bibfnamefont {G.~D.}\ \bibnamefont {Megias}},\ and\
  \bibinfo {author} {\bibfnamefont {I.}~\bibnamefont {Ruiz~Simo}},\ }\bibfield
  {title} {\bibinfo {title} {{Electron- versus neutrino-nucleus scattering}},\
  }\href@noop {} {\bibfield  {journal} {\bibinfo  {journal} {J.\ Phys.\ G}\
  }\textbf {\bibinfo {volume} {47}},\ \bibinfo {pages} {124001} (\bibinfo
  {year} {2020})},\ \Eprint {https://arxiv.org/abs/1912.10612}
  {arXiv:1912.10612 [nucl-th]} \BibitemShut {NoStop}%
\bibitem [{\citenamefont {Sajjad~Athar}\ and\ \citenamefont
  {Morfin}(2021)}]{SajjadAthar:2020nvy}%
  \BibitemOpen
  \bibfield  {author} {\bibinfo {author} {\bibfnamefont {M.}~\bibnamefont
  {Sajjad~Athar}}\ and\ \bibinfo {author} {\bibfnamefont {J.~G.}\ \bibnamefont
  {Morfin}},\ }\bibfield  {title} {\bibinfo {title}
  {{Neutrino(Antineutrino)--nucleus interactions in the shallow- and
  deep-inelastic scattering regions}},\ }\href
  {https://doi.org/10.1088/1361-6471/abbb11} {\bibfield  {journal} {\bibinfo
  {journal} {J.\ Phys.\ G}\ }\textbf {\bibinfo {volume} {48}},\ \bibinfo
  {pages} {034001} (\bibinfo {year} {2021})},\ \Eprint
  {https://arxiv.org/abs/2006.08603} {arXiv:2006.08603 [hep-ph]} \BibitemShut
  {NoStop}%
\bibitem [{\citenamefont {Coloma}\ \emph {et~al.}(2020)\citenamefont {Coloma},
  \citenamefont {Esteban}, \citenamefont {Gonzalez-Garcia},\ and\ \citenamefont
  {Menendez}}]{Coloma:2020nhf}%
  \BibitemOpen
  \bibfield  {author} {\bibinfo {author} {\bibfnamefont {P.}~\bibnamefont
  {Coloma}}, \bibinfo {author} {\bibfnamefont {I.}~\bibnamefont {Esteban}},
  \bibinfo {author} {\bibfnamefont {M.~C.}\ \bibnamefont {Gonzalez-Garcia}},\
  and\ \bibinfo {author} {\bibfnamefont {J.}~\bibnamefont {Menendez}},\
  }\bibfield  {title} {\bibinfo {title} {{Determining the nuclear neutron
  distribution from coherent elastic neutrino-nucleus scattering: current
  results and future prospects}},\ }\href
  {https://doi.org/10.1007/JHEP08(2020)030} {\bibfield  {journal} {\bibinfo
  {journal} {\relax{JHEP}}\ }\textbf {\bibinfo {volume} {08}},\ \bibinfo
  {pages} {030} (\bibinfo {year} {2020})},\ \Eprint
  {https://arxiv.org/abs/2006.08624} {arXiv:2006.08624 [hep-ph]} \BibitemShut
  {NoStop}%
\bibitem [{\citenamefont {Kitagaki}\ \emph {et~al.}(1983)\citenamefont
  {Kitagaki} \emph {et~al.}}]{Kitagaki:1983px}%
  \BibitemOpen
  \bibfield  {author} {\bibinfo {author} {\bibfnamefont {T.}~\bibnamefont
  {Kitagaki}} \emph {et~al.},\ }\bibfield  {title} {\bibinfo {title}
  {{High-energy quasielastic $\nu_{\mu}n\to\mu^-p$ scattering in deuterium}},\
  }\href {https://doi.org/10.1103/PhysRevD.28.436} {\bibfield  {journal}
  {\bibinfo  {journal} {Phys.\ Rev.\ D}\ }\textbf {\bibinfo {volume} {28}},\
  \bibinfo {pages} {436} (\bibinfo {year} {1983})}\BibitemShut {NoStop}%
\bibitem [{\citenamefont {Belikov}\ \emph {et~al.}(1985)\citenamefont {Belikov}
  \emph {et~al.}}]{Belikov:1983kg}%
  \BibitemOpen
  \bibfield  {author} {\bibinfo {author} {\bibfnamefont {S.~V.}\ \bibnamefont
  {Belikov}} \emph {et~al.},\ }\bibfield  {title} {\bibinfo {title}
  {{Quasielastic neutrino and antineutrinos scattering: total cross-sections,
  axial-vector form-factor}},\ }\href {https://doi.org/10.1007/BF01411863}
  {\bibfield  {journal} {\bibinfo  {journal} {Z.\ Phys.\ A}\ }\textbf {\bibinfo
  {volume} {320}},\ \bibinfo {pages} {625} (\bibinfo {year}
  {1985})}\BibitemShut {NoStop}%
\bibitem [{\citenamefont {Abe}\ \emph {et~al.}(2011)\citenamefont {Abe} \emph
  {et~al.}}]{Abe:2011ks}%
  \BibitemOpen
  \bibfield  {author} {\bibinfo {author} {\bibfnamefont {K.}~\bibnamefont
  {Abe}} \emph {et~al.} (\bibinfo {collaboration} {T2K Collaboration}),\
  }\bibfield  {title} {\bibinfo {title} {{The T2K Experiment}},\ }\href
  {https://doi.org/10.1016/j.nima.2011.06.067} {\bibfield  {journal} {\bibinfo
  {journal} {Nucl.\ Instrum.\ Meth.\ A}\ }\textbf {\bibinfo {volume} {659}},\
  \bibinfo {pages} {106} (\bibinfo {year} {2011})},\ \Eprint
  {https://arxiv.org/abs/1106.1238} {arXiv:1106.1238 [physics.ins-det]}
  \BibitemShut {NoStop}%
\bibitem [{\citenamefont {Abe}\ \emph {et~al.}(2014)\citenamefont {Abe} \emph
  {et~al.}}]{Abe:2014ugx}%
  \BibitemOpen
  \bibfield  {author} {\bibinfo {author} {\bibfnamefont {K.}~\bibnamefont
  {Abe}} \emph {et~al.} (\bibinfo {collaboration} {T2K Collaboration}),\
  }\bibfield  {title} {\bibinfo {title} {{Precise measurement of the neutrino
  mixing parameter $\theta_{23}$ from muon neutrino disappearance in an
  off-axis beam}},\ }\href {https://doi.org/10.1103/PhysRevLett.112.181801}
  {\bibfield  {journal} {\bibinfo  {journal} {Phys.\ Rev.\ Lett.}\ }\textbf
  {\bibinfo {volume} {112}},\ \bibinfo {pages} {181801} (\bibinfo {year}
  {2014})},\ \Eprint {https://arxiv.org/abs/1403.1532} {arXiv:1403.1532
  [hep-ex]} \BibitemShut {NoStop}%
\bibitem [{\citenamefont {Abe}\ \emph {et~al.}(2015{\natexlab{a}})\citenamefont
  {Abe} \emph {et~al.}}]{Abe:2014iza}%
  \BibitemOpen
  \bibfield  {author} {\bibinfo {author} {\bibfnamefont {K.}~\bibnamefont
  {Abe}} \emph {et~al.} (\bibinfo {collaboration} {T2K Collaboration}),\
  }\bibfield  {title} {\bibinfo {title} {{Measurement of the $\nu_\mu$
  charged-current quasielastic cross section on carbon with the ND280 detector
  at T2K}},\ }\href {https://doi.org/10.1103/PhysRevD.92.112003} {\bibfield
  {journal} {\bibinfo  {journal} {Phys.\ Rev.\ D}\ }\textbf {\bibinfo {volume}
  {92}},\ \bibinfo {pages} {112003} (\bibinfo {year} {2015}{\natexlab{a}})},\
  \Eprint {https://arxiv.org/abs/1411.6264} {arXiv:1411.6264 [hep-ex]}
  \BibitemShut {NoStop}%
\bibitem [{\citenamefont {Abe}\ \emph {et~al.}(2015{\natexlab{b}})\citenamefont
  {Abe} \emph {et~al.}}]{Abe:2015oar}%
  \BibitemOpen
  \bibfield  {author} {\bibinfo {author} {\bibfnamefont {K.}~\bibnamefont
  {Abe}} \emph {et~al.} (\bibinfo {collaboration} {T2K Collaboration}),\
  }\bibfield  {title} {\bibinfo {title} {{Measurement of the $\nu_\mu$ charged
  current quasielastic cross section on carbon with the T2K on-axis neutrino
  beam}},\ }\href {https://doi.org/10.1103/PhysRevD.91.112002} {\bibfield
  {journal} {\bibinfo  {journal} {Phys.\ Rev.\ D}\ }\textbf {\bibinfo {volume}
  {91}},\ \bibinfo {pages} {112002} (\bibinfo {year} {2015}{\natexlab{b}})},\
  \Eprint {https://arxiv.org/abs/1503.07452} {arXiv:1503.07452 [hep-ex]}
  \BibitemShut {NoStop}%
\bibitem [{\citenamefont {Abe}\ \emph {et~al.}(2016)\citenamefont {Abe} \emph
  {et~al.}}]{Abe:2015biq}%
  \BibitemOpen
  \bibfield  {author} {\bibinfo {author} {\bibfnamefont {K.}~\bibnamefont
  {Abe}} \emph {et~al.} (\bibinfo {collaboration} {T2K Collaboration}),\
  }\bibfield  {title} {\bibinfo {title} {{Measurement of the muon neutrino
  inclusive charged-current cross section in the energy range of 1--3~GeV with
  the T2K INGRID detector}},\ }\href
  {https://doi.org/10.1103/PhysRevD.93.072002} {\bibfield  {journal} {\bibinfo
  {journal} {Phys.\ Rev.\ D}\ }\textbf {\bibinfo {volume} {93}},\ \bibinfo
  {pages} {072002} (\bibinfo {year} {2016})},\ \Eprint
  {https://arxiv.org/abs/1509.06940} {arXiv:1509.06940 [hep-ex]} \BibitemShut
  {NoStop}%
\bibitem [{\citenamefont {Nakajima}\ \emph {et~al.}(2011)\citenamefont
  {Nakajima} \emph {et~al.}}]{Nakajima:2010fp}%
  \BibitemOpen
  \bibfield  {author} {\bibinfo {author} {\bibfnamefont {Y.}~\bibnamefont
  {Nakajima}} \emph {et~al.} (\bibinfo {collaboration} {SciBooNE
  Collaboration}),\ }\bibfield  {title} {\bibinfo {title} {{Measurement of
  inclusive charged current interactions on Carbon in a few-GeV neutrino
  beam}},\ }\href {https://doi.org/10.1103/PhysRevD.83.012005} {\bibfield
  {journal} {\bibinfo  {journal} {Phys.\ Rev. D}\ }\textbf {\bibinfo {volume}
  {83}},\ \bibinfo {pages} {012005} (\bibinfo {year} {2011})},\ \Eprint
  {https://arxiv.org/abs/1011.2131} {arXiv:1011.2131 [hep-ex]} \BibitemShut
  {NoStop}%
\bibitem [{\citenamefont {Alcaraz-Aunion}\ and\ \citenamefont
  {Walding}(2009)}]{AlcarazAunion:2009ku}%
  \BibitemOpen
  \bibfield  {author} {\bibinfo {author} {\bibfnamefont {J.~L.}\ \bibnamefont
  {Alcaraz-Aunion}}\ and\ \bibinfo {author} {\bibfnamefont {J.}~\bibnamefont
  {Walding}} (\bibinfo {collaboration} {SciBooNE Collaboration}),\ }\bibfield
  {title} {\bibinfo {title} {{Measurement of the $\nu_{mu}$-CCQE cross section
  in the SciBooNE experiment}},\ }\bibfield  {booktitle} {\emph {\bibinfo
  {booktitle} {{Proceedings of the 6th International Workshop on
  Neutrino-Nucleus Interactions in the Few GeV Region (NuInt\,2009), Sitges,
  Spain, May 18--22, 2009}}},\ }\href {https://doi.org/10.1063/1.3274145}
  {\bibfield  {journal} {\bibinfo  {journal} {AIP Conf.\ Proc.}\ }\textbf
  {\bibinfo {volume} {1189}},\ \bibinfo {pages} {145} (\bibinfo {year}
  {2009})},\ \Eprint {https://arxiv.org/abs/0909.5647} {arXiv:0909.5647
  [hep-ex]} \BibitemShut {NoStop}%
\bibitem [{\citenamefont {Aguilar-Arevalo}\ \emph {et~al.}(2018)\citenamefont
  {Aguilar-Arevalo} \emph {et~al.}}]{Aguilar-Arevalo:2018ylq}%
  \BibitemOpen
  \bibfield  {author} {\bibinfo {author} {\bibfnamefont {A.~A.}\ \bibnamefont
  {Aguilar-Arevalo}} \emph {et~al.} (\bibinfo {collaboration} {MiniBooNE
  Collaboration}),\ }\bibfield  {title} {\bibinfo {title} {{First measurement
  of monoenergetic muon neutrino charged current interactions}},\ }\href
  {https://doi.org/10.1103/PhysRevLett.120.141802} {\bibfield  {journal}
  {\bibinfo  {journal} {Phys.\ Rev.\ Lett.}\ }\textbf {\bibinfo {volume}
  {120}},\ \bibinfo {pages} {141802} (\bibinfo {year} {2018})},\ \Eprint
  {https://arxiv.org/abs/1801.03848} {arXiv:1801.03848 [hep-ex]} \BibitemShut
  {NoStop}%
\bibitem [{\citenamefont {Aguilar-Arevalo}\ \emph {et~al.}(2013)\citenamefont
  {Aguilar-Arevalo} \emph {et~al.}}]{Aguilar-Arevalo:2013dva}%
  \BibitemOpen
  \bibfield  {author} {\bibinfo {author} {\bibfnamefont {A.~A.}\ \bibnamefont
  {Aguilar-Arevalo}} \emph {et~al.} (\bibinfo {collaboration} {MiniBooNE
  Collaboration}),\ }\bibfield  {title} {\bibinfo {title} {{First measurement
  of the muon antineutrino double-differential charged-current quasielastic
  cross section}},\ }\href {https://doi.org/10.1103/PhysRevD.88.032001}
  {\bibfield  {journal} {\bibinfo  {journal} {Phys.\ Rev.\ D}\ }\textbf
  {\bibinfo {volume} {88}},\ \bibinfo {pages} {032001} (\bibinfo {year}
  {2013})},\ \Eprint {https://arxiv.org/abs/1301.7067} {arXiv:1301.7067
  [hep-ex]} \BibitemShut {NoStop}%
\bibitem [{\citenamefont {Aguilar-Arevalo}\ \emph {et~al.}(2010)\citenamefont
  {Aguilar-Arevalo} \emph {et~al.}}]{Aguilar-Arevalo:2010zc}%
  \BibitemOpen
  \bibfield  {author} {\bibinfo {author} {\bibfnamefont {A.~A.}\ \bibnamefont
  {Aguilar-Arevalo}} \emph {et~al.} (\bibinfo {collaboration} {MiniBooNE
  Collaboration}),\ }\bibfield  {title} {\bibinfo {title} {{First measurement
  of the muon neutrino charged current quasielastic double differential cross
  section}},\ }\href {https://doi.org/10.1103/PhysRevD.81.092005} {\bibfield
  {journal} {\bibinfo  {journal} {Phys.\ Rev.\ D}\ }\textbf {\bibinfo {volume}
  {81}},\ \bibinfo {pages} {092005} (\bibinfo {year} {2010})},\ \Eprint
  {https://arxiv.org/abs/1002.2680} {arXiv:1002.2680 [hep-ex]} \BibitemShut
  {NoStop}%
\bibitem [{\citenamefont {Aguilar-Arevalo}\ \emph {et~al.}(2008)\citenamefont
  {Aguilar-Arevalo} \emph {et~al.}}]{Aguilar-Arevalo:2007ab}%
  \BibitemOpen
  \bibfield  {author} {\bibinfo {author} {\bibfnamefont {A.~A.}\ \bibnamefont
  {Aguilar-Arevalo}} \emph {et~al.} (\bibinfo {collaboration} {MiniBooNE
  Collaboration}),\ }\bibfield  {title} {\bibinfo {title} {{Measurement of muon
  neutrino quasi-elastic scattering on carbon}},\ }\href
  {https://doi.org/10.1103/PhysRevLett.100.032301} {\bibfield  {journal}
  {\bibinfo  {journal} {Phys.\ Rev.\ Lett.}\ }\textbf {\bibinfo {volume}
  {100}},\ \bibinfo {pages} {032301} (\bibinfo {year} {2008})},\ \Eprint
  {https://arxiv.org/abs/0706.0926} {arXiv:0706.0926 [hep-ex]} \BibitemShut
  {NoStop}%
\bibitem [{\citenamefont {Adamson}\ \emph {et~al.}(2015)\citenamefont {Adamson}
  \emph {et~al.}}]{Adamson:2014pgc}%
  \BibitemOpen
  \bibfield  {author} {\bibinfo {author} {\bibfnamefont {P.}~\bibnamefont
  {Adamson}} \emph {et~al.} (\bibinfo {collaboration} {MINOS Collaboration}),\
  }\bibfield  {title} {\bibinfo {title} {{Study of quasielastic scattering
  using charged-current $\nu_\mu$-iron interactions in the MINOS near
  detector}},\ }\href {https://doi.org/10.1103/PhysRevD.91.012005} {\bibfield
  {journal} {\bibinfo  {journal} {Phys.\ Rev.\ D}\ }\textbf {\bibinfo {volume}
  {91}},\ \bibinfo {pages} {012005} (\bibinfo {year} {2015})},\ \Eprint
  {https://arxiv.org/abs/1410.8613} {arXiv:1410.8613 [hep-ex]} \BibitemShut
  {NoStop}%
\bibitem [{\citenamefont {Carneiro}\ \emph {et~al.}(2020)\citenamefont
  {Carneiro} \emph {et~al.}}]{Carneiro:2019jds}%
  \BibitemOpen
  \bibfield  {author} {\bibinfo {author} {\bibfnamefont {M.~F.}\ \bibnamefont
  {Carneiro}} \emph {et~al.} (\bibinfo {collaboration} {MINER$\nu$A
  Collaboration}),\ }\bibfield  {title} {\bibinfo {title} {{High-statistics
  measurement of neutrino quasielasticlike scattering at 6 GeV on a hydrocarbon
  target}},\ }\href {https://doi.org/10.1103/PhysRevLett.124.121801} {\bibfield
   {journal} {\bibinfo  {journal} {Phys.\ Rev.\ Lett.}\ }\textbf {\bibinfo
  {volume} {124}},\ \bibinfo {pages} {121801} (\bibinfo {year} {2020})},\
  \Eprint {https://arxiv.org/abs/1912.09890} {arXiv:1912.09890 [hep-ex]}
  \BibitemShut {NoStop}%
\bibitem [{\citenamefont {Ruterbories}\ \emph {et~al.}(2019)\citenamefont
  {Ruterbories} \emph {et~al.}}]{Ruterbories:2018gub}%
  \BibitemOpen
  \bibfield  {author} {\bibinfo {author} {\bibfnamefont {D.}~\bibnamefont
  {Ruterbories}} \emph {et~al.} (\bibinfo {collaboration} {MINER$\nu$A
  Collaboration}),\ }\bibfield  {title} {\bibinfo {title} {{Measurement of
  quasielastic-like neutrino scattering at $\langle{E_\nu}\rangle\sim3.5$~GeV
  on a hydrocarbon target}},\ }\href
  {https://doi.org/10.1103/PhysRevD.99.012004} {\bibfield  {journal} {\bibinfo
  {journal} {Phys.\ Rev.\ D}\ }\textbf {\bibinfo {volume} {99}},\ \bibinfo
  {pages} {012004} (\bibinfo {year} {2019})},\ \Eprint
  {https://arxiv.org/abs/1811.02774} {arXiv:1811.02774 [hep-ex]} \BibitemShut
  {NoStop}%
\bibitem [{\citenamefont {Patrick}\ \emph {et~al.}(2018)\citenamefont {Patrick}
  \emph {et~al.}}]{Patrick:2018gvi}%
  \BibitemOpen
  \bibfield  {author} {\bibinfo {author} {\bibfnamefont {C.~E.}\ \bibnamefont
  {Patrick}} \emph {et~al.} (\bibinfo {collaboration} {MINER$\nu$A
  Collaboration}),\ }\bibfield  {title} {\bibinfo {title} {{Measurement of the
  muon antineutrino double-differential cross section for quasielastic-like
  scattering on hydrocarbon at $E_\nu \sim 3.5$~GeV}},\ }\href
  {https://doi.org/10.1103/PhysRevD.97.052002} {\bibfield  {journal} {\bibinfo
  {journal} {Phys.\ Rev.\ D}\ }\textbf {\bibinfo {volume} {97}},\ \bibinfo
  {pages} {052002} (\bibinfo {year} {2018})},\ \Eprint
  {https://arxiv.org/abs/1801.01197} {arXiv:1801.01197 [hep-ex]} \BibitemShut
  {NoStop}%
\bibitem [{\citenamefont {Wolcott}\ \emph {et~al.}(2016)\citenamefont {Wolcott}
  \emph {et~al.}}]{Wolcott:2015hda}%
  \BibitemOpen
  \bibfield  {author} {\bibinfo {author} {\bibfnamefont {J.}~\bibnamefont
  {Wolcott}} \emph {et~al.} (\bibinfo {collaboration} {MINER$\nu$A
  Collaboration}),\ }\bibfield  {title} {\bibinfo {title} {{Measurement of
  electron neutrino quasielastic and quasielasticlike scattering on hydrocarbon
  at $\langle{E_{\nu}}\rangle=3.6$~GeV}},\ }\href
  {https://doi.org/10.1103/PhysRevLett.116.081802} {\bibfield  {journal}
  {\bibinfo  {journal} {Phys.\ Rev.\ Lett.}\ }\textbf {\bibinfo {volume}
  {116}},\ \bibinfo {pages} {081802} (\bibinfo {year} {2016})},\ \Eprint
  {https://arxiv.org/abs/1509.05729} {arXiv:1509.05729 [hep-ex]} \BibitemShut
  {NoStop}%
\bibitem [{\citenamefont {Fiorentini}\ \emph {et~al.}(2013)\citenamefont
  {Fiorentini} \emph {et~al.}}]{Fiorentini:2013ezn}%
  \BibitemOpen
  \bibfield  {author} {\bibinfo {author} {\bibfnamefont {G.~A.}\ \bibnamefont
  {Fiorentini}} \emph {et~al.} (\bibinfo {collaboration} {MINER$\nu$A
  Collaboration}),\ }\bibfield  {title} {\bibinfo {title} {{Measurement of muon
  neutrino quasielastic scattering on a hydrocarbon target at $E_{\nu} \sim
  3.5$~GeV}},\ }\href {https://doi.org/10.1103/PhysRevLett.111.022502}
  {\bibfield  {journal} {\bibinfo  {journal} {Phys.\ Rev.\ Lett.}\ }\textbf
  {\bibinfo {volume} {111}},\ \bibinfo {pages} {022502} (\bibinfo {year}
  {2013})},\ \Eprint {https://arxiv.org/abs/1305.2243} {arXiv:1305.2243
  [hep-ex]} \BibitemShut {NoStop}%
\bibitem [{\citenamefont {Fields}\ \emph {et~al.}(2013)\citenamefont {Fields}
  \emph {et~al.}}]{Fields:2013zhk}%
  \BibitemOpen
  \bibfield  {author} {\bibinfo {author} {\bibfnamefont {L.}~\bibnamefont
  {Fields}} \emph {et~al.} (\bibinfo {collaboration} {MINER$\nu$A
  Collaboration}),\ }\bibfield  {title} {\bibinfo {title} {{Measurement of muon
  antineutrino quasielastic scattering on a hydrocarbon target at $E_{\nu} \sim
  3.5$~GeV}},\ }\href {https://doi.org/10.1103/PhysRevLett.111.022501}
  {\bibfield  {journal} {\bibinfo  {journal} {Phys.\ Rev.\ Lett.}\ }\textbf
  {\bibinfo {volume} {111}},\ \bibinfo {pages} {022501} (\bibinfo {year}
  {2013})},\ \Eprint {https://arxiv.org/abs/1305.2234} {arXiv:1305.2234
  [hep-ex]} \BibitemShut {NoStop}%
\bibitem [{\citenamefont {Abratenko}\ \emph {et~al.}(2020)\citenamefont
  {Abratenko} \emph {et~al.}}]{Abratenko:2020acr}%
  \BibitemOpen
  \bibfield  {author} {\bibinfo {author} {\bibfnamefont {P.}~\bibnamefont
  {Abratenko}} \emph {et~al.} (\bibinfo {collaboration} {MicroBooNE
  Collaboration}),\ }\bibfield  {title} {\bibinfo {title} {{First measurement
  of differential charged current quasielastic-like $\nu_\mu$-Argon Scattering
  cross sections with the MicroBooNE Detector}},\ }\href
  {https://doi.org/10.1103/PhysRevLett.125.201803} {\bibfield  {journal}
  {\bibinfo  {journal} {Phys.\ Rev.\ Lett.}\ }\textbf {\bibinfo {volume}
  {125}},\ \bibinfo {pages} {201803} (\bibinfo {year} {2020})},\ \Eprint
  {https://arxiv.org/abs/2006.00108} {arXiv:2006.00108 [hep-ex]} \BibitemShut
  {NoStop}%
\bibitem [{\citenamefont {Acciarri}\ \emph {et~al.}(2020)\citenamefont
  {Acciarri} \emph {et~al.}}]{Acciarri:2020lhp}%
  \BibitemOpen
  \bibfield  {author} {\bibinfo {author} {\bibfnamefont {R.}~\bibnamefont
  {Acciarri}} \emph {et~al.} (\bibinfo {collaboration} {ArgoNeuT
  Collaboration}),\ }\bibfield  {title} {\bibinfo {title} {{First measurement
  of electron neutrino scattering cross section on argon}},\ }\href
  {https://doi.org/10.1103/PhysRevD.102.011101} {\bibfield  {journal} {\bibinfo
   {journal} {Phys.\ Rev.\ D}\ }\textbf {\bibinfo {volume} {102}},\ \bibinfo
  {pages} {011101} (\bibinfo {year} {2020})},\ \Eprint
  {https://arxiv.org/abs/2004.01956} {arXiv:2004.01956 [hep-ex]} \BibitemShut
  {NoStop}%
\bibitem [{\citenamefont {Acciarri}\ \emph {et~al.}(2014)\citenamefont
  {Acciarri} \emph {et~al.}}]{Acciarri:2014isz}%
  \BibitemOpen
  \bibfield  {author} {\bibinfo {author} {\bibfnamefont {R.}~\bibnamefont
  {Acciarri}} \emph {et~al.} (\bibinfo {collaboration} {ArgoNeuT
  Collaboration}),\ }\bibfield  {title} {\bibinfo {title} {{Measurements of
  inclusive muon neutrino and antineutrino charged current differential cross
  sections on Argon in the NuMI antineutrino beam}},\ }\href
  {https://doi.org/10.1103/PhysRevD.89.112003} {\bibfield  {journal} {\bibinfo
  {journal} {Phys.\ Rev.\ D}\ }\textbf {\bibinfo {volume} {89}},\ \bibinfo
  {pages} {112003} (\bibinfo {year} {2014})},\ \Eprint
  {https://arxiv.org/abs/1404.4809} {arXiv:1404.4809 [hep-ex]} \BibitemShut
  {NoStop}%
\bibitem [{\citenamefont {Adamson}\ \emph {et~al.}(2017)\citenamefont {Adamson}
  \emph {et~al.}}]{Adamson:2017gxd}%
  \BibitemOpen
  \bibfield  {author} {\bibinfo {author} {\bibfnamefont {P.}~\bibnamefont
  {Adamson}} \emph {et~al.} (\bibinfo {collaboration} {NO$\nu$A
  Collaboration}),\ }\bibfield  {title} {\bibinfo {title} {{Constraints on
  oscillation parameters from $\nu_e$ appearance and $\nu_\mu$ disappearance in
  NO$\nu$A}},\ }\href {https://doi.org/10.1103/PhysRevLett.118.231801}
  {\bibfield  {journal} {\bibinfo  {journal} {Phys. Rev. Lett.}\ }\textbf
  {\bibinfo {volume} {118}},\ \bibinfo {pages} {231801} (\bibinfo {year}
  {2017})},\ \Eprint {https://arxiv.org/abs/1703.03328} {arXiv:1703.03328
  [hep-ex]} \BibitemShut {NoStop}%
\bibitem [{\citenamefont {Ankowski}(2006)}]{Ankowski:2005jf}%
  \BibitemOpen
  \bibfield  {author} {\bibinfo {author} {\bibfnamefont {A.~M.}\ \bibnamefont
  {Ankowski}},\ }\bibfield  {title} {\bibinfo {title} {{High-energy limit of
  neutrino quasielastic cross section}},\ }\href@noop {} {\bibfield  {journal}
  {\bibinfo  {journal} {Acta Phys.\ Polon.\ B}\ }\textbf {\bibinfo {volume}
  {37}},\ \bibinfo {pages} {377} (\bibinfo {year} {2006})},\ \Eprint
  {https://arxiv.org/abs/hep-ph/0503187} {hep-ph/0503187} \BibitemShut
  {NoStop}%
\bibitem [{\citenamefont {Ankowski}(2019)}]{Ankowski:2019yll}%
  \BibitemOpen
  \bibfield  {author} {\bibinfo {author} {\bibfnamefont {A.~M.}\ \bibnamefont
  {Ankowski}},\ }\bibfield  {title} {\bibinfo {title} {{How different can the
  $\nu_\mu$ and $\nu_e$ cross sections be?}},\ }in\ \href
  {https://doi.org/10.22323/1.341.0092} {\emph {\bibinfo {booktitle}
  {{Proceedings of the 20th International Workshop on Neutrinos from
  Accelerators (NUFACT\,2018), Blacksburg, Virginia, USA, August 13--18,
  2018}}}},\ Vol.\ \bibinfo {volume} {NuFACT2018}\ (\bibinfo {year} {2019})\
  p.\ \bibinfo {pages} {092}\BibitemShut {NoStop}%
\bibitem [{\citenamefont {Abe}\ \emph {et~al.}(2020)\citenamefont {Abe} \emph
  {et~al.}}]{Abe:2019vii}%
  \BibitemOpen
  \bibfield  {author} {\bibinfo {author} {\bibfnamefont {K.}~\bibnamefont
  {Abe}} \emph {et~al.} (\bibinfo {collaboration} {T2K Collaboration}),\
  }\bibfield  {title} {\bibinfo {title} {{Constraint on the matter-antimatter
  symmetry-violating phase in neutrino oscillations}},\ }\href
  {https://doi.org/10.1038/s41586-020-2177-0} {\bibfield  {journal} {\bibinfo
  {journal} {Nature}\ }\textbf {\bibinfo {volume} {580}},\ \bibinfo {pages}
  {339} (\bibinfo {year} {2020})}\BibitemShut {NoStop}%
\bibitem [{Abe(2020)}]{Abe:2019viiErratum}%
  \BibitemOpen
  \href {https://doi.org/10.1038/s41586-020-2415-5} {\bibfield  {journal}
  {\bibinfo  {journal} {Erratum: {\it ibid.}}\ }\textbf {\bibinfo {volume}
  {583}},\ \bibinfo {pages} {E16} (\bibinfo {year} {2020})},\ \Eprint
  {https://arxiv.org/abs/1910.03887} {arXiv:1910.03887 [hep-ex]} \BibitemShut
  {NoStop}%
\bibitem [{\citenamefont {Acciarri}\ \emph {et~al.}(2015)\citenamefont
  {Acciarri} \emph {et~al.}}]{Acciarri:2015uup}%
  \BibitemOpen
  \bibfield  {author} {\bibinfo {author} {\bibfnamefont {R.}~\bibnamefont
  {Acciarri}} \emph {et~al.} (\bibinfo {collaboration} {DUNE Collaboration}),\
  }\href@noop {} {\bibinfo {title} {{Long-Baseline Neutrino Facility (LBNF) and
  Deep Underground Neutrino Experiment (DUNE). Conceptual design report. Volume
  2: The physics program for DUNE at LBNF}}} (\bibinfo {year} {2015}),\ \Eprint
  {https://arxiv.org/abs/1512.06148} {arXiv:1512.06148 [physics.ins-det]}
  \BibitemShut {NoStop}%
\bibitem [{\citenamefont {Ankowski}\ and\ \citenamefont
  {Mariani}(2017)}]{Ankowski:2016jdd}%
  \BibitemOpen
  \bibfield  {author} {\bibinfo {author} {\bibfnamefont {A.~M.}\ \bibnamefont
  {Ankowski}}\ and\ \bibinfo {author} {\bibfnamefont {C.}~\bibnamefont
  {Mariani}},\ }\bibfield  {title} {\bibinfo {title} {{Systematic uncertainties
  in long-baseline neutrino-oscillation experiments}},\ }\href
  {https://doi.org/10.1088/1361-6471/aa61b2} {\bibfield  {journal} {\bibinfo
  {journal} {J.\ Phys.\ G}\ }\textbf {\bibinfo {volume} {44}},\ \bibinfo
  {pages} {054001} (\bibinfo {year} {2017})},\ \Eprint
  {https://arxiv.org/abs/1609.00258} {arXiv:1609.00258 [hep-ph]} \BibitemShut
  {NoStop}%
\bibitem [{\citenamefont {Ankowski}(2017)}]{Ankowski:2017yvm}%
  \BibitemOpen
  \bibfield  {author} {\bibinfo {author} {\bibfnamefont {A.~M.}\ \bibnamefont
  {Ankowski}},\ }\bibfield  {title} {\bibinfo {title} {{Effect of the
  charged-lepton's mass on the quasielastic neutrino cross sections}},\ }\href
  {https://doi.org/10.1103/PhysRevC.96.035501} {\bibfield  {journal} {\bibinfo
  {journal} {Phys.\ Rev.\ C}\ }\textbf {\bibinfo {volume} {96}},\ \bibinfo
  {pages} {035501} (\bibinfo {year} {2017})},\ \Eprint
  {https://arxiv.org/abs/1707.01014} {arXiv:1707.01014 [nucl-th]} \BibitemShut
  {NoStop}%
\bibitem [{\citenamefont {Martini}\ \emph {et~al.}(2016)\citenamefont
  {Martini}, \citenamefont {Jachowicz}, \citenamefont {Ericson}, \citenamefont
  {Pandey}, \citenamefont {Van~Cuyck},\ and\ \citenamefont
  {Van~Dessel}}]{Martini:2016eec}%
  \BibitemOpen
  \bibfield  {author} {\bibinfo {author} {\bibfnamefont {M.}~\bibnamefont
  {Martini}}, \bibinfo {author} {\bibfnamefont {N.}~\bibnamefont {Jachowicz}},
  \bibinfo {author} {\bibfnamefont {M.}~\bibnamefont {Ericson}}, \bibinfo
  {author} {\bibfnamefont {V.}~\bibnamefont {Pandey}}, \bibinfo {author}
  {\bibfnamefont {T.}~\bibnamefont {Van~Cuyck}},\ and\ \bibinfo {author}
  {\bibfnamefont {N.}~\bibnamefont {Van~Dessel}},\ }\bibfield  {title}
  {\bibinfo {title} {{Electron-neutrino scattering off nuclei from two
  different theoretical perspectives}},\ }\href
  {https://doi.org/10.1103/PhysRevC.94.015501} {\bibfield  {journal} {\bibinfo
  {journal} {Phys.\ Rev.\ C}\ }\textbf {\bibinfo {volume} {94}},\ \bibinfo
  {pages} {015501} (\bibinfo {year} {2016})},\ \Eprint
  {https://arxiv.org/abs/1602.00230} {arXiv:1602.00230 [nucl-th]} \BibitemShut
  {NoStop}%
\bibitem [{\citenamefont {Nikolakopoulos}\ \emph {et~al.}(2019)\citenamefont
  {Nikolakopoulos}, \citenamefont {Jachowicz}, \citenamefont {Van~Dessel},
  \citenamefont {Niewczas}, \citenamefont {Gonz\'{a}lez-Jim\'{e}nez},
  \citenamefont {Ud\'{i}as},\ and\ \citenamefont
  {Pandey}}]{Nikolakopoulos:2019qcr}%
  \BibitemOpen
  \bibfield  {author} {\bibinfo {author} {\bibfnamefont {A.}~\bibnamefont
  {Nikolakopoulos}}, \bibinfo {author} {\bibfnamefont {N.}~\bibnamefont
  {Jachowicz}}, \bibinfo {author} {\bibfnamefont {N.}~\bibnamefont
  {Van~Dessel}}, \bibinfo {author} {\bibfnamefont {K.}~\bibnamefont
  {Niewczas}}, \bibinfo {author} {\bibfnamefont {R.}~\bibnamefont
  {Gonz\'{a}lez-Jim\'{e}nez}}, \bibinfo {author} {\bibfnamefont {J.~M.}\
  \bibnamefont {Ud\'{i}as}},\ and\ \bibinfo {author} {\bibfnamefont
  {V.}~\bibnamefont {Pandey}},\ }\bibfield  {title} {\bibinfo {title}
  {{Electron versus muon neutrino induced cross sections in charged current
  quasielastic processes}},\ }\href
  {https://doi.org/10.1103/PhysRevLett.123.052501} {\bibfield  {journal}
  {\bibinfo  {journal} {Phys.\ Rev.\ Lett.}\ }\textbf {\bibinfo {volume}
  {123}},\ \bibinfo {pages} {052501} (\bibinfo {year} {2019})},\ \Eprint
  {https://arxiv.org/abs/1901.08050} {arXiv:1901.08050 [nucl-th]} \BibitemShut
  {NoStop}%
\bibitem [{\citenamefont {Alberico}\ and\ \citenamefont
  {Molinari}(1981)}]{Alberico:1981xd}%
  \BibitemOpen
  \bibfield  {author} {\bibinfo {author} {\bibfnamefont {W.~M.}\ \bibnamefont
  {Alberico}}\ and\ \bibinfo {author} {\bibfnamefont {A.}~\bibnamefont
  {Molinari}},\ }\bibfield  {title} {\bibinfo {title} {{Relativistic response
  of a Fermi gas}},\ }\href {https://doi.org/10.1088/0305-4616/7/5/003}
  {\bibfield  {journal} {\bibinfo  {journal} {J.\ Phys.\ G}\ }\textbf {\bibinfo
  {volume} {7}},\ \bibinfo {pages} {L93} (\bibinfo {year} {1981})}\BibitemShut
  {NoStop}%
\bibitem [{\citenamefont {Smith}\ and\ \citenamefont
  {Moniz}(1972{\natexlab{a}})}]{Smith:1972xh}%
  \BibitemOpen
  \bibfield  {author} {\bibinfo {author} {\bibfnamefont {R.~A.}\ \bibnamefont
  {Smith}}\ and\ \bibinfo {author} {\bibfnamefont {E.~J.}\ \bibnamefont
  {Moniz}},\ }\bibfield  {title} {\bibinfo {title} {{Neutrino reactions on
  nuclear targets}},\ }\href {https://doi.org/10.1016/0550-3213(72)90040-5}
  {\bibfield  {journal} {\bibinfo  {journal} {Nucl.\ Phys.\ B}\ }\textbf
  {\bibinfo {volume} {43}},\ \bibinfo {pages} {605} (\bibinfo {year}
  {1972}{\natexlab{a}})}\BibitemShut {NoStop}%
\bibitem [{Smi(1975)}]{Smith:1972xhErratum}%
  \BibitemOpen
  \href@noop {} {\bibfield  {journal} {\bibinfo  {journal} {Erratum: {\it
  ibid.}}\ }\textbf {\bibinfo {volume} {101}},\ \bibinfo {pages} {547}
  (\bibinfo {year} {1975})}\BibitemShut {NoStop}%
\bibitem [{\citenamefont {Moniz}\ \emph {et~al.}(1971)\citenamefont {Moniz},
  \citenamefont {Sick}, \citenamefont {Whitney}, \citenamefont {Ficenec},
  \citenamefont {Kephart},\ and\ \citenamefont {Trower}}]{Moniz:1971mt}%
  \BibitemOpen
  \bibfield  {author} {\bibinfo {author} {\bibfnamefont {E.~J.}\ \bibnamefont
  {Moniz}}, \bibinfo {author} {\bibfnamefont {I.}~\bibnamefont {Sick}},
  \bibinfo {author} {\bibfnamefont {R.~R.}\ \bibnamefont {Whitney}}, \bibinfo
  {author} {\bibfnamefont {J.~R.}\ \bibnamefont {Ficenec}}, \bibinfo {author}
  {\bibfnamefont {R.~D.}\ \bibnamefont {Kephart}},\ and\ \bibinfo {author}
  {\bibfnamefont {W.~P.}\ \bibnamefont {Trower}},\ }\bibfield  {title}
  {\bibinfo {title} {{Nuclear Fermi momenta from quasielastic electron
  scattering}},\ }\href {https://doi.org/10.1103/PhysRevLett.26.445} {\bibfield
   {journal} {\bibinfo  {journal} {Phys.\ Rev.\ Lett.}\ }\textbf {\bibinfo
  {volume} {26}},\ \bibinfo {pages} {445} (\bibinfo {year} {1971})}\BibitemShut
  {NoStop}%
\bibitem [{\citenamefont {Moniz}(1969)}]{Moniz:1969sr}%
  \BibitemOpen
  \bibfield  {author} {\bibinfo {author} {\bibfnamefont {E.~J.}\ \bibnamefont
  {Moniz}},\ }\bibfield  {title} {\bibinfo {title} {{Pion electroproduction
  from nuclei}},\ }\href {https://doi.org/10.1103/PhysRev.184.1154} {\bibfield
  {journal} {\bibinfo  {journal} {Phys.\ Rev.}\ }\textbf {\bibinfo {volume}
  {184}},\ \bibinfo {pages} {1154} (\bibinfo {year} {1969})}\BibitemShut
  {NoStop}%
\bibitem [{\citenamefont {Kuzmin}\ \emph {et~al.}(2008)\citenamefont {Kuzmin},
  \citenamefont {Lyubushkin},\ and\ \citenamefont {Naumov}}]{Kuzmin:2007kr}%
  \BibitemOpen
  \bibfield  {author} {\bibinfo {author} {\bibfnamefont {K.~S.}\ \bibnamefont
  {Kuzmin}}, \bibinfo {author} {\bibfnamefont {V.~V.}\ \bibnamefont
  {Lyubushkin}},\ and\ \bibinfo {author} {\bibfnamefont {V.~A.}\ \bibnamefont
  {Naumov}},\ }\bibfield  {title} {\bibinfo {title} {{Quasielastic axial-vector
  mass from experiments on neutrino-nucleus scattering}},\ }\href
  {https://doi.org/10.1140/epjc/s10052-008-0582-x} {\bibfield  {journal}
  {\bibinfo  {journal} {Eur.\ Phys.\ J.\ C}\ }\textbf {\bibinfo {volume}
  {54}},\ \bibinfo {pages} {517} (\bibinfo {year} {2008})},\ \Eprint
  {https://arxiv.org/abs/0712.4384} {arXiv:0712.4384 [hep-ph]} \BibitemShut
  {NoStop}%
\bibitem [{\citenamefont {Amaro}\ \emph {et~al.}(2015)\citenamefont {Amaro},
  \citenamefont {Ruiz~Arriola},\ and\ \citenamefont
  {Ruiz~Simo}}]{Amaro:2015zja}%
  \BibitemOpen
  \bibfield  {author} {\bibinfo {author} {\bibfnamefont {J.~E.}\ \bibnamefont
  {Amaro}}, \bibinfo {author} {\bibfnamefont {E.}~\bibnamefont
  {Ruiz~Arriola}},\ and\ \bibinfo {author} {\bibfnamefont {I.}~\bibnamefont
  {Ruiz~Simo}},\ }\bibfield  {title} {\bibinfo {title} {{Scaling violation and
  relativistic effective mass from quasi-elastic electron scattering:
  Implications for neutrino reactions}},\ }\href
  {https://doi.org/10.1103/PhysRevC.92.054607} {\bibfield  {journal} {\bibinfo
  {journal} {Phys.\ Rev.\ C}\ }\textbf {\bibinfo {volume} {92}},\ \bibinfo
  {pages} {054607} (\bibinfo {year} {2015})}\BibitemShut {NoStop}%
\bibitem [{Ama(2019)}]{Amaro:2015zjaErratum}%
  \BibitemOpen
  \href {https://doi.org/10.1103/PhysRevC.92.054607} {\bibfield  {journal}
  {\bibinfo  {journal} {Erratum: {\it ibid.}}\ }\textbf {\bibinfo {volume}
  {100}},\ \bibinfo {pages} {019904} (\bibinfo {year} {2019})},\ \Eprint
  {https://arxiv.org/abs/1505.05415} {arXiv:1505.05415 [nucl-th]} \BibitemShut
  {NoStop}%
\bibitem [{\citenamefont {Amaro}\ \emph {et~al.}(2017)\citenamefont {Amaro},
  \citenamefont {Ruiz~Arriola},\ and\ \citenamefont
  {Ruiz~Simo}}]{Amaro:2017pkd}%
  \BibitemOpen
  \bibfield  {author} {\bibinfo {author} {\bibfnamefont {J.~E.}\ \bibnamefont
  {Amaro}}, \bibinfo {author} {\bibfnamefont {E.}~\bibnamefont
  {Ruiz~Arriola}},\ and\ \bibinfo {author} {\bibfnamefont {I.}~\bibnamefont
  {Ruiz~Simo}},\ }\bibfield  {title} {\bibinfo {title} {{Superscaling analysis
  of quasielastic electron scattering with relativistic effective mass}},\
  }\href {https://doi.org/10.1103/PhysRevD.95.076009} {\bibfield  {journal}
  {\bibinfo  {journal} {Phys.\ Rev.\ D}\ }\textbf {\bibinfo {volume} {95}},\
  \bibinfo {pages} {076009} (\bibinfo {year} {2017})},\ \Eprint
  {https://arxiv.org/abs/1701.05417} {arXiv:1701.05417 [nucl-th]} \BibitemShut
  {NoStop}%
\bibitem [{\citenamefont {Martinez-Consentino}\ \emph
  {et~al.}(2017)\citenamefont {Martinez-Consentino}, \citenamefont {Ruiz~Simo},
  \citenamefont {Amaro},\ and\ \citenamefont
  {Ruiz~Arriola}}]{Martinez-Consentino:2017ryk}%
  \BibitemOpen
  \bibfield  {author} {\bibinfo {author} {\bibfnamefont {V.~L.}\ \bibnamefont
  {Martinez-Consentino}}, \bibinfo {author} {\bibfnamefont {I.}~\bibnamefont
  {Ruiz~Simo}}, \bibinfo {author} {\bibfnamefont {J.~E.}\ \bibnamefont
  {Amaro}},\ and\ \bibinfo {author} {\bibfnamefont {E.}~\bibnamefont
  {Ruiz~Arriola}},\ }\bibfield  {title} {\bibinfo {title} {{Fermi-momentum
  dependence of relativistic effective mass below saturation from superscaling
  of quasielastic electron scattering}},\ }\href
  {https://doi.org/10.1103/PhysRevC.96.064612} {\bibfield  {journal} {\bibinfo
  {journal} {Phys.\ Rev.\ C}\ }\textbf {\bibinfo {volume} {96}},\ \bibinfo
  {pages} {064612} (\bibinfo {year} {2017})},\ \Eprint
  {https://arxiv.org/abs/1710.04988} {arXiv:1710.04988 [nucl-th]} \BibitemShut
  {NoStop}%
\bibitem [{\citenamefont {Ruiz~Simo}\ \emph
  {et~al.}(2018{\natexlab{a}})\citenamefont {Ruiz~Simo}, \citenamefont
  {Martinez-Consentino}, \citenamefont {Amaro},\ and\ \citenamefont
  {Ruiz~Arriola}}]{RuizSimo:2018kdl}%
  \BibitemOpen
  \bibfield  {author} {\bibinfo {author} {\bibfnamefont {I.}~\bibnamefont
  {Ruiz~Simo}}, \bibinfo {author} {\bibfnamefont {V.~L.}\ \bibnamefont
  {Martinez-Consentino}}, \bibinfo {author} {\bibfnamefont {J.~E.}\
  \bibnamefont {Amaro}},\ and\ \bibinfo {author} {\bibfnamefont
  {E.}~\bibnamefont {Ruiz~Arriola}},\ }\bibfield  {title} {\bibinfo {title}
  {{Quasielastic charged-current neutrino scattering in the scaling model with
  relativistic effective mass}},\ }\href
  {https://doi.org/10.1103/PhysRevD.97.116006} {\bibfield  {journal} {\bibinfo
  {journal} {Phys.\ Rev.\ D}\ }\textbf {\bibinfo {volume} {97}},\ \bibinfo
  {pages} {116006} (\bibinfo {year} {2018}{\natexlab{a}})},\ \Eprint
  {https://arxiv.org/abs/1804.07548} {arXiv:1804.07548 [nucl-th]} \BibitemShut
  {NoStop}%
\bibitem [{\citenamefont {Amaro}\ \emph {et~al.}(2018)\citenamefont {Amaro},
  \citenamefont {Martinez-Consentino}, \citenamefont {Ruiz~Arriola},\ and\
  \citenamefont {Ruiz~Simo}}]{Amaro:2018xdi}%
  \BibitemOpen
  \bibfield  {author} {\bibinfo {author} {\bibfnamefont {J.~E.}\ \bibnamefont
  {Amaro}}, \bibinfo {author} {\bibfnamefont {V.~L.}\ \bibnamefont
  {Martinez-Consentino}}, \bibinfo {author} {\bibfnamefont {E.}~\bibnamefont
  {Ruiz~Arriola}},\ and\ \bibinfo {author} {\bibfnamefont {I.}~\bibnamefont
  {Ruiz~Simo}},\ }\bibfield  {title} {\bibinfo {title} {{Global superscaling
  analysis of quasielastic electron scattering with relativistic effective
  mass}},\ }\href {https://doi.org/10.1103/PhysRevC.98.024627} {\bibfield
  {journal} {\bibinfo  {journal} {Phys.\ Rev.\ C}\ }\textbf {\bibinfo {volume}
  {98}},\ \bibinfo {pages} {024627} (\bibinfo {year} {2018})},\ \Eprint
  {https://arxiv.org/abs/1806.09512} {arXiv:1806.09512 [nucl-th]} \BibitemShut
  {NoStop}%
\bibitem [{\citenamefont {Alberico}\ \emph {et~al.}(1988)\citenamefont
  {Alberico}, \citenamefont {Molinari}, \citenamefont {Donnelly}, \citenamefont
  {Kronenberg},\ and\ \citenamefont {Van~Orden}}]{Alberico:1988bv}%
  \BibitemOpen
  \bibfield  {author} {\bibinfo {author} {\bibfnamefont {W.~M.}\ \bibnamefont
  {Alberico}}, \bibinfo {author} {\bibfnamefont {A.}~\bibnamefont {Molinari}},
  \bibinfo {author} {\bibfnamefont {T.~W.}\ \bibnamefont {Donnelly}}, \bibinfo
  {author} {\bibfnamefont {E.~L.}\ \bibnamefont {Kronenberg}},\ and\ \bibinfo
  {author} {\bibfnamefont {J.~W.}\ \bibnamefont {Van~Orden}},\ }\bibfield
  {title} {\bibinfo {title} {{Scaling in electron scattering from a
  relativistic Fermi gas}},\ }\href {https://doi.org/10.1103/PhysRevC.38.1801}
  {\bibfield  {journal} {\bibinfo  {journal} {Phys.\ Rev.\ C}\ }\textbf
  {\bibinfo {volume} {38}},\ \bibinfo {pages} {1801} (\bibinfo {year}
  {1988})}\BibitemShut {NoStop}%
\bibitem [{\citenamefont {Barbaro}\ \emph {et~al.}(2004)\citenamefont
  {Barbaro}, \citenamefont {Caballero}, \citenamefont {Donnelly},\ and\
  \citenamefont {Maieron}}]{Barbaro:2003ie}%
  \BibitemOpen
  \bibfield  {author} {\bibinfo {author} {\bibfnamefont {M.~B.}\ \bibnamefont
  {Barbaro}}, \bibinfo {author} {\bibfnamefont {J.~A.}\ \bibnamefont
  {Caballero}}, \bibinfo {author} {\bibfnamefont {T.~W.}\ \bibnamefont
  {Donnelly}},\ and\ \bibinfo {author} {\bibfnamefont {C.}~\bibnamefont
  {Maieron}},\ }\bibfield  {title} {\bibinfo {title} {{Inelastic electron
  nucleus scattering and scaling at high inelasticity}},\ }\href
  {https://doi.org/10.1103/PhysRevC.69.035502} {\bibfield  {journal} {\bibinfo
  {journal} {Phys.\ Rev.\ C}\ }\textbf {\bibinfo {volume} {69}},\ \bibinfo
  {pages} {035502} (\bibinfo {year} {2004})},\ \Eprint
  {https://arxiv.org/abs/nucl-th/0311088} {nucl-th/0311088} \BibitemShut
  {NoStop}%
\bibitem [{\citenamefont {Barbaro}\ \emph {et~al.}(1998)\citenamefont
  {Barbaro}, \citenamefont {Cenni}, \citenamefont {De~Pace}, \citenamefont
  {Donnelly},\ and\ \citenamefont {Molinari}}]{Barbaro:1998gu}%
  \BibitemOpen
  \bibfield  {author} {\bibinfo {author} {\bibfnamefont {M.~B.}\ \bibnamefont
  {Barbaro}}, \bibinfo {author} {\bibfnamefont {R.}~\bibnamefont {Cenni}},
  \bibinfo {author} {\bibfnamefont {A.}~\bibnamefont {De~Pace}}, \bibinfo
  {author} {\bibfnamefont {T.~W.}\ \bibnamefont {Donnelly}},\ and\ \bibinfo
  {author} {\bibfnamefont {A.}~\bibnamefont {Molinari}},\ }\bibfield  {title}
  {\bibinfo {title} {{Relativistic $y$-scaling and the Coulomb sum rule in
  nuclei}},\ }\href {https://doi.org/10.1016/S0375-9474(98)00443-6} {\bibfield
  {journal} {\bibinfo  {journal} {Nucl.\ Phys.\ A}\ }\textbf {\bibinfo {volume}
  {643}},\ \bibinfo {pages} {137} (\bibinfo {year} {1998})},\ \Eprint
  {https://arxiv.org/abs/nucl-th/9804054} {nucl-th/9804054} \BibitemShut
  {NoStop}%
\bibitem [{\citenamefont {Day}\ \emph {et~al.}(1990)\citenamefont {Day},
  \citenamefont {McCarthy}, \citenamefont {Donnelly},\ and\ \citenamefont
  {Sick}}]{Day:1990mf}%
  \BibitemOpen
  \bibfield  {author} {\bibinfo {author} {\bibfnamefont {D.~B.}\ \bibnamefont
  {Day}}, \bibinfo {author} {\bibfnamefont {J.~S.}\ \bibnamefont {McCarthy}},
  \bibinfo {author} {\bibfnamefont {T.~W.}\ \bibnamefont {Donnelly}},\ and\
  \bibinfo {author} {\bibfnamefont {I.}~\bibnamefont {Sick}},\ }\bibfield
  {title} {\bibinfo {title} {{Scaling in inclusive electron-nucleus
  scattering}},\ }\href {https://doi.org/10.1146/annurev.ns.40.120190.002041}
  {\bibfield  {journal} {\bibinfo  {journal} {Ann.\ Rev.\ Nucl.\ Part.\ Sci.}\
  }\textbf {\bibinfo {volume} {40}},\ \bibinfo {pages} {357} (\bibinfo {year}
  {1990})}\BibitemShut {NoStop}%
\bibitem [{\citenamefont {Caballero}\ \emph {et~al.}(2007)\citenamefont
  {Caballero}, \citenamefont {Amaro}, \citenamefont {Barbaro}, \citenamefont
  {Donnelly},\ and\ \citenamefont {Ud\'{i}as}}]{Caballero:2007tz}%
  \BibitemOpen
  \bibfield  {author} {\bibinfo {author} {\bibfnamefont {J.~A.}\ \bibnamefont
  {Caballero}}, \bibinfo {author} {\bibfnamefont {J.~E.}\ \bibnamefont
  {Amaro}}, \bibinfo {author} {\bibfnamefont {M.~B.}\ \bibnamefont {Barbaro}},
  \bibinfo {author} {\bibfnamefont {T.~W.}\ \bibnamefont {Donnelly}},\ and\
  \bibinfo {author} {\bibfnamefont {J.~M.}\ \bibnamefont {Ud\'{i}as}},\
  }\bibfield  {title} {\bibinfo {title} {{Scaling and isospin effects in
  quasielastic lepton-nucleus scattering in the Relativistic Mean Field
  Approach}},\ }\href {https://doi.org/10.1016/j.physletb.2007.08.018}
  {\bibfield  {journal} {\bibinfo  {journal} {Phys.\ Lett.\ B}\ }\textbf
  {\bibinfo {volume} {653}},\ \bibinfo {pages} {366} (\bibinfo {year}
  {2007})},\ \Eprint {https://arxiv.org/abs/0705.1429} {arXiv:0705.1429
  [nucl-th]} \BibitemShut {NoStop}%
\bibitem [{San()}]{SanchezNieto:2021}%
  \BibitemOpen
  \href@noop {} {}\bibinfo {note} {{\relax F. Sanchez}, private
  communication.}\BibitemShut {Stop}%
\bibitem [{\citenamefont {Amaro}\ \emph
  {et~al.}(2005{\natexlab{a}})\citenamefont {Amaro}, \citenamefont {Barbaro},
  \citenamefont {Caballero}, \citenamefont {Donnelly}, \citenamefont
  {Molinari},\ and\ \citenamefont {Sick}}]{Amaro:2004bs}%
  \BibitemOpen
  \bibfield  {author} {\bibinfo {author} {\bibfnamefont {J.~E.}\ \bibnamefont
  {Amaro}}, \bibinfo {author} {\bibfnamefont {M.~B.}\ \bibnamefont {Barbaro}},
  \bibinfo {author} {\bibfnamefont {J.~A.}\ \bibnamefont {Caballero}}, \bibinfo
  {author} {\bibfnamefont {T.~W.}\ \bibnamefont {Donnelly}}, \bibinfo {author}
  {\bibfnamefont {A.}~\bibnamefont {Molinari}},\ and\ \bibinfo {author}
  {\bibfnamefont {I.}~\bibnamefont {Sick}},\ }\bibfield  {title} {\bibinfo
  {title} {{Using electron scattering superscaling to predict charge-changing
  neutrino cross sections in nuclei}},\ }\href
  {https://doi.org/10.1103/PhysRevC.71.015501} {\bibfield  {journal} {\bibinfo
  {journal} {Phys.\ Rev.\ C}\ }\textbf {\bibinfo {volume} {71}},\ \bibinfo
  {pages} {015501} (\bibinfo {year} {2005}{\natexlab{a}})},\ \Eprint
  {https://arxiv.org/abs/nucl-th/0409078} {nucl-th/0409078} \BibitemShut
  {NoStop}%
\bibitem [{\citenamefont {Amaro}\ \emph
  {et~al.}(2005{\natexlab{b}})\citenamefont {Amaro}, \citenamefont {Barbaro},
  \citenamefont {Caballero}, \citenamefont {Donnelly},\ and\ \citenamefont
  {Maieron}}]{Amaro:2005dn}%
  \BibitemOpen
  \bibfield  {author} {\bibinfo {author} {\bibfnamefont {J.~E.}\ \bibnamefont
  {Amaro}}, \bibinfo {author} {\bibfnamefont {M.~B.}\ \bibnamefont {Barbaro}},
  \bibinfo {author} {\bibfnamefont {J.~A.}\ \bibnamefont {Caballero}}, \bibinfo
  {author} {\bibfnamefont {T.~W.}\ \bibnamefont {Donnelly}},\ and\ \bibinfo
  {author} {\bibfnamefont {C.}~\bibnamefont {Maieron}},\ }\bibfield  {title}
  {\bibinfo {title} {{Semi-relativistic description of quasielastic neutrino
  reactions and superscaling in a continuum shell model}},\ }\href
  {https://doi.org/10.1103/PhysRevC.71.065501} {\bibfield  {journal} {\bibinfo
  {journal} {Phys.\ Rev.\ C}\ }\textbf {\bibinfo {volume} {71}},\ \bibinfo
  {pages} {065501} (\bibinfo {year} {2005}{\natexlab{b}})},\ \Eprint
  {https://arxiv.org/abs/nucl-th/0503062} {nucl-th/0503062} \BibitemShut
  {NoStop}%
\bibitem [{\citenamefont {Amaro}\ and\ \citenamefont
  {Ruiz~Arriola}(2016)}]{Amaro:2015lga}%
  \BibitemOpen
  \bibfield  {author} {\bibinfo {author} {\bibfnamefont {J.~E.}\ \bibnamefont
  {Amaro}}\ and\ \bibinfo {author} {\bibfnamefont {E.}~\bibnamefont
  {Ruiz~Arriola}},\ }\bibfield  {title} {\bibinfo {title} {{Axial-vector
  dominance predictions in quasielastic neutrino-nucleus scattering}},\ }\href
  {https://doi.org/10.1103/PhysRevD.93.053002} {\bibfield  {journal} {\bibinfo
  {journal} {Phys.\ Rev.\ D}\ }\textbf {\bibinfo {volume} {93}},\ \bibinfo
  {pages} {053002} (\bibinfo {year} {2016})},\ \Eprint
  {https://arxiv.org/abs/1510.07532} {arXiv:1510.07532 [nucl-th]} \BibitemShut
  {NoStop}%
\bibitem [{Note1()}]{Note1}%
  \BibitemOpen
  \bibinfo {note} {Notice that in Eq.\ \protect \textup {\hbox {\mathsurround
  \z@ \protect \normalfont (\ignorespaces \ref {d2sigma_dtmu_dcosmu}\unskip
  \@@italiccorr )}} we have particularized the general expression for the
  scattering of muon neutrinos/antineutrinos, but the expression is general for
  other kind of neutrino species, just by changing the final lepton
  mass.}\BibitemShut {Stop}%
\bibitem [{\citenamefont {Maieron}\ \emph {et~al.}(2002)\citenamefont
  {Maieron}, \citenamefont {Donnelly},\ and\ \citenamefont
  {Sick}}]{Maieron:2001it}%
  \BibitemOpen
  \bibfield  {author} {\bibinfo {author} {\bibfnamefont {C.}~\bibnamefont
  {Maieron}}, \bibinfo {author} {\bibfnamefont {T.~W.}\ \bibnamefont
  {Donnelly}},\ and\ \bibinfo {author} {\bibfnamefont {I.}~\bibnamefont
  {Sick}},\ }\bibfield  {title} {\bibinfo {title} {{Extended superscaling of
  electron scattering from nuclei}},\ }\href
  {https://doi.org/10.1103/PhysRevC.65.025502} {\bibfield  {journal} {\bibinfo
  {journal} {Phys. Rev. C}\ }\textbf {\bibinfo {volume} {65}},\ \bibinfo
  {pages} {025502} (\bibinfo {year} {2002})},\ \Eprint
  {https://arxiv.org/abs/nucl-th/0109032} {arXiv:nucl-th/0109032} \BibitemShut
  {NoStop}%
\bibitem [{\citenamefont {Gonz\'{a}lez-Jim\'{e}nez}\ \emph
  {et~al.}(2014)\citenamefont {Gonz\'{a}lez-Jim\'{e}nez}, \citenamefont
  {Megias}, \citenamefont {Barbaro}, \citenamefont {Caballero},\ and\
  \citenamefont {Donnelly}}]{Gonzalez-Jimenez:2014eqa}%
  \BibitemOpen
  \bibfield  {author} {\bibinfo {author} {\bibfnamefont {R.}~\bibnamefont
  {Gonz\'{a}lez-Jim\'{e}nez}}, \bibinfo {author} {\bibfnamefont {G.~D.}\
  \bibnamefont {Megias}}, \bibinfo {author} {\bibfnamefont {M.~B.}\
  \bibnamefont {Barbaro}}, \bibinfo {author} {\bibfnamefont {J.~A.}\
  \bibnamefont {Caballero}},\ and\ \bibinfo {author} {\bibfnamefont {T.~W.}\
  \bibnamefont {Donnelly}},\ }\bibfield  {title} {\bibinfo {title} {{Extensions
  of superscaling from relativistic mean field theory: the SuSAv2 model}},\
  }\href {https://doi.org/10.1103/PhysRevC.90.035501} {\bibfield  {journal}
  {\bibinfo  {journal} {Phys.\ Rev.\ C}\ }\textbf {\bibinfo {volume} {90}},\
  \bibinfo {pages} {035501} (\bibinfo {year} {2014})},\ \Eprint
  {https://arxiv.org/abs/1407.8346} {arXiv:1407.8346 [nucl-th]} \BibitemShut
  {NoStop}%
\bibitem [{Note2()}]{Note2}%
  \BibitemOpen
  \bibinfo {note} {If the kind of neutrino is a distinct one, one can use the
  formulae developed in sects.\ \ref {lepton-kinem-bound} and appendix\ \ref
  {appendix-a4} and change them for other final charged lepton
  masses.}\BibitemShut {Stop}%
\bibitem [{\citenamefont {Walecka}(1974)}]{Walecka:1974qa}%
  \BibitemOpen
  \bibfield  {author} {\bibinfo {author} {\bibfnamefont {J.~D.}\ \bibnamefont
  {Walecka}},\ }\bibfield  {title} {\bibinfo {title} {{A theory of highly
  condensed matter}},\ }\href {https://doi.org/10.1016/0003-4916(74)90208-5}
  {\bibfield  {journal} {\bibinfo  {journal} {Annals Phys.}\ }\textbf {\bibinfo
  {volume} {83}},\ \bibinfo {pages} {491} (\bibinfo {year} {1974})}\BibitemShut
  {NoStop}%
\bibitem [{\citenamefont {Serot}\ and\ \citenamefont
  {Walecka}(1986)}]{Serot:1984ey}%
  \BibitemOpen
  \bibfield  {author} {\bibinfo {author} {\bibfnamefont {B.~D.}\ \bibnamefont
  {Serot}}\ and\ \bibinfo {author} {\bibfnamefont {J.~D.}\ \bibnamefont
  {Walecka}},\ }\bibfield  {title} {\bibinfo {title} {{The relativistic nuclear
  many body problem}},\ }\href@noop {} {\bibfield  {journal} {\bibinfo
  {journal} {Adv.\ Nucl.\ Phys.}\ }\textbf {\bibinfo {volume} {16}},\ \bibinfo
  {pages} {1} (\bibinfo {year} {1986})}\BibitemShut {NoStop}%
\bibitem [{\citenamefont {Wehrberger}(1993)}]{Wehrberger:1993zu}%
  \BibitemOpen
  \bibfield  {author} {\bibinfo {author} {\bibfnamefont {K.}~\bibnamefont
  {Wehrberger}},\ }\bibfield  {title} {\bibinfo {title} {{Electromagnetic
  response functions in quantum hadrodynamics}},\ }\href
  {https://doi.org/10.1016/0370-1573(93)90102-J} {\bibfield  {journal}
  {\bibinfo  {journal} {Phys.\ Rept.}\ }\textbf {\bibinfo {volume} {225}},\
  \bibinfo {pages} {273} (\bibinfo {year} {1993})}\BibitemShut {NoStop}%
\bibitem [{\citenamefont {Rosenfelder}(1980)}]{Rosenfelder:1980nd}%
  \BibitemOpen
  \bibfield  {author} {\bibinfo {author} {\bibfnamefont {R.}~\bibnamefont
  {Rosenfelder}},\ }\bibfield  {title} {\bibinfo {title} {{Quasielastic
  electron scattering from nuclei}},\ }\href
  {https://doi.org/10.1016/0003-4916(80)90059-7} {\bibfield  {journal}
  {\bibinfo  {journal} {Annals Phys.}\ }\textbf {\bibinfo {volume} {128}},\
  \bibinfo {pages} {188} (\bibinfo {year} {1980})}\BibitemShut {NoStop}%
\bibitem [{\citenamefont {Benhar}\ \emph {et~al.}(2006)\citenamefont {Benhar},
  \citenamefont {Day},\ and\ \citenamefont {Sick}}]{Benhar:2006er}%
  \BibitemOpen
  \bibfield  {author} {\bibinfo {author} {\bibfnamefont {O.}~\bibnamefont
  {Benhar}}, \bibinfo {author} {\bibfnamefont {D.~B.}\ \bibnamefont {Day}},\
  and\ \bibinfo {author} {\bibfnamefont {I.}~\bibnamefont {Sick}},\ }\href
  {http://faculty.virginia.edu/qes-archive} {\bibinfo {title} {{An archive for
  quasi-elastic electron-nucleus scattering data}}} (\bibinfo {year} {2006}),\
  \Eprint {https://arxiv.org/abs/nucl-ex/0603032} {nucl-ex/0603032}
  \BibitemShut {NoStop}%
\bibitem [{\citenamefont {Benhar}\ \emph {et~al.}(2008)\citenamefont {Benhar},
  \citenamefont {Day},\ and\ \citenamefont {Sick}}]{Benhar:2006wy}%
  \BibitemOpen
  \bibfield  {author} {\bibinfo {author} {\bibfnamefont {O.}~\bibnamefont
  {Benhar}}, \bibinfo {author} {\bibfnamefont {D.~B.}\ \bibnamefont {Day}},\
  and\ \bibinfo {author} {\bibfnamefont {I.}~\bibnamefont {Sick}},\ }\bibfield
  {title} {\bibinfo {title} {{Inclusive quasi-elastic electron-nucleus
  scattering}},\ }\href {https://doi.org/10.1103/RevModPhys.80.189} {\bibfield
  {journal} {\bibinfo  {journal} {Rev.\ Mod.\ Phys.}\ }\textbf {\bibinfo
  {volume} {80}},\ \bibinfo {pages} {189} (\bibinfo {year} {2008})},\ \Eprint
  {https://arxiv.org/abs/nucl-ex/0603029} {nucl-ex/0603029} \BibitemShut
  {NoStop}%
\bibitem [{\citenamefont {Ruiz~Simo}\ \emph
  {et~al.}(2017{\natexlab{a}})\citenamefont {Ruiz~Simo}, \citenamefont
  {Navarro~P\'{e}rez}, \citenamefont {Amaro},\ and\ \citenamefont
  {Ruiz~Arriola}}]{RuizSimo:2017tcb}%
  \BibitemOpen
  \bibfield  {author} {\bibinfo {author} {\bibfnamefont {I.}~\bibnamefont
  {Ruiz~Simo}}, \bibinfo {author} {\bibfnamefont {R.}~\bibnamefont
  {Navarro~P\'{e}rez}}, \bibinfo {author} {\bibfnamefont {J.~E.}\ \bibnamefont
  {Amaro}},\ and\ \bibinfo {author} {\bibfnamefont {E.}~\bibnamefont
  {Ruiz~Arriola}},\ }\bibfield  {title} {\bibinfo {title} {{Coarse graining the
  Bethe-Goldstone equation: nucleon-nucleon high-momentum components}},\ }\href
  {https://doi.org/10.1103/PhysRevC.96.054006} {\bibfield  {journal} {\bibinfo
  {journal} {Phys.\ Rev.\ C}\ }\textbf {\bibinfo {volume} {96}},\ \bibinfo
  {pages} {054006} (\bibinfo {year} {2017}{\natexlab{a}})},\ \Eprint
  {https://arxiv.org/abs/1708.00878} {arXiv:1708.00878 [nucl-th]} \BibitemShut
  {NoStop}%
\bibitem [{\citenamefont {Sick}\ \emph {et~al.}(1980)\citenamefont {Sick},
  \citenamefont {Day},\ and\ \citenamefont {McCarthy}}]{Sick:1980ey}%
  \BibitemOpen
  \bibfield  {author} {\bibinfo {author} {\bibfnamefont {I.}~\bibnamefont
  {Sick}}, \bibinfo {author} {\bibfnamefont {D.~B.}\ \bibnamefont {Day}},\ and\
  \bibinfo {author} {\bibfnamefont {J.~S.}\ \bibnamefont {McCarthy}},\
  }\bibfield  {title} {\bibinfo {title} {{Nuclear high momentum components and
  y scaling in electron scattering}},\ }\href
  {https://doi.org/10.1103/PhysRevLett.45.871} {\bibfield  {journal} {\bibinfo
  {journal} {Phys.\ Rev.\ Lett.}\ }\textbf {\bibinfo {volume} {45}},\ \bibinfo
  {pages} {871} (\bibinfo {year} {1980})}\BibitemShut {NoStop}%
\bibitem [{\citenamefont {Ramos}\ \emph {et~al.}(1989)\citenamefont {Ramos},
  \citenamefont {Polls},\ and\ \citenamefont {Dickhoff}}]{Ramos:1989hqs}%
  \BibitemOpen
  \bibfield  {author} {\bibinfo {author} {\bibfnamefont {A.}~\bibnamefont
  {Ramos}}, \bibinfo {author} {\bibfnamefont {A.}~\bibnamefont {Polls}},\ and\
  \bibinfo {author} {\bibfnamefont {W.~H.}\ \bibnamefont {Dickhoff}},\
  }\bibfield  {title} {\bibinfo {title} {{Single-particle properties and
  short-range correlations in nuclear matter}},\ }\href
  {https://doi.org/10.1016/0375-9474(89)90252-2} {\bibfield  {journal}
  {\bibinfo  {journal} {Nucl.\ Phys.\ A}\ }\textbf {\bibinfo {volume} {503}},\
  \bibinfo {pages} {1} (\bibinfo {year} {1989})}\BibitemShut {NoStop}%
\bibitem [{\citenamefont {Arrington}\ \emph {et~al.}(2012)\citenamefont
  {Arrington}, \citenamefont {Higinbotham}, \citenamefont {Rosner},\ and\
  \citenamefont {Sargsian}}]{Arrington:2011xs}%
  \BibitemOpen
  \bibfield  {author} {\bibinfo {author} {\bibfnamefont {J.}~\bibnamefont
  {Arrington}}, \bibinfo {author} {\bibfnamefont {D.~W.}\ \bibnamefont
  {Higinbotham}}, \bibinfo {author} {\bibfnamefont {G.}~\bibnamefont
  {Rosner}},\ and\ \bibinfo {author} {\bibfnamefont {M.}~\bibnamefont
  {Sargsian}},\ }\bibfield  {title} {\bibinfo {title} {{Hard probes of
  short-range nucleon-nucleon correlations}},\ }\href
  {https://doi.org/10.1016/j.ppnp.2012.04.002} {\bibfield  {journal} {\bibinfo
  {journal} {Prog.\ Part.\ Nucl.\ Phys.}\ }\textbf {\bibinfo {volume} {67}},\
  \bibinfo {pages} {898} (\bibinfo {year} {2012})},\ \Eprint
  {https://arxiv.org/abs/1104.1196} {arXiv:1104.1196 [nucl-ex]} \BibitemShut
  {NoStop}%
\bibitem [{\citenamefont {Muther}\ \emph {et~al.}(1995)\citenamefont {Muther},
  \citenamefont {Polls},\ and\ \citenamefont {Dickhoff}}]{Muther:1995zz}%
  \BibitemOpen
  \bibfield  {author} {\bibinfo {author} {\bibfnamefont {H.}~\bibnamefont
  {Muther}}, \bibinfo {author} {\bibfnamefont {A.}~\bibnamefont {Polls}},\ and\
  \bibinfo {author} {\bibfnamefont {W.~H.}\ \bibnamefont {Dickhoff}},\
  }\bibfield  {title} {\bibinfo {title} {{Momentum and energy distributions of
  nucleons in finite nuclei due to short-range correlations}},\ }\href
  {https://doi.org/10.1103/PhysRevC.51.3040} {\bibfield  {journal} {\bibinfo
  {journal} {Phys.\ Rev.\ C}\ }\textbf {\bibinfo {volume} {51}},\ \bibinfo
  {pages} {3040} (\bibinfo {year} {1995})},\ \Eprint
  {https://arxiv.org/abs/nucl-th/9411005} {nucl-th/9411005} \BibitemShut
  {NoStop}%
\bibitem [{\citenamefont {Giusti}\ \emph {et~al.}(1999)\citenamefont {Giusti},
  \citenamefont {Muther}, \citenamefont {Pacati},\ and\ \citenamefont
  {Stauf}}]{Giusti:1999sv}%
  \BibitemOpen
  \bibfield  {author} {\bibinfo {author} {\bibfnamefont {C.}~\bibnamefont
  {Giusti}}, \bibinfo {author} {\bibfnamefont {H.}~\bibnamefont {Muther}},
  \bibinfo {author} {\bibfnamefont {F.~D.}\ \bibnamefont {Pacati}},\ and\
  \bibinfo {author} {\bibfnamefont {M.}~\bibnamefont {Stauf}},\ }\bibfield
  {title} {\bibinfo {title} {{Short range and tensor correlations in the
  ${}^{16}\text{O}\left(e,e'p\,n\right)$ reaction}},\ }\href
  {https://doi.org/10.1103/PhysRevC.60.054608} {\bibfield  {journal} {\bibinfo
  {journal} {Phys.\ Rev.\ C}\ }\textbf {\bibinfo {volume} {60}},\ \bibinfo
  {pages} {054608} (\bibinfo {year} {1999})},\ \Eprint
  {https://arxiv.org/abs/nucl-th/9903065} {nucl-th/9903065} \BibitemShut
  {NoStop}%
\bibitem [{\citenamefont {Stoitsov}\ \emph {et~al.}(1993)\citenamefont
  {Stoitsov}, \citenamefont {Antonov},\ and\ \citenamefont
  {Dimitrova}}]{Stoitsov:1993zz}%
  \BibitemOpen
  \bibfield  {author} {\bibinfo {author} {\bibfnamefont {M.~V.}\ \bibnamefont
  {Stoitsov}}, \bibinfo {author} {\bibfnamefont {A.~N.}\ \bibnamefont
  {Antonov}},\ and\ \bibinfo {author} {\bibfnamefont {S.~S.}\ \bibnamefont
  {Dimitrova}},\ }\bibfield  {title} {\bibinfo {title} {{Natural orbital
  representation and short-range correlations in nuclei}},\ }\href
  {https://doi.org/10.1103/PhysRevC.48.74} {\bibfield  {journal} {\bibinfo
  {journal} {Phys.\ Rev.\ C}\ }\textbf {\bibinfo {volume} {48}},\ \bibinfo
  {pages} {74} (\bibinfo {year} {1993})}\BibitemShut {NoStop}%
\bibitem [{\citenamefont {Alvioli}\ \emph {et~al.}(2013)\citenamefont
  {Alvioli}, \citenamefont {Ciofi~degli Atti}, \citenamefont {Kaptari},
  \citenamefont {Mezzetti},\ and\ \citenamefont {Morita}}]{Alvioli:2012qa}%
  \BibitemOpen
  \bibfield  {author} {\bibinfo {author} {\bibfnamefont {M.}~\bibnamefont
  {Alvioli}}, \bibinfo {author} {\bibfnamefont {C.}~\bibnamefont {Ciofi~degli
  Atti}}, \bibinfo {author} {\bibfnamefont {L.~P.}\ \bibnamefont {Kaptari}},
  \bibinfo {author} {\bibfnamefont {C.~B.}\ \bibnamefont {Mezzetti}},\ and\
  \bibinfo {author} {\bibfnamefont {H.}~\bibnamefont {Morita}},\ }\bibfield
  {title} {\bibinfo {title} {{Nucleon momentum distributions, their
  spin-isospin dependence and short-range correlations}},\ }\href
  {https://doi.org/10.1103/PhysRevC.87.034603} {\bibfield  {journal} {\bibinfo
  {journal} {Phys.\ Rev.\ C}\ }\textbf {\bibinfo {volume} {87}},\ \bibinfo
  {pages} {034603} (\bibinfo {year} {2013})},\ \Eprint
  {https://arxiv.org/abs/1211.0134} {arXiv:1211.0134 [nucl-th]} \BibitemShut
  {NoStop}%
\bibitem [{\citenamefont {Vanhalst}\ \emph {et~al.}(2012)\citenamefont
  {Vanhalst}, \citenamefont {Ryckebusch},\ and\ \citenamefont
  {Cosyn}}]{Vanhalst:2012ur}%
  \BibitemOpen
  \bibfield  {author} {\bibinfo {author} {\bibfnamefont {M.}~\bibnamefont
  {Vanhalst}}, \bibinfo {author} {\bibfnamefont {J.}~\bibnamefont
  {Ryckebusch}},\ and\ \bibinfo {author} {\bibfnamefont {W.}~\bibnamefont
  {Cosyn}},\ }\bibfield  {title} {\bibinfo {title} {{Quantifying short-range
  correlations in nuclei}},\ }\href
  {https://doi.org/10.1103/PhysRevC.86.044619} {\bibfield  {journal} {\bibinfo
  {journal} {Phys.\ Rev.\ C}\ }\textbf {\bibinfo {volume} {86}},\ \bibinfo
  {pages} {044619} (\bibinfo {year} {2012})},\ \Eprint
  {https://arxiv.org/abs/1206.5151} {arXiv:1206.5151 [nucl-th]} \BibitemShut
  {NoStop}%
\bibitem [{\citenamefont {Van~Cuyck}\ \emph {et~al.}(2016)\citenamefont
  {Van~Cuyck}, \citenamefont {Jachowicz}, \citenamefont
  {Gonz\'{a}lez-Jim\'enez}, \citenamefont {Martini}, \citenamefont {Pandey},
  \citenamefont {Ryckebusch},\ and\ \citenamefont
  {Van~Dessel}}]{VanCuyck:2016fab}%
  \BibitemOpen
  \bibfield  {author} {\bibinfo {author} {\bibfnamefont {T.}~\bibnamefont
  {Van~Cuyck}}, \bibinfo {author} {\bibfnamefont {N.}~\bibnamefont
  {Jachowicz}}, \bibinfo {author} {\bibfnamefont {R.}~\bibnamefont
  {Gonz\'{a}lez-Jim\'enez}}, \bibinfo {author} {\bibfnamefont {M.}~\bibnamefont
  {Martini}}, \bibinfo {author} {\bibfnamefont {V.}~\bibnamefont {Pandey}},
  \bibinfo {author} {\bibfnamefont {J.}~\bibnamefont {Ryckebusch}},\ and\
  \bibinfo {author} {\bibfnamefont {N.}~\bibnamefont {Van~Dessel}},\ }\bibfield
   {title} {\bibinfo {title} {{Influence of short-range correlations in
  neutrino-nucleus scattering}},\ }\href
  {https://doi.org/10.1103/PhysRevC.94.024611} {\bibfield  {journal} {\bibinfo
  {journal} {Phys.\ Rev.\ C}\ }\textbf {\bibinfo {volume} {94}},\ \bibinfo
  {pages} {024611} (\bibinfo {year} {2016})},\ \Eprint
  {https://arxiv.org/abs/1606.00273} {arXiv:1606.00273 [nucl-th]} \BibitemShut
  {NoStop}%
\bibitem [{\citenamefont {Fomin}\ \emph {et~al.}(2017)\citenamefont {Fomin},
  \citenamefont {Higinbotham}, \citenamefont {Sargsian},\ and\ \citenamefont
  {Solvignon}}]{Fomin:2017ydn}%
  \BibitemOpen
  \bibfield  {author} {\bibinfo {author} {\bibfnamefont {N.}~\bibnamefont
  {Fomin}}, \bibinfo {author} {\bibfnamefont {D.}~\bibnamefont {Higinbotham}},
  \bibinfo {author} {\bibfnamefont {M.}~\bibnamefont {Sargsian}},\ and\
  \bibinfo {author} {\bibfnamefont {P.}~\bibnamefont {Solvignon}},\ }\bibfield
  {title} {\bibinfo {title} {{New results on short-range correlations in
  nuclei}},\ }\href {https://doi.org/10.1146/annurev-nucl-102115-044939}
  {\bibfield  {journal} {\bibinfo  {journal} {Ann.\ Rev.\ Nucl.\ Part.\ Sci.}\
  }\textbf {\bibinfo {volume} {67}},\ \bibinfo {pages} {129} (\bibinfo {year}
  {2017})},\ \Eprint {https://arxiv.org/abs/1708.08581} {arXiv:1708.08581
  [nucl-th]} \BibitemShut {NoStop}%
\bibitem [{\citenamefont {Amaro}\ \emph {et~al.}(1998)\citenamefont {Amaro},
  \citenamefont {Lallena}, \citenamefont {Co'},\ and\ \citenamefont
  {Fabrocini}}]{Amaro:1998rr}%
  \BibitemOpen
  \bibfield  {author} {\bibinfo {author} {\bibfnamefont {J.~E.}\ \bibnamefont
  {Amaro}}, \bibinfo {author} {\bibfnamefont {A.~M.}\ \bibnamefont {Lallena}},
  \bibinfo {author} {\bibfnamefont {G.}~\bibnamefont {Co'}},\ and\ \bibinfo
  {author} {\bibfnamefont {A.}~\bibnamefont {Fabrocini}},\ }\bibfield  {title}
  {\bibinfo {title} {{A model of short range correlations in the charge
  response}},\ }\href {https://doi.org/10.1103/PhysRevC.57.3473} {\bibfield
  {journal} {\bibinfo  {journal} {Phys.\ Rev.\ C}\ }\textbf {\bibinfo {volume}
  {57}},\ \bibinfo {pages} {3473} (\bibinfo {year} {1998})},\ \Eprint
  {https://arxiv.org/abs/nucl-th/9803013} {nucl-th/9803013} \BibitemShut
  {NoStop}%
\bibitem [{\citenamefont {Mazziotta}\ \emph {et~al.}(2002)\citenamefont
  {Mazziotta}, \citenamefont {Amaro},\ and\ \citenamefont {Arias~de
  Saavedra}}]{Mazziotta:2001qh}%
  \BibitemOpen
  \bibfield  {author} {\bibinfo {author} {\bibfnamefont {M.}~\bibnamefont
  {Mazziotta}}, \bibinfo {author} {\bibfnamefont {J.~E.}\ \bibnamefont
  {Amaro}},\ and\ \bibinfo {author} {\bibfnamefont {F.}~\bibnamefont {Arias~de
  Saavedra}},\ }\bibfield  {title} {\bibinfo {title} {{Effects of short range
  correlations in $(e,e'p)$ reactions and nuclear overlap functions}},\ }\href
  {https://doi.org/10.1103/PhysRevC.65.034602} {\bibfield  {journal} {\bibinfo
  {journal} {Phys.\ Rev.\ C}\ }\textbf {\bibinfo {volume} {65}},\ \bibinfo
  {pages} {034602} (\bibinfo {year} {2002})},\ \Eprint
  {https://arxiv.org/abs/nucl-th/0107018} {nucl-th/0107018} \BibitemShut
  {NoStop}%
\bibitem [{\citenamefont {Weise}(1972)}]{Weise:1972klw}%
  \BibitemOpen
  \bibfield  {author} {\bibinfo {author} {\bibfnamefont {W.}~\bibnamefont
  {Weise}},\ }\bibfield  {title} {\bibinfo {title} {{Effects of short-range
  correlations on the quasielastic scattering of electrons}},\ }\href
  {https://doi.org/10.1016/0375-9474(72)90344-2} {\bibfield  {journal}
  {\bibinfo  {journal} {Nucl.\ Phys.\ A}\ }\textbf {\bibinfo {volume} {193}},\
  \bibinfo {pages} {625} (\bibinfo {year} {1972})}\BibitemShut {NoStop}%
\bibitem [{\citenamefont {Ruiz~Simo}\ \emph
  {et~al.}(2017{\natexlab{b}})\citenamefont {Ruiz~Simo}, \citenamefont
  {Navarro~P\'{e}rez}, \citenamefont {Amaro},\ and\ \citenamefont
  {Ruiz~Arriola}}]{RuizSimo:2016vsh}%
  \BibitemOpen
  \bibfield  {author} {\bibinfo {author} {\bibfnamefont {I.}~\bibnamefont
  {Ruiz~Simo}}, \bibinfo {author} {\bibfnamefont {R.}~\bibnamefont
  {Navarro~P\'{e}rez}}, \bibinfo {author} {\bibfnamefont {J.~E.}\ \bibnamefont
  {Amaro}},\ and\ \bibinfo {author} {\bibfnamefont {E.}~\bibnamefont
  {Ruiz~Arriola}},\ }\bibfield  {title} {\bibinfo {title} {{Coarse grained
  short-range correlations}},\ }\href
  {https://doi.org/10.1103/PhysRevC.95.054003} {\bibfield  {journal} {\bibinfo
  {journal} {Phys.\ Rev.\ C}\ }\textbf {\bibinfo {volume} {95}},\ \bibinfo
  {pages} {054003} (\bibinfo {year} {2017}{\natexlab{b}})},\ \Eprint
  {https://arxiv.org/abs/1612.06228} {arXiv:1612.06228 [nucl-th]} \BibitemShut
  {NoStop}%
\bibitem [{\citenamefont {Wiringa}\ \emph {et~al.}(2014)\citenamefont
  {Wiringa}, \citenamefont {Schiavilla}, \citenamefont {Pieper},\ and\
  \citenamefont {Carlson}}]{Wiringa:2013ala}%
  \BibitemOpen
  \bibfield  {author} {\bibinfo {author} {\bibfnamefont {R.~B.}\ \bibnamefont
  {Wiringa}}, \bibinfo {author} {\bibfnamefont {R.}~\bibnamefont {Schiavilla}},
  \bibinfo {author} {\bibfnamefont {S.~C.}\ \bibnamefont {Pieper}},\ and\
  \bibinfo {author} {\bibfnamefont {J.}~\bibnamefont {Carlson}},\ }\bibfield
  {title} {\bibinfo {title} {{Nucleon and nucleon-pair momentum distributions
  in $A \le 12$ nuclei}},\ }\href {https://doi.org/10.1103/PhysRevC.89.024305}
  {\bibfield  {journal} {\bibinfo  {journal} {Phys.\ Rev.\ C}\ }\textbf
  {\bibinfo {volume} {89}},\ \bibinfo {pages} {024305} (\bibinfo {year}
  {2014})},\ \Eprint {https://arxiv.org/abs/1309.3794} {arXiv:1309.3794
  [nucl-th]} \BibitemShut {NoStop}%
\bibitem [{\citenamefont {Schiavilla}\ \emph {et~al.}(1987)\citenamefont
  {Schiavilla}, \citenamefont {Lewart}, \citenamefont {Pandharipande},
  \citenamefont {Pieper}, \citenamefont {Wiringa},\ and\ \citenamefont
  {Fantoni}}]{Schiavilla:1987ziw}%
  \BibitemOpen
  \bibfield  {author} {\bibinfo {author} {\bibfnamefont {R.}~\bibnamefont
  {Schiavilla}}, \bibinfo {author} {\bibfnamefont {D.~S.}\ \bibnamefont
  {Lewart}}, \bibinfo {author} {\bibfnamefont {V.~R.}\ \bibnamefont
  {Pandharipande}}, \bibinfo {author} {\bibfnamefont {S.~C.}\ \bibnamefont
  {Pieper}}, \bibinfo {author} {\bibfnamefont {R.~B.}\ \bibnamefont
  {Wiringa}},\ and\ \bibinfo {author} {\bibfnamefont {S.}~\bibnamefont
  {Fantoni}},\ }\bibfield  {title} {\bibinfo {title} {{Structure functions and
  correlations in nuclei}},\ }\href
  {https://doi.org/10.1016/0375-9474(87)90145-X} {\bibfield  {journal}
  {\bibinfo  {journal} {Nucl.\ Phys.\ A}\ }\textbf {\bibinfo {volume} {473}},\
  \bibinfo {pages} {267} (\bibinfo {year} {1987})}\BibitemShut {NoStop}%
\bibitem [{\citenamefont {Berardo}\ \emph {et~al.}(2011)\citenamefont
  {Berardo}, \citenamefont {Barbaro}, \citenamefont {Cenni}, \citenamefont
  {Donnelly},\ and\ \citenamefont {Molinari}}]{Berardo:2011fx}%
  \BibitemOpen
  \bibfield  {author} {\bibinfo {author} {\bibfnamefont {D.}~\bibnamefont
  {Berardo}}, \bibinfo {author} {\bibfnamefont {M.~B.}\ \bibnamefont
  {Barbaro}}, \bibinfo {author} {\bibfnamefont {R.}~\bibnamefont {Cenni}},
  \bibinfo {author} {\bibfnamefont {T.~W.}\ \bibnamefont {Donnelly}},\ and\
  \bibinfo {author} {\bibfnamefont {A.}~\bibnamefont {Molinari}},\ }\bibfield
  {title} {\bibinfo {title} {{Connecting scaling with short-range
  correlations}},\ }\href {https://doi.org/10.1103/PhysRevC.84.054315}
  {\bibfield  {journal} {\bibinfo  {journal} {Phys.\ Rev.\ C}\ }\textbf
  {\bibinfo {volume} {84}},\ \bibinfo {pages} {054315} (\bibinfo {year}
  {2011})},\ \Eprint {https://arxiv.org/abs/1105.0358} {arXiv:1105.0358
  [nucl-th]} \BibitemShut {NoStop}%
\bibitem [{\citenamefont {Tornow}\ \emph {et~al.}(1981)\citenamefont {Tornow},
  \citenamefont {Drechsel}, \citenamefont {Orlandini},\ and\ \citenamefont
  {Traini}}]{Tornow:1981tb}%
  \BibitemOpen
  \bibfield  {author} {\bibinfo {author} {\bibfnamefont {V.}~\bibnamefont
  {Tornow}}, \bibinfo {author} {\bibfnamefont {D.}~\bibnamefont {Drechsel}},
  \bibinfo {author} {\bibfnamefont {G.}~\bibnamefont {Orlandini}},\ and\
  \bibinfo {author} {\bibfnamefont {M.}~\bibnamefont {Traini}},\ }\bibfield
  {title} {\bibinfo {title} {{Effects of wave function correlations on scaling
  violation in quasifree electron scattering}},\ }\href
  {https://doi.org/10.1016/0370-2693(81)90825-X} {\bibfield  {journal}
  {\bibinfo  {journal} {Phys.\ Lett.\ B}\ }\textbf {\bibinfo {volume} {107}},\
  \bibinfo {pages} {259} (\bibinfo {year} {1981})}\BibitemShut {NoStop}%
\bibitem [{\citenamefont {Day}(2008)}]{Day:2008zz}%
  \BibitemOpen
  \bibfield  {author} {\bibinfo {author} {\bibfnamefont {D.}~\bibnamefont
  {Day}},\ }\bibfield  {title} {\bibinfo {title} {{Short range correlations,
  inclusive electron-nucleus scattering, and scaling}},\ }\bibfield
  {booktitle} {\emph {\bibinfo {booktitle} {{Proceedings of the 6th
  International Conference on Perspectives in hadronic physics (Hadron\,2008),
  Trieste, Italy, May 12--16, 2008}}},\ }\href
  {https://doi.org/10.1063/1.3013058} {\bibfield  {journal} {\bibinfo
  {journal} {AIP Conf.\ Proc.}\ }\textbf {\bibinfo {volume} {1056}},\ \bibinfo
  {pages} {315} (\bibinfo {year} {2008})}\BibitemShut {NoStop}%
\bibitem [{\citenamefont {Martinez-Consentino}\ \emph
  {et~al.}(2021)\citenamefont {Martinez-Consentino}, \citenamefont {Simo},\
  and\ \citenamefont {Amaro}}]{Martinez-Consentino:2021avr}%
  \BibitemOpen
  \bibfield  {author} {\bibinfo {author} {\bibfnamefont {V.~L.}\ \bibnamefont
  {Martinez-Consentino}}, \bibinfo {author} {\bibfnamefont {I.~R.}\
  \bibnamefont {Simo}},\ and\ \bibinfo {author} {\bibfnamefont {J.~E.}\
  \bibnamefont {Amaro}},\ }\bibfield  {title} {\bibinfo {title}
  {{Meson-exchange currents and superscaling analysis with relativistic
  effective mass of quasielastic electron scattering from C12}},\ }\href
  {https://doi.org/10.1103/PhysRevC.104.025501} {\bibfield  {journal} {\bibinfo
   {journal} {Phys.\ Rev.\ C}\ }\textbf {\bibinfo {volume} {104}},\ \bibinfo
  {pages} {025501} (\bibinfo {year} {2021})},\ \Eprint
  {https://arxiv.org/abs/2105.09079} {arXiv:2105.09079 [nucl-th]} \BibitemShut
  {NoStop}%
\bibitem [{\citenamefont {Megias}\ \emph {et~al.}(2014)\citenamefont {Megias},
  \citenamefont {Ivanov}, \citenamefont {Gonz\'{a}lez-Jim\'{e}nez},
  \citenamefont {Barbaro}, \citenamefont {Caballero}, \citenamefont
  {Donnelly},\ and\ \citenamefont {Ud\'{i}as}}]{Megias:2014kiaWithErratum}%
  \BibitemOpen
  \bibfield  {author} {\bibinfo {author} {\bibfnamefont {G.~D.}\ \bibnamefont
  {Megias}}, \bibinfo {author} {\bibfnamefont {M.~V.}\ \bibnamefont {Ivanov}},
  \bibinfo {author} {\bibfnamefont {R.}~\bibnamefont
  {Gonz\'{a}lez-Jim\'{e}nez}}, \bibinfo {author} {\bibfnamefont {M.~B.}\
  \bibnamefont {Barbaro}}, \bibinfo {author} {\bibfnamefont {J.~A.}\
  \bibnamefont {Caballero}}, \bibinfo {author} {\bibfnamefont {T.~W.}\
  \bibnamefont {Donnelly}},\ and\ \bibinfo {author} {\bibfnamefont {J.~M.}\
  \bibnamefont {Ud\'{i}as}},\ }\bibfield  {title} {\bibinfo {title} {{Nuclear
  effects in neutrino and antineutrino charged-current quasielastic scattering
  at MINER$\nu$A kinematics}},\ }\href
  {https://doi.org/10.1103/PhysRevD.89.093002} {\bibfield  {journal} {\bibinfo
  {journal} {Phys.\ Rev.\ D}\ }\textbf {\bibinfo {volume} {89}},\ \bibinfo
  {pages} {093002} (\bibinfo {year} {2014})},\ \bibinfo {note} {[Erratum: {\it
  ibid.} {\bf 91}, 039903(E) (2015)]},\ \Eprint
  {https://arxiv.org/abs/1402.1611} {arXiv:1402.1611 [nucl-th]} \BibitemShut
  {NoStop}%
\bibitem [{Note3()}]{Note3}%
  \BibitemOpen
  \bibinfo {note} {The case of $\tau $ neutrinos should be taken with care and
  studied in depth because both solutions could be positive if the nucleon mass
  were the relativistic effective mass of the Walecka model \protect \cite
  {Walecka:1974qa,Serot:1984ey,Wehrberger:1993zu,Rosenfelder:1980nd}, which is
  the underlying theoretical model in which the SuSAM* is based
  on.}\BibitemShut {Stop}%
\bibitem [{\citenamefont {Dolan}\ \emph {et~al.}(2020)\citenamefont {Dolan},
  \citenamefont {Megias},\ and\ \citenamefont {Bolognesi}}]{Dolan:2019bxf}%
  \BibitemOpen
  \bibfield  {author} {\bibinfo {author} {\bibfnamefont {S.}~\bibnamefont
  {Dolan}}, \bibinfo {author} {\bibfnamefont {G.~D.}\ \bibnamefont {Megias}},\
  and\ \bibinfo {author} {\bibfnamefont {S.}~\bibnamefont {Bolognesi}},\
  }\bibfield  {title} {\bibinfo {title} {{Implementation of the SuSAv2-meson
  exchange current 1p1h and 2p2h models in GENIE and analysis of nuclear
  effects in T2K measurements}},\ }\href
  {https://doi.org/10.1103/PhysRevD.101.033003} {\bibfield  {journal} {\bibinfo
   {journal} {Phys.\ Rev.\ D}\ }\textbf {\bibinfo {volume} {101}},\ \bibinfo
  {pages} {033003} (\bibinfo {year} {2020})},\ \Eprint
  {https://arxiv.org/abs/1905.08556} {arXiv:1905.08556 [hep-ex]} \BibitemShut
  {NoStop}%
\bibitem [{\citenamefont {Megias}\ \emph
  {et~al.}(2016{\natexlab{a}})\citenamefont {Megias}, \citenamefont {Amaro},
  \citenamefont {Barbaro}, \citenamefont {Caballero}, \citenamefont
  {Donnelly},\ and\ \citenamefont {Ruiz~Simo}}]{Megias:2016fjk}%
  \BibitemOpen
  \bibfield  {author} {\bibinfo {author} {\bibfnamefont {G.~D.}\ \bibnamefont
  {Megias}}, \bibinfo {author} {\bibfnamefont {J.~E.}\ \bibnamefont {Amaro}},
  \bibinfo {author} {\bibfnamefont {M.~B.}\ \bibnamefont {Barbaro}}, \bibinfo
  {author} {\bibfnamefont {J.~A.}\ \bibnamefont {Caballero}}, \bibinfo {author}
  {\bibfnamefont {T.~W.}\ \bibnamefont {Donnelly}},\ and\ \bibinfo {author}
  {\bibfnamefont {I.}~\bibnamefont {Ruiz~Simo}},\ }\bibfield  {title} {\bibinfo
  {title} {{Charged-current neutrino-nucleus reactions within the superscaling
  meson-exchange current approach}},\ }\href
  {https://doi.org/10.1103/PhysRevD.94.093004} {\bibfield  {journal} {\bibinfo
  {journal} {Phys.\ Rev.\ D}\ }\textbf {\bibinfo {volume} {94}},\ \bibinfo
  {pages} {093004} (\bibinfo {year} {2016}{\natexlab{a}})},\ \Eprint
  {https://arxiv.org/abs/1607.08565} {arXiv:1607.08565 [nucl-th]} \BibitemShut
  {NoStop}%
\bibitem [{\citenamefont {Lyubushkin}\ \emph {et~al.}(2009)\citenamefont
  {Lyubushkin} \emph {et~al.}}]{Lyubushkin:2008pe}%
  \BibitemOpen
  \bibfield  {author} {\bibinfo {author} {\bibfnamefont {V.~V.}\ \bibnamefont
  {Lyubushkin}} \emph {et~al.} (\bibinfo {collaboration} {NOMAD
  Collaboration}),\ }\bibfield  {title} {\bibinfo {title} {{A study of
  quasi-elastic muon neutrino and antineutrino scattering in the NOMAD
  experiment}},\ }\href {https://doi.org/10.1140/epjc/s10052-009-1113-0}
  {\bibfield  {journal} {\bibinfo  {journal} {Eur.\ Phys.\ J.\ C}\ }\textbf
  {\bibinfo {volume} {63}},\ \bibinfo {pages} {355} (\bibinfo {year} {2009})},\
  \Eprint {https://arxiv.org/abs/0812.4543} {arXiv:0812.4543 [hep-ex]}
  \BibitemShut {NoStop}%
\bibitem [{\citenamefont {Megias}\ \emph
  {et~al.}(2016{\natexlab{b}})\citenamefont {Megias}, \citenamefont {Amaro},
  \citenamefont {Barbaro}, \citenamefont {Caballero},\ and\ \citenamefont
  {Donnelly}}]{Megias:2016lke}%
  \BibitemOpen
  \bibfield  {author} {\bibinfo {author} {\bibfnamefont {G.~D.}\ \bibnamefont
  {Megias}}, \bibinfo {author} {\bibfnamefont {J.~E.}\ \bibnamefont {Amaro}},
  \bibinfo {author} {\bibfnamefont {M.~B.}\ \bibnamefont {Barbaro}}, \bibinfo
  {author} {\bibfnamefont {J.~A.}\ \bibnamefont {Caballero}},\ and\ \bibinfo
  {author} {\bibfnamefont {T.~W.}\ \bibnamefont {Donnelly}},\ }\bibfield
  {title} {\bibinfo {title} {{Inclusive electron scattering within the SuSAv2
  meson-exchange current approach}},\ }\href
  {https://doi.org/10.1103/PhysRevD.94.013012} {\bibfield  {journal} {\bibinfo
  {journal} {Phys.\ Rev.\ D}\ }\textbf {\bibinfo {volume} {94}},\ \bibinfo
  {pages} {013012} (\bibinfo {year} {2016}{\natexlab{b}})},\ \Eprint
  {https://arxiv.org/abs/1603.08396} {arXiv:1603.08396 [nucl-th]} \BibitemShut
  {NoStop}%
\bibitem [{\citenamefont {Ruiz~Simo}\ \emph
  {et~al.}(2017{\natexlab{c}})\citenamefont {Ruiz~Simo}, \citenamefont {Amaro},
  \citenamefont {Barbaro}, \citenamefont {De~Pace}, \citenamefont {Caballero},\
  and\ \citenamefont {Donnelly}}]{Simo:2016ikv}%
  \BibitemOpen
  \bibfield  {author} {\bibinfo {author} {\bibfnamefont {I.}~\bibnamefont
  {Ruiz~Simo}}, \bibinfo {author} {\bibfnamefont {J.~E.}\ \bibnamefont
  {Amaro}}, \bibinfo {author} {\bibfnamefont {M.~B.}\ \bibnamefont {Barbaro}},
  \bibinfo {author} {\bibfnamefont {A.}~\bibnamefont {De~Pace}}, \bibinfo
  {author} {\bibfnamefont {J.~A.}\ \bibnamefont {Caballero}},\ and\ \bibinfo
  {author} {\bibfnamefont {T.~W.}\ \bibnamefont {Donnelly}},\ }\bibfield
  {title} {\bibinfo {title} {{Relativistic model of 2p-2h meson exchange
  currents in (anti)neutrino scattering}},\ }\href
  {https://doi.org/10.1088/1361-6471/aa6a06} {\bibfield  {journal} {\bibinfo
  {journal} {J.\ Phys.\ G}\ }\textbf {\bibinfo {volume} {44}},\ \bibinfo
  {pages} {065105} (\bibinfo {year} {2017}{\natexlab{c}})},\ \Eprint
  {https://arxiv.org/abs/1604.08423} {arXiv:1604.08423 [nucl-th]} \BibitemShut
  {NoStop}%
\bibitem [{\citenamefont {Butkevich}(2008)}]{Butkevich:2008ef}%
  \BibitemOpen
  \bibfield  {author} {\bibinfo {author} {\bibfnamefont {A.~V.}\ \bibnamefont
  {Butkevich}},\ }\bibfield  {title} {\bibinfo {title} {{Analysis of
  quasi-elastic neutrino charged-current scattering off ${}^{16}\text{O}$ and
  neutrino energy reconstruction}},\ }\href
  {https://doi.org/10.1103/PhysRevC.78.015501} {\bibfield  {journal} {\bibinfo
  {journal} {Phys.\ Rev.\ C}\ }\textbf {\bibinfo {volume} {78}},\ \bibinfo
  {pages} {015501} (\bibinfo {year} {2008})},\ \Eprint
  {https://arxiv.org/abs/0804.4102} {arXiv:0804.4102 [nucl-th]} \BibitemShut
  {NoStop}%
\bibitem [{\citenamefont {Martini}\ \emph {et~al.}(2012)\citenamefont
  {Martini}, \citenamefont {Ericson},\ and\ \citenamefont
  {Chanfray}}]{Martini:2012fa}%
  \BibitemOpen
  \bibfield  {author} {\bibinfo {author} {\bibfnamefont {M.}~\bibnamefont
  {Martini}}, \bibinfo {author} {\bibfnamefont {M.}~\bibnamefont {Ericson}},\
  and\ \bibinfo {author} {\bibfnamefont {G.}~\bibnamefont {Chanfray}},\
  }\bibfield  {title} {\bibinfo {title} {{Neutrino energy reconstruction
  problems and neutrino oscillations}},\ }\href
  {https://doi.org/10.1103/PhysRevD.85.093012} {\bibfield  {journal} {\bibinfo
  {journal} {Phys.\ Rev.\ D}\ }\textbf {\bibinfo {volume} {85}},\ \bibinfo
  {pages} {093012} (\bibinfo {year} {2012})},\ \Eprint
  {https://arxiv.org/abs/1202.4745} {arXiv:1202.4745 [hep-ph]} \BibitemShut
  {NoStop}%
\bibitem [{\citenamefont {Nieves}\ \emph {et~al.}(2012)\citenamefont {Nieves},
  \citenamefont {Sanchez}, \citenamefont {Ruiz~Simo},\ and\ \citenamefont
  {Vicente~Vacas}}]{Nieves:2012yz}%
  \BibitemOpen
  \bibfield  {author} {\bibinfo {author} {\bibfnamefont {J.}~\bibnamefont
  {Nieves}}, \bibinfo {author} {\bibfnamefont {F.}~\bibnamefont {Sanchez}},
  \bibinfo {author} {\bibfnamefont {I.}~\bibnamefont {Ruiz~Simo}},\ and\
  \bibinfo {author} {\bibfnamefont {M.~J.}\ \bibnamefont {Vicente~Vacas}},\
  }\bibfield  {title} {\bibinfo {title} {{Neutrino energy reconstruction and
  the shape of the CCQE-like total cross section}},\ }\href
  {https://doi.org/10.1103/PhysRevD.85.113008} {\bibfield  {journal} {\bibinfo
  {journal} {Phys.\ Rev.\ D}\ }\textbf {\bibinfo {volume} {85}},\ \bibinfo
  {pages} {113008} (\bibinfo {year} {2012})},\ \Eprint
  {https://arxiv.org/abs/1204.5404} {arXiv:1204.5404 [hep-ph]} \BibitemShut
  {NoStop}%
\bibitem [{\citenamefont {Martini}\ \emph {et~al.}(2013)\citenamefont
  {Martini}, \citenamefont {Ericson},\ and\ \citenamefont
  {Chanfray}}]{Martini:2012uc}%
  \BibitemOpen
  \bibfield  {author} {\bibinfo {author} {\bibfnamefont {M.}~\bibnamefont
  {Martini}}, \bibinfo {author} {\bibfnamefont {M.}~\bibnamefont {Ericson}},\
  and\ \bibinfo {author} {\bibfnamefont {G.}~\bibnamefont {Chanfray}},\
  }\bibfield  {title} {\bibinfo {title} {{Energy reconstruction effects in
  neutrino oscillation experiments and implications for the analysis}},\ }\href
  {https://doi.org/10.1103/PhysRevD.87.013009} {\bibfield  {journal} {\bibinfo
  {journal} {Phys.\ Rev.\ D}\ }\textbf {\bibinfo {volume} {87}},\ \bibinfo
  {pages} {013009} (\bibinfo {year} {2013})},\ \Eprint
  {https://arxiv.org/abs/1211.1523} {arXiv:1211.1523 [hep-ph]} \BibitemShut
  {NoStop}%
\bibitem [{\citenamefont {Leitner}\ and\ \citenamefont
  {Mosel}(2010)}]{Leitner:2010kp}%
  \BibitemOpen
  \bibfield  {author} {\bibinfo {author} {\bibfnamefont {T.}~\bibnamefont
  {Leitner}}\ and\ \bibinfo {author} {\bibfnamefont {U.}~\bibnamefont
  {Mosel}},\ }\bibfield  {title} {\bibinfo {title} {{Neutrino-nucleus
  scattering reexamined: Quasielastic scattering and pion production
  entanglement and implications for neutrino energy reconstruction}},\ }\href
  {https://doi.org/10.1103/PhysRevC.81.064614} {\bibfield  {journal} {\bibinfo
  {journal} {Phys.\ Rev.\ C}\ }\textbf {\bibinfo {volume} {81}},\ \bibinfo
  {pages} {064614} (\bibinfo {year} {2010})},\ \Eprint
  {https://arxiv.org/abs/1004.4433} {arXiv:1004.4433 [nucl-th]} \BibitemShut
  {NoStop}%
\bibitem [{\citenamefont {Ankowski}\ \emph
  {et~al.}(2015{\natexlab{a}})\citenamefont {Ankowski}, \citenamefont
  {Benhar},\ and\ \citenamefont {Sakuda}}]{Ankowski:2014yfa}%
  \BibitemOpen
  \bibfield  {author} {\bibinfo {author} {\bibfnamefont {A.~M.}\ \bibnamefont
  {Ankowski}}, \bibinfo {author} {\bibfnamefont {O.}~\bibnamefont {Benhar}},\
  and\ \bibinfo {author} {\bibfnamefont {M.}~\bibnamefont {Sakuda}},\
  }\bibfield  {title} {\bibinfo {title} {{Improving the accuracy of neutrino
  energy reconstruction in charged-current quasielastic scattering off nuclear
  targets}},\ }\href {https://doi.org/10.1103/PhysRevD.91.033005} {\bibfield
  {journal} {\bibinfo  {journal} {Phys.\ Rev.\ D}\ }\textbf {\bibinfo {volume}
  {91}},\ \bibinfo {pages} {033005} (\bibinfo {year} {2015}{\natexlab{a}})},\
  \Eprint {https://arxiv.org/abs/1404.5687} {arXiv:1404.5687 [nucl-th]}
  \BibitemShut {NoStop}%
\bibitem [{\citenamefont {Mosel}\ \emph {et~al.}(2014)\citenamefont {Mosel},
  \citenamefont {Lalakulich},\ and\ \citenamefont
  {Gallmeister}}]{Mosel:2013fxa}%
  \BibitemOpen
  \bibfield  {author} {\bibinfo {author} {\bibfnamefont {U.}~\bibnamefont
  {Mosel}}, \bibinfo {author} {\bibfnamefont {O.}~\bibnamefont {Lalakulich}},\
  and\ \bibinfo {author} {\bibfnamefont {K.}~\bibnamefont {Gallmeister}},\
  }\bibfield  {title} {\bibinfo {title} {{Energy reconstruction in the
  long-baseline neutrino experiment}},\ }\href
  {https://doi.org/10.1103/PhysRevLett.112.151802} {\bibfield  {journal}
  {\bibinfo  {journal} {Phys.\ Rev.\ Lett.}\ }\textbf {\bibinfo {volume}
  {112}},\ \bibinfo {pages} {151802} (\bibinfo {year} {2014})},\ \Eprint
  {https://arxiv.org/abs/1311.7288} {arXiv:1311.7288 [nucl-th]} \BibitemShut
  {NoStop}%
\bibitem [{\citenamefont {De~Romeri}\ \emph {et~al.}(2016)\citenamefont
  {De~Romeri}, \citenamefont {Fernandez-Martinez},\ and\ \citenamefont
  {Sorel}}]{DeRomeri:2016qwo}%
  \BibitemOpen
  \bibfield  {author} {\bibinfo {author} {\bibfnamefont {V.}~\bibnamefont
  {De~Romeri}}, \bibinfo {author} {\bibfnamefont {E.}~\bibnamefont
  {Fernandez-Martinez}},\ and\ \bibinfo {author} {\bibfnamefont
  {M.}~\bibnamefont {Sorel}},\ }\bibfield  {title} {\bibinfo {title} {{Neutrino
  oscillations at DUNE with improved energy reconstruction}},\ }\href
  {https://doi.org/10.1007/JHEP09(2016)030} {\bibfield  {journal} {\bibinfo
  {journal} {\relax{JHEP}}\ }\textbf {\bibinfo {volume} {09}},\ \bibinfo
  {pages} {030} (\bibinfo {year} {2016})},\ \Eprint
  {https://arxiv.org/abs/1607.00293} {arXiv:1607.00293 [hep-ph]} \BibitemShut
  {NoStop}%
\bibitem [{\citenamefont {Lu}\ \emph {et~al.}(2015)\citenamefont {Lu},
  \citenamefont {Coplowe}, \citenamefont {Shah}, \citenamefont {Barr},
  \citenamefont {Wark},\ and\ \citenamefont {Weber}}]{Lu:2015hea}%
  \BibitemOpen
  \bibfield  {author} {\bibinfo {author} {\bibfnamefont {X.~G.}\ \bibnamefont
  {Lu}}, \bibinfo {author} {\bibfnamefont {D.}~\bibnamefont {Coplowe}},
  \bibinfo {author} {\bibfnamefont {R.}~\bibnamefont {Shah}}, \bibinfo {author}
  {\bibfnamefont {G.}~\bibnamefont {Barr}}, \bibinfo {author} {\bibfnamefont
  {D.}~\bibnamefont {Wark}},\ and\ \bibinfo {author} {\bibfnamefont
  {A.}~\bibnamefont {Weber}},\ }\bibfield  {title} {\bibinfo {title}
  {{Reconstruction of energy spectra of neutrino beams independent of nuclear
  effects}},\ }\href {https://doi.org/10.1103/PhysRevD.92.051302} {\bibfield
  {journal} {\bibinfo  {journal} {Phys.\ Rev.\ D}\ }\textbf {\bibinfo {volume}
  {92}},\ \bibinfo {pages} {051302} (\bibinfo {year} {2015})},\ \Eprint
  {https://arxiv.org/abs/1507.00967} {arXiv:1507.00967 [hep-ex]} \BibitemShut
  {NoStop}%
\bibitem [{\citenamefont {Ankowski}\ \emph
  {et~al.}(2015{\natexlab{b}})\citenamefont {Ankowski}, \citenamefont {Benhar},
  \citenamefont {Coloma}, \citenamefont {Huber}, \citenamefont {Jen},
  \citenamefont {Mariani}, \citenamefont {Meloni},\ and\ \citenamefont
  {Vagnoni}}]{Ankowski:2015jya}%
  \BibitemOpen
  \bibfield  {author} {\bibinfo {author} {\bibfnamefont {A.~M.}\ \bibnamefont
  {Ankowski}}, \bibinfo {author} {\bibfnamefont {O.}~\bibnamefont {Benhar}},
  \bibinfo {author} {\bibfnamefont {P.}~\bibnamefont {Coloma}}, \bibinfo
  {author} {\bibfnamefont {P.}~\bibnamefont {Huber}}, \bibinfo {author}
  {\bibfnamefont {C.-M.}\ \bibnamefont {Jen}}, \bibinfo {author} {\bibfnamefont
  {C.}~\bibnamefont {Mariani}}, \bibinfo {author} {\bibfnamefont
  {D.}~\bibnamefont {Meloni}},\ and\ \bibinfo {author} {\bibfnamefont
  {E.}~\bibnamefont {Vagnoni}},\ }\bibfield  {title} {\bibinfo {title}
  {{Comparison of the calorimetric and kinematic methods of neutrino energy
  reconstruction in disappearance experiments}},\ }\href
  {https://doi.org/10.1103/PhysRevD.92.073014} {\bibfield  {journal} {\bibinfo
  {journal} {Phys.\ Rev.\ D}\ }\textbf {\bibinfo {volume} {92}},\ \bibinfo
  {pages} {073014} (\bibinfo {year} {2015}{\natexlab{b}})},\ \Eprint
  {https://arxiv.org/abs/1507.08560} {arXiv:1507.08560 [hep-ph]} \BibitemShut
  {NoStop}%
\bibitem [{\citenamefont {Munteanu}\ \emph {et~al.}(2020)\citenamefont
  {Munteanu}, \citenamefont {Suvorov}, \citenamefont {Dolan}, \citenamefont
  {Sgalaberna}, \citenamefont {Bolognesi}, \citenamefont {Manly}, \citenamefont
  {Yang}, \citenamefont {Giganti}, \citenamefont {Iwamoto},\ and\ \citenamefont
  {Jes\'{u}s-Valls}}]{Munteanu:2019llq}%
  \BibitemOpen
  \bibfield  {author} {\bibinfo {author} {\bibfnamefont {L.}~\bibnamefont
  {Munteanu}}, \bibinfo {author} {\bibfnamefont {S.}~\bibnamefont {Suvorov}},
  \bibinfo {author} {\bibfnamefont {S.}~\bibnamefont {Dolan}}, \bibinfo
  {author} {\bibfnamefont {D.}~\bibnamefont {Sgalaberna}}, \bibinfo {author}
  {\bibfnamefont {S.}~\bibnamefont {Bolognesi}}, \bibinfo {author}
  {\bibfnamefont {S.}~\bibnamefont {Manly}}, \bibinfo {author} {\bibfnamefont
  {G.}~\bibnamefont {Yang}}, \bibinfo {author} {\bibfnamefont {C.}~\bibnamefont
  {Giganti}}, \bibinfo {author} {\bibfnamefont {K.}~\bibnamefont {Iwamoto}},\
  and\ \bibinfo {author} {\bibfnamefont {C.}~\bibnamefont {Jes\'{u}s-Valls}},\
  }\bibfield  {title} {\bibinfo {title} {{New method for an improved
  antineutrino energy reconstruction with charged-current interactions in
  next-generation detectors}},\ }\href
  {https://doi.org/10.1103/PhysRevD.101.092003} {\bibfield  {journal} {\bibinfo
   {journal} {Phys.\ Rev.\ D}\ }\textbf {\bibinfo {volume} {101}},\ \bibinfo
  {pages} {092003} (\bibinfo {year} {2020})},\ \Eprint
  {https://arxiv.org/abs/1912.01511} {arXiv:1912.01511 [physics.ins-det]}
  \BibitemShut {NoStop}%
\bibitem [{\citenamefont {Furmanski}\ and\ \citenamefont
  {Sobczyk}(2017)}]{Furmanski:2016wqo}%
  \BibitemOpen
  \bibfield  {author} {\bibinfo {author} {\bibfnamefont {A.~P.}\ \bibnamefont
  {Furmanski}}\ and\ \bibinfo {author} {\bibfnamefont {J.~T.}\ \bibnamefont
  {Sobczyk}},\ }\bibfield  {title} {\bibinfo {title} {{Neutrino energy
  reconstruction from one muon and one proton events}},\ }\href
  {https://doi.org/10.1103/PhysRevC.95.065501} {\bibfield  {journal} {\bibinfo
  {journal} {Phys.\ Rev.\ C}\ }\textbf {\bibinfo {volume} {95}},\ \bibinfo
  {pages} {065501} (\bibinfo {year} {2017})},\ \Eprint
  {https://arxiv.org/abs/1609.03530} {arXiv:1609.03530 [hep-ex]} \BibitemShut
  {NoStop}%
\bibitem [{\citenamefont {Galster}\ \emph {et~al.}(1971)\citenamefont
  {Galster}, \citenamefont {Klein}, \citenamefont {Moritz}, \citenamefont
  {Schmidt}, \citenamefont {Wegener},\ and\ \citenamefont
  {Bleckwenn}}]{Galster:1971kv}%
  \BibitemOpen
  \bibfield  {author} {\bibinfo {author} {\bibfnamefont {S.}~\bibnamefont
  {Galster}}, \bibinfo {author} {\bibfnamefont {H.}~\bibnamefont {Klein}},
  \bibinfo {author} {\bibfnamefont {J.}~\bibnamefont {Moritz}}, \bibinfo
  {author} {\bibfnamefont {K.~H.}\ \bibnamefont {Schmidt}}, \bibinfo {author}
  {\bibfnamefont {D.}~\bibnamefont {Wegener}},\ and\ \bibinfo {author}
  {\bibfnamefont {J.}~\bibnamefont {Bleckwenn}},\ }\bibfield  {title} {\bibinfo
  {title} {{Elastic electron-deuteron scattering and the electric neutron form
  factor at four-momentum transfers 5 fm$^{-2} < q^2 < 14$ fm$^{-2}$}},\ }\href
  {https://doi.org/10.1016/0550-3213(71)90068-X} {\bibfield  {journal}
  {\bibinfo  {journal} {Nucl.\ Phys.\ B}\ }\textbf {\bibinfo {volume} {32}},\
  \bibinfo {pages} {221} (\bibinfo {year} {1971})}\BibitemShut {NoStop}%
\bibitem [{\citenamefont {Smith}\ and\ \citenamefont
  {Moniz}(1972{\natexlab{b}})}]{Smith:1972xhWithErratum}%
  \BibitemOpen
  \bibfield  {author} {\bibinfo {author} {\bibfnamefont {R.~A.}\ \bibnamefont
  {Smith}}\ and\ \bibinfo {author} {\bibfnamefont {E.~J.}\ \bibnamefont
  {Moniz}},\ }\bibfield  {title} {\bibinfo {title} {{Neutrino reactions on
  nuclear targets}},\ }\href {https://doi.org/10.1016/0550-3213(72)90040-5}
  {\bibfield  {journal} {\bibinfo  {journal} {Nucl.\ Phys.\ B}\ }\textbf
  {\bibinfo {volume} {43}},\ \bibinfo {pages} {605} (\bibinfo {year}
  {1972}{\natexlab{b}})},\ \bibinfo {note} {[Erratum: {\it ibid.} {\bf 101},
  547 (1975)]}\BibitemShut {NoStop}%
\bibitem [{\citenamefont {Kakorin}\ \emph {et~al.}(2020)\citenamefont
  {Kakorin}, \citenamefont {Kuzmin},\ and\ \citenamefont
  {Naumov}}]{Kakorin:2020atz}%
  \BibitemOpen
  \bibfield  {author} {\bibinfo {author} {\bibfnamefont {I.~D.}\ \bibnamefont
  {Kakorin}}, \bibinfo {author} {\bibfnamefont {K.~S.}\ \bibnamefont
  {Kuzmin}},\ and\ \bibinfo {author} {\bibfnamefont {V.~A.}\ \bibnamefont
  {Naumov}},\ }\bibfield  {title} {\bibinfo {title} {{A unified empirical model
  for quasielastic interactions of neutrino and antineutrino with nuclei}},\
  }\href {https://doi.org/10.3390/sym12081285} {\bibfield  {journal} {\bibinfo
  {journal} {Phys.\ Part.\ Nucl.\ Lett.}\ }\textbf {\bibinfo {volume} {17}},\
  \bibinfo {pages} {265} (\bibinfo {year} {2020})}\BibitemShut {NoStop}%
\bibitem [{\citenamefont {Suwonjandee}(2004)}]{Suwonjandee:2004aw}%
  \BibitemOpen
  \bibfield  {author} {\bibinfo {author} {\bibfnamefont {N.}~\bibnamefont
  {Suwonjandee}},\ }\emph {\bibinfo {title} {{The measurement of the
  quasi-elastic neutrino-nucleon scattering cross section at the Tevatron}}},\
  \href {http://wwwlib.umi.com/dissertations/fullcit?p3120857} {Ph.D. thesis},\
  \bibinfo  {school} {Cincinnati University} (\bibinfo {year}
  {2004})\BibitemShut {NoStop}%
\bibitem [{Note4()}]{Note4}%
  \BibitemOpen
  \bibinfo {note} {Note that $\epsilon _0=\kappa \protect \sqrt {1+1/\tau
  }-\lambda $ can be also hold in the region where $\kappa <\eta _F$ if $\kappa
  \protect \sqrt {1+1/\tau }>\epsilon _F-\lambda $ (see Eq.~\protect \textup
  {\hbox {\mathsurround \z@ \protect \normalfont (\ignorespaces \ref
  {e0_cond1}\unskip \@@italiccorr )}} and below).}\BibitemShut {Stop}%
\bibitem [{Note5()}]{Note5}%
  \BibitemOpen
  \bibinfo {note} {This statement can be rigorously and mathematically proved,
  but the easiest way to convince any reader of it is to have a look at Fig. 1
  of Ref.\ \cite {Alberico:1988bv}}\BibitemShut {NoStop}%
\bibitem [{\citenamefont {Ruiz~Simo}\ \emph
  {et~al.}(2018{\natexlab{b}})\citenamefont {Ruiz~Simo}, \citenamefont {Amaro},
  \citenamefont {Barbaro}, \citenamefont {Caballero}, \citenamefont {Megias},\
  and\ \citenamefont {Donnelly}}]{RuizSimo:2017hlc}%
  \BibitemOpen
  \bibfield  {author} {\bibinfo {author} {\bibfnamefont {I.}~\bibnamefont
  {Ruiz~Simo}}, \bibinfo {author} {\bibfnamefont {J.~E.}\ \bibnamefont
  {Amaro}}, \bibinfo {author} {\bibfnamefont {M.~B.}\ \bibnamefont {Barbaro}},
  \bibinfo {author} {\bibfnamefont {J.~A.}\ \bibnamefont {Caballero}}, \bibinfo
  {author} {\bibfnamefont {G.~D.}\ \bibnamefont {Megias}},\ and\ \bibinfo
  {author} {\bibfnamefont {T.~W.}\ \bibnamefont {Donnelly}},\ }\bibfield
  {title} {\bibinfo {title} {{Two-nucleon emission in neutrino and electron
  scattering from nuclei: the modified convolution approximation}},\ }\href
  {https://doi.org/10.1016/j.aop.2017.11.029} {\bibfield  {journal} {\bibinfo
  {journal} {Annals Phys.}\ }\textbf {\bibinfo {volume} {388}},\ \bibinfo
  {pages} {323} (\bibinfo {year} {2018}{\natexlab{b}})},\ \Eprint
  {https://arxiv.org/abs/1706.06377} {arXiv:1706.06377 [nucl-th]} \BibitemShut
  {NoStop}%
\bibitem [{Note6()}]{Note6}%
  \BibitemOpen
  \bibinfo {note} {The other possibility, i.e, that $\lambda \ge \lambda _{+}
  >\lambda _{-}$ can be ruled out because then $\lambda \ge (\epsilon
  _F+1)/2>1>\eta _F\equiv k_F/m_N$ (even with effective nucleon masses as in
  the SuSAM* model, we will always have that the Fermi momentum is smaller than
  the nucleon mass, regardless of this mass being the free nucleon mass or the
  relativistic effective one), and we are seeking solutions in the region where
  $\lambda < \eta _F$.}\BibitemShut {Stop}%
\bibitem [{Note7()}]{Note7}%
  \BibitemOpen
  \bibinfo {note} {This last feature can be stated because the discriminant of
  Eq.\ \protect \textup {\hbox {\mathsurround \z@ \protect \normalfont
  (\ignorespaces \ref {upm_sols1}\unskip \@@italiccorr )}} is positive and
  lesser than $\rho ^2$. Thus the square root of the discriminant is also
  lesser than $\rho $, and then $u_{-}(\lambda )$ is necessarily positive in
  the region where $0\le \lambda \le \lambda _{-}$.}\BibitemShut {Stop}%
\bibitem [{Note8()}]{Note8}%
  \BibitemOpen
  \bibinfo {note} {Another way to arrive to the same solution would have been
  to solve $\kappa ^\protect \text {NPB}_{+}(0)= \kappa ^{\protect \text
  {lepton}}_{\protect \qopname \relax m{max}}(0)$ for $\epsilon _{\nu }$. As
  $\kappa ^{\protect \text {NPB}}_{+}(0)=\eta _F$ (see equation \protect
  \textup {\hbox {\mathsurround \z@ \protect \normalfont (\ignorespaces \ref
  {kappa_plus_lambda_func}\unskip \@@italiccorr )}}), and $\kappa ^{\protect
  \text {lepton}}_{\protect \qopname \relax m{max}}(0)=\epsilon _{\nu
  }+\protect \sqrt {\epsilon ^2_{\nu } - \setbox \z@ \hbox {\mathsurround \z@
  $\textstyle m$}\mathaccent "0365{m}^2_{\mu }}$ (see definition given in
  \protect \textup {\hbox {\mathsurround \z@ \protect \normalfont
  (\ignorespaces \ref {kappa_max_lepton}\unskip \@@italiccorr )}}), the
  solution $\epsilon _{\nu _{+}}$ would have been obtained in a much simpler
  way.}\BibitemShut {Stop}%
\bibitem [{Note9()}]{Note9}%
  \BibitemOpen
  \bibinfo {note} {The reason for this statement is because $\kappa ^\protect
  \text {lepton}_{\protect \qopname \relax m{max}}(\lambda )$ is monotonically
  decreasing with $\lambda $ and $\kappa ^\protect \text {NPB}_{-}(\lambda )$
  is a monotonically increasing function of $\lambda $, but smaller than
  $\kappa ^\protect \text {NPB}_{+}(\lambda )$.}\BibitemShut {Stop}%
\bibitem [{Note10()}]{Note10}%
  \BibitemOpen
  \bibinfo {note} {For the demonstrations provided here, we drop the label NPB
  from all the expressions in order to shorten the already cumbersome
  notation}\BibitemShut {NoStop}%
\end{thebibliography}%
\end{document}